\documentclass[12pt,a4paper]{article} 

\usepackage[cp1251]{inputenc}
\usepackage[russian,ukrainian]{babel}

\usepackage{amsmath}
\usepackage{amssymb}

\usepackage[bookmarks=true,colorlinks=true,citecolor=blue,linkcolor=blue,urlcolor=magenta]{hyperref}

\usepackage[numbers,sort&compress]{natbib}

\usepackage{geometry} 
\usepackage{graphicx}

\begin{document}
\title{Багаточастинкові поля на підмножині одночасності}

\author{Пташинський Д.А.}
\author{Зеленцова Т.М.}
\author{Чудак Н.О.}
\author{Меркотан К.К.}
\author{Потієнко О.С.}
\author{Войтенко В.В.}
\author{Березовський О.Д.}
\author{Опятюк  В.В.}
\author{Жарова  О.В.}
\author{Юшкевич  О.В.}
\author{Шарф І.В.}
\author{Русов В.Д.}


%

\author{Пташинський~Д.А., Зеленцова~Т.М., Чудак~Н.О., Меркотан~К.К.,\\
        Потієнко~О.С., Войтенко~В.В., Березовський~О.Д., Опятюк~В.В.,\\
        Жарова~О.В., Юшкевич~Т.В., Шарф~І.В., Русов~В.Д.}

\date{}

\setcounter{page}{1}%

\maketitle

\begin{center}
\textit{Одеській Національний політехнічний університет,\\
Проспект Шевченка 1, Одеса, 65000, Україна}
\end{center}

\begin{abstract}
В роботі пропонується модель для опису процесів розсіяння адронів як зв'язаних станів конституентних кварків. Проведено аналіз ролі одночасності вимірів характеристик різних частинок при описі релятивістських квантових систем і проблеми неможливості однієї й тієї реалізації таких вимірів для різних інерційних систем відліку внаслідок відносності одночасності в релятивістській ситуації. На підставі цього налізу робиться висновок, що релятивістське розсіяння зв'язаних станів частинок потрібно описувати полями, що визначені не на просторі Мінковського, а на підмножині одночасних подій, яка виділяється з тензорного добутку двох просторів Мінковського. Такі поля названі багаточастинковими полями. При цьому аргументи польових функцій в різних системах відліку неможливо пов'язати ані перетвореннями Лоренца, ані якимось іншим чином. Показано, що таке положення не містить протиріччя із принципом відносності. На підмножині одночасності розглядається побудова динамічних рівнянь для багаточастинкових полів за допомогою методу Лагранжа, аналогічно тому як це робиться для \mbox{<<звичайних>>} одночастинкових полів. Розглянуто калібрувальні поля, які відновлюють локальну внутрішню симетрію на підмножині одночасності. Оскільки багаточастинкові поля, що описують мезони як зв'язані стани кварка і антикварка є двоіндексними тензорами відносно локальної калібрувальної групи, виникає можливість розглядати модель з двома різними калібрувальними полями, кожне з яких пов'язане із своїм індексом. Такі поля перетворюються за однаковим законом при локальному калібрувальному перетворенні і задовольняють однаковим динамічним рівнянням, але на них накладаються різні граничні умови. Показано, що ці поля задовольняють рівнянням зв'язку, які призводять до фізично зрозумілого наслідку, який полягає в тому, що паралельний зсув безкольорового адрона як цілого не повинен породжувати поля. Однак при зсуві окремих частинок адрона таке поле може виникати. Показано що динамічні рівняння для багаточастинкових калібрувальних полів при певних граничних умовах описують такі фізичні явища як конфайнмент і асимптотичну свободу кольорових об'єктів, а при інших граничних умовах - механізм спонтанного порушення симетрії. Ці динамічні рівняння дозволяють в межах однієї й тієї ж моделі описати як утримання кварків всередині адронів, так і їх взаємодію в процесах розсіяння адронів, шляхом обміном зв'язаними станами глюонів -- глюболами.
\end{abstract}

\section{Вступ}

Мабуть вперше, ідея багаточастинкових полів була запропонована Х.Юкавою \cite{Yukawa_1949_PhysRev.76.300.2, Yukawa_p1_PhysRev.77.219, Yukawa_p2_PhysRev.80.1047}. Юкава називав ці поля  <<нелокальними>>. Ми використовуємо інший термін <<багаточастинквові поля>> щоб підкреслити відмінності моделі що пропонується від моделі Юкави. Найбільш принциповою відмінністю моделі що пропонується, не тільки від моделі Юкави а й від моделей на світловому конусі \cite{Dirac:1949cp, Heinzl:2000ht}, квазіпотенціальних моделей   \cite{Logunov1963_NovCim, Logunov1955_NovCim, Faustov1973176}, моделей з багаточасовими амплітудами ймовірностей \cite{Tomonaga,DiracFockPodolsky,PETRAT201417} є те, що на нашу думку, внутрішні змінні таких полів в різних інерційних системах відліку неможливо пов'язати між собою ніяким чином, в той час як в згаданих моделях вважається що вони пов'язані перетвореннями Лоренца. Свою точку зору ми вже частково обгрунтовували в попередній роботі \cite{Chudak_2016UJP}. Іншими словами цю відмінність можна сформулювати таким чином, що наш підхід передбачає, що амплітуда ймовірності, віднесена до будь-якої інерційної системи відліку, повинна описувати вимірювання одночасні відносно цієї системи відліку, в той час як в згаданих моделях ця одночасність забезпечується лише по відношенню до певних систем відліку і не забезпечується відносно решти систем.
На нашу думку використання багаточасових амплітуд ймовірності, яке є явним, наприклад, в роботах  \cite{Tomonaga, DiracFockPodolsky, Marx1972, SazdjianPhysRevD.33.3401, PETRAT201417, Petrat20130632} і інших роботах цього напряму,і неявним в роботах тих напрямів, про які йшла мова вище, і, зокрема, в згаданих роботах Юкави, означає перехід до якоїсь \mbox{<<нової>>} квантової механіки з якимись \mbox{<<новими>>} постулатами, які не містяться в \mbox{<<звичайній>>} квантовій механіці \cite{DiracPrincipi}. Окрім того, на наш погляд така \mbox{<<нова>>} квантова механіка суперечить \mbox{<<звичайній>>}, бо не враховує, на нашу думку, впливу вимірювального приладу на стан мікросистеми. Пояснимо це більш докладно.

Розглянемо спочатку систему, що складається з однієї релятивістської мікрочастинки. Такий стан позначатимемо $\left| {{\Psi }^{\left( 1 \right)}} \right\rangle $. Він описуватиметься фоківським стовпцем, у якого відмінна від нуля лише одночастинкова компонента. Якщо розглядати цей стан в координатному представленні, то ця компонента представляється функцією ${{\Psi }^{\left( 1 \right)}}\left( t,\vec{r} \right)$, значення якої є коефіцієнтами розкладу:
\begin{equation}\label{Rozclad_odnochastincovogo_stanu}
\left| {{\Psi }^{\left( 1 \right)}} \right\rangle =\int{{{\Psi }^{\left( 1 \right)}}\left( t,\vec{r} \right)}\left| {\vec{r}} \right\rangle d\vec{r}.
\end{equation}
Тут $\left| {\vec{r}} \right\rangle $, позначено одночастинковий стан, власний для радіус-вектору частинки і відповідаючий власному значенню  $\vec{r}$.  Якщо в деякий момент часу $t$ ми внаслідок вимірювання спостерігаємо частинку в околі $d{{\vec{r}}^{\left( 0 \right)}}$ точки з радіус-вектором $\vec{r}^{\left( 0 \right)},$ то ми миттєво, тобто в той же момет часу $t$, узнаємо, що частинки немає в околах решти точок області, в якій вона могла спостерігатися. Це означає  \cite{DiracPrincipi}, що з розкладу (\ref{Rozclad_odnochastincovogo_stanu}), впливом вимірювального приладу миттєво вилучаються всі власні стани, окрім тих, які відповідають власним значенням що потрапляють в об'єм $d{{\vec{r}}^{\left( 0 \right)}}$ навколо точки з радіус-вектором $\vec{r}^{\left( 0 \right)}.$ Така миттєвість не суперечить принципам теорії відносності внаслідок суттєво нелокального характеру взаємодії стану частинки з вимірювальним приладом. Дійсно, якщо мова йде про опис стану мікрочастинки в координатному представленні, то цей опис передбачає результат такої взаємодії цього стану з вимірювальним приладом, яка надає потенційну можливість виявити частинку при вимірюванні в околі довільної точки деякої області. Тому така взаємодія не може бути локалізована в околі якоїсь однієї точки. Оскільки ми розглядаємо релятивістську мікрочастинку, то будь-яка взаємодія, в якій вона бере участь, може здійснюватись лише посередництвом відповідного поля, яке знову ж таки не локалізоване в деякій точці. В тому числі й взаємодія з вимірювальним приладом.  Оскільки вимірювальна взаємодія не локалізована, кожен з власних станів $\left| {\vec{r}} \right\rangle $ з розкладу (\ref{Rozclad_odnochastincovogo_stanu}) взаємодіє з вимірювальним полем в \mbox{<<своїй>>} точці, і ця взаємодія нікуди не повинна розповсюджуватися. Відповідно, скінченність швидкості розповсюдження взаємодії не відіграє в процесі нелокального вимірювання такої ролі яка б заперечувала можливість миттєвого вилучення з розкладу (\ref{Rozclad_odnochastincovogo_stanu}), внаслідок вимірювального впливу, тих власних станів, які не  відповідають власним значенням що потрапляють в об'єм $d{{\vec{r}}^{\left( 0 \right)}}$ навколо точки з радіус-вектором $\vec{r}^{\left( 0 \right)}.$ Зазначимо, що якщо замість координатного представлення стану розглянути імпульсне, то просторова локалізація взаємодії мікрочастинки із вимірювальним приладом взагалі є виключеною.

Розглянемо тепер для прикладу, систему двох нетотожних взаємодіючих між собою мікрочастинок, які для зручності назвемо $A$ і $B.$ Припустимо, що по відношенню до деякої системи відліку, ми провели одночасне вимірювання радіус-векторів цих частинок і спостерігали їх в околах точок з радіус-векторами відповідно ${\vec{r}}^{\left( 0 \right)}_{A}$ і ${\vec{r}}^{\left( 0 \right)}_{B}$   в момент часу $t.$ Внаслідок перетворень Лоренца, з точки зору спостерігача, що користується іншою інерційною системою відліку, ці виміри будуть неодночасними і характеризуватимуться різними моментами часу ${{{t}'}_{A}}$ і ${{{t}'}_{B}}$ і іншими координатами точок ${\vec{r}}'^{\left( 0 \right)}_{A}$ і ${\vec{r}}'^{\left( 0 \right)}_{B}.$ При цьому відношення відстані $\left|{\vec{r}}'^{\left( 0 \right)}_{A}-{\vec{r}}'^{\left( 0 \right)}_{B} \right|$ до проміжку часу $\left| {{{{t}'}}_{A}}-{{{{t}'}}_{B}} \right|$ перевищуватиме граничну швидкість розповсюдження взаємодії $c$ (тут і далі ми користуватимемось системою одиниць в якій $c=1$). Але внаслідок принципів квантової механіки ми не можемо робити висновок що більш раннє вимірювання не впливає на більш пізнє. Дійсно, припустимо, наприклад, що ${{{t}'}_{A}}<{{{t}'}_{B}}$. Якщо в момент ${{{t}'}_{A}}$ спостерігач виявляє частинку $A$ в околі точки  ${\vec{r}}'^{\left( 0 \right)}_{A},$ то, як вже зазначалося, вимірювальна взаємодія миттєво вилучить з розкладу стану деякі базисні елементи. А саме,  якщо до вимірювання стан системи 
$\left| {{\Psi }^{\left( 2 \right)}} \right\rangle $можна було виразити через власні для радіус-векторів стани $\left| {{{\vec{r}'}}_{A}} \right\rangle $ і $\left| {{{\vec{r}'}}_{B}} \right\rangle $ таким чим чином  
\begin{equation}\label{Rozklad_po_vlasnih_stanah_do_vimiruvanna}
\begin{split}
& \left| {{\Psi }^{\left( 2 \right)}} \right\rangle =\int{\Psi \left( {t}',{{{{\vec{r}}'}}_{A}},{{{{\vec{r}}'}}_{B}} \right)\left| {{{{\vec{r}}'}}_{A}} \right\rangle \left| {{{{\vec{r}}'}}_{B}} \right\rangle }d{{{\vec{r}}'}_{A}}d{{{\vec{r}}'}_{B}},\\
& {t}'\le {{{t}'}_{A}}.
\end{split}
\end{equation}
 Тут $d{{\vec{r}'}_{A}}$ і $d{{\vec{r}'}_{B}}$ позначені елементи об'ємів конфігураційних просторів частинок  $A$ і $B.$ То після вимірювання, вплив цього вимірювання вилучає з лінійної комбінації (\ref{Rozklad_po_vlasnih_stanah_do_vimiruvanna}) всі внески що містять стани ${{\vec{r}'}_{A}}$, шо не відповідають власним значенням, які потрапляють в окіл $d{{{\vec{r}}'}_{A}}$ навколо значення  ${\vec{r}}'^{\left( 0 \right)}_{A},$ тобто той окіл, в якому при вимірюванні була виявлена частинка $A$:
 \begin{equation}\label{Rozklad_po_vlasnih_stanah_posle_vimiruvanna}
 \begin{split}
&\left| {{\Psi }^{\left( 2 \right)}} \right\rangle =\left| {{\vec{r}}'^{\left( 0 \right)}_{A}} \right\rangle d{{{\vec{r}}'}_{A}}\int{\Psi \left( {t}',{\vec{r}}'^{\left( 0 \right)}_{A},{{{{r}'}}_{B}} \right)\left| {{{{\vec{r}}'}}_{B}} \right\rangle }d{{{\vec{r}}'}_{B}},\\
& {{{t}'}_{A}}<{t}'\le {{{t}'}_{B}}
 \end{split}
 \end{equation}
Навіть якщо інтервал між подіями $\left( {{{{t}'}}_{A}},{{{{\vec{r}}'}}_{A}} \right)$ і $\left( {{{{t}'}}_{B}},{{{{\vec{r}}'}}_{B}} \right)$ є простороподібний, з лінійної комбінації (\ref{Rozklad_po_vlasnih_stanah_do_vimiruvanna}) вимірювальним впливом миттєво вилучаються власні стани 
$\left| {{{{\vec{r}}'}}_{A}} \right\rangle $ з ${{{\vec{r}}'}_{A}}$ достатньо близькими до $ {{\vec{r}}'^{\left( 0 \right)}_{B}}  $ так що 
\begin{equation}\label{Umova}
{\left| {{{{\vec{r}}'}}_{A}}-{{\vec{r}}'^{\left( 0 \right)}_{B}} \right|}/{\left| {{{{t}'}}_{A}}-{{{{t}'}}_{B}} \right|}\;\le 1
\end{equation}	
Тому у випадку коли б вимірювання не було і власні стани ${{{\vec{r}}'}_{A}}$ були б присутні в розкладі (\ref{Rozklad_po_vlasnih_stanah_posle_vimiruvanna}), то можливі були б віртуальні обміни переносниками взаємодії між точками 
${{{\vec{r}}'}_{A}}$ і $ {{\vec{r}}'^{\left( 0 \right)}_{B}} $ і ці обміни впливали б на розвиток стану в часі. А внаслідок вимірювання такі стани вилучені з розкладу (\ref{Rozklad_po_vlasnih_stanah_posle_vimiruvanna}) і відповідних обмінів не відбувається і, таким чином розвиток стану із часом внаслідок впливу вимірювання стає іншим. З огляду на наведені міркуання щодо нелокальності взаємодії стану із вимірювальним приладом нам видається що виділення просторово-подібних конфігурацій \cite{Tomonaga, PETRAT201417} при багаточасовому описі не має сенсу. До того ж, виникає питання як побудувати подібний опис в імпульсному представленні, де розглянута вище просторова нелокальність проявляється вже навіть на формальному рівні. В такому випадку неможливо сформулювати умову, аналогічну умові простороподібності. Окрім того, незрозуміло як за виконання такої умови переходити до імпульсного представлення. Якщо за фіксованих часових змінних проводити перетворення Фур'є по просторових змінних, то інтегрування не обмежиться областю, на якій всі аргументи багаточасової амплітуди ймовірності розділені просторово-подібними інтервалами і тоді на частині такої області інтегрування підінтегральний вираз буде невизначений. Окрім того, якщо розглядати амплітуду ймовірності, залежну від часів і імпульсів окремих частинок, то зникає навіть формальна можливість пов'язувати аргументи такої амплітуди ймовірності в різних системах відліку перетвореннями Лренца. З іншого боку, якщо ми маємо систему квантових взаємодіючих релятивістських частинок, то ніщо не заважає кожному інерційному спостерігачеві проводити вимір імпульсів окремих частинок. Якщо ж розглянути перетворення Фур'є для багаточасової амплітуди ймовірності не тільки по просторових змінних а й по часових, то окрім того, що знов на частині області інтегрування підінтегральний вираз не буде визначений, поряд із часовими змінними для кожної частинки ми ще матимемо енергетичні змінні також для кожної частинки окремо, в той час як в системі взаємодіючих частинок маємо лише власні значення енергії всієї системи, а не окремих частинок.

Отже, у зв'язку з розглядом багаточасових вимірювань, на наш погляд, виникають наступні проблеми. В \mbox{<<звичайній>>} квантовій теорії взаємодія з вимірювальним приладом проявляється лише в результатах реалізацій процесу вимірювання і не впливає на розвиток стану в часі і, відтак, вона не повинна явно описуватися в динамічних рівняннях, що описують цей розвиток. У випадку розгляду бгаточасових вимірювань маємо протилежну ситуацію - взаємодія із приладом змінюватиме часову еволюцію стану і повинна явно описуватися в динамічних рівняннях. Окрім того, результат багаточасового вимірювання можна передбачити за допомогою \mbox{<<звичайної>>} одночасової амплітуди ймовірності, послідовно діючи на неї оператором часової еволюції, який враховуватиме в тому числі й взаємодію із приладом. Дійсно, розглянемо систему з $n$ нетотожних взаємодіючих частинок. Якщо в момент часу ${{t}_{1}}$, стан такої системи описується одночасовою амплітудою ймовірності $\Psi \left( {{t}_{1}},{{{\vec{r}}}_{1}},\ldots ,{{{\vec{r}}}_{n}} \right)$, де ${{\vec{r}}_{1}},\ldots ,{{\vec{r}}_{n}}$- радіус-вектори частинок, то проінтегрувавши квадрат модуля амплітуди ймовірності по всіх радіус-векторах окрім ${{\vec{r}}_{1}}$ і множачи результат на малий елемент об'єму $d{{V}_{1}}$ отримаємо ймовірність виявити при вимірюванні в момент часу ${{t}_{1}}$ першу частинку в об'ємі  $d{{V}_{1}}$, описаному навколо  точки із радіус-вектором ${{\vec{r}}_{1}}$. Розглянемо стан системи після впливу цього вимірювання, і подіємо на амплітуду ймовірності цього стану оператором часової еволюції до моменту ${{t}_{2}}$.  Розрахуємо квадрат модуля  отриманої таким чином функції, проінтегруємо його по всіх радіус-векторах за винятком  ${{\vec{r}}_{2}}$ і помножимо на малий об'єм $d{{V}_{2}}$. Отриману таким чином величину можна інтерпретувати як умовну ймовірність виявити при вимірюванні в момент часу ${{t}_{2}}$ другу частинку в малому об'ємі $d{{V}_{2}}$ навколо точки з радіус-вектором ${{\vec{r}}_{2}}$ за умови що в момент часу  ${{t}_{1}}$ перша частинка була виявлена в об'ємі $d{{V}_{1}}$  навколо вектора ${{\vec{r}}_{1}}.$ Множачи цю умовну ймовірність на ймовірність виявити в момент часу ${{t}_{1}}$ першу частинку в об'ємі $d{{V}_{1}}$ отримаємо сумісну ймовірність результатів вимірювання в моменти часу  ${{t}_{1}}$ і ${{t}_{2}}.$ Далі врахуємо вимірювальний плив на стан в момент  ${{t}_{2}}$ і на отриману таким чином амплітуду ймовірності подіємо оператором часової еволюції до моменту часу ${{t}_{3}}$ і продовжуючи описану процедуру матимемо сумісну ймовірність отримати частинки в малих об'ємах  $d{{V}_{1}},d{{V}_{2}},\ldots ,d{{V}_{n}}$, описаних навколо радіус-векторів ${{\vec{r}}_{1}},\ldots ,{{\vec{r}}_{n}}$ при вимірюваннях у  відповідні моменти часу ${{t}_{1}},{{t}_{2}},\ldots ,{{t}_{n}}.$ Ми детально описали цю процедуру, щоб підкреслити, що вона не зводиться до розрахунку квадрату модуля якоїсь багаточасової амплітуди ймовірності. Відтак, введення таких амплітуд ймовірності, на нашу думку, означає перехід до якоїсь \mbox{<<нової>>} квантової механіки.   

Наведені міркування показують, що в системах відліку, відносно яких вимірювання проводяться одночасно і в тих системах, в яких ци вимірювання неодночасні, маємо суттєво різний квантово-механічний опис. В цьому ми вбачаємо порушення принципу відносності. З огляду на цей принцип виникає ще одне, важливе на наш погляд, питання. Якщо деякий інерційний спостерігач проведе одночасне відносно себе вимірювання координат частинок, то він згенерує своїм вимірюанням певні події, які полягатимуть в тому що в момент він виявив частинки в околах точок із радіус-векторами ${{\vec{r}}_{1}},{{\vec{r}}_{2}},\ldots ,{{\vec{r}}_{n}}.$ Ці події в системі відліку, повязаній із спостерігачем, що розглядаються, мають однакову часову координату $t$ і просторові координати, що відповідають переліченим радіус-векторам. Розглянемо тепер іншого інерційного спостерігача. Якщо він скористується тією ж реалізацією вимірювань, то в його системі відліку події матимуть різні часові координати. Але ящо він не користуватиметья результатами вимірювання іншого інерційного спостерігача, а проведе свої вимірювання, то чи повинен він їх проводити одночасно відносно себе, або він повинен проводити різночасові вимірювання? На наш погляд і з наведених вище міркувань і з принипу відносності випливає, що цей спсостерігач повинен проводити вимірювання одночасно відносно своєї системи відліку. Але тоді кожен спостерігач повинен користуватися своїми реалізаціями процесу вимірювання які генерують різні події для різних спостерігачів. Оскільки перетворення Лоренца пов'язують просторово-часові координати однієї події в різних системах відліку, то аргументи функцій що складають компоненти фоківського стовпця в різних системах відліку не пов'язані перетворенням Лоренца і взагалі не пов'язані якимось чином бо мова йде про просторово-часові координати різних подій. При цьому, як було показано в роботах \cite{Chudak_2016UJP, Chudak:2016} відповідність із принципом відносності досягається не за рахунок Лоренц-інваріантності, а за рахунок того, що залежність від внутрішніх змінних є однаковою в різних інерційних системах відліку. Натомість залежність від змінних центру мас залишається Лоренц-інваріантною, але не за рахунок зв'язку між координатами в різних системах відліку, а за рахунок формальної аналогії \cite{Chudak_2016UJP} перетворення оератору часової еволюції з перетворенням польових операторів  \cite{Bogolyubov_rus}.

Отже, якщо ми розглянемо $n-$частинкову компоненту фоківського стовпця відносно певної інерційної системи відліку в момент часу $t$, то для того, щоб побудувати множину на якій вона визначена, ми можемо спочатку розглянути декартовий  добуток  $n$ просторів Мінковського, а потім виділити підмножину цього добутку, яка утворена сукупностями $n$ подій у яких чаова координата однакова і дорівнює $t.$ Такі підмножини вже розглядалися в роботі \cite{Chudak:2016}. Називатимемо їх далі підмножинами одночасності.

Наведені міркування не мають суттєвого прояву в існуючих польових теоріях внаслідок того, що всі ці теорії формулюються в термінах чисел заповнення одночастинкових станів невзаємодіючих частинок. Дійсно, для отримання комутаційних співвідношень, за допомогою яких польовим операторам в імпульсному представленні  надається сенс операторів народження і знищення, суттєво щоб оператор енергії-імпульсу був квадратичний по цих операторах \cite{Bogolyubov_rus}. Квантування за допомогою функціональних інтегралів \cite{Berezin:1986VtorKv} не змінює цієї ситуації, бо для побудови виробляючого функціоналу для оператора часової еволюції потрібен виробляючий функціонал для лагранжіану взаємодії в нормальній формі, тобто знову ж таки потрібна інтерпретація польових операторів як операторів народження і знищення. Тому кванування одночастинкових польових теорій відбувається в представленні взаємодії \cite{Bogolyubov_rus}, в якому залежність польових операторів від їх аргументів така сама як в операторів вільних полів. Тож всі ефекти, пов'язані із взаємодією зводяться лише до зміни чисел заповнення одночастинкових станів вільних частинок. Такий підхід пристосовний для опису процесів в яких взаємодію між полями можна вважати такою що відсутня в початковому і кінцевому станах. Тоді  числа заповнення одночастинкових станів наближаються до певних граничних значень в початковому і кінцевому станах. Натомість для процесів розсіяння адронів в початковому і кінцевому станах взаємодія відсутня саме для адронів. Тому числа заповнення одночастинкових кваркових і глюонних станів не наближатимуться до певних границь. 

Окрім того, в наявних теоріях польові оператори задані на просторі Мінковського і їх аргументи в різних системах відліку пов'язані перетвореннями Лоренца. Внаслідок цього ми не можемо розглядати оператор, який би народжував або знищував частинку із певним імпульсом але не з певною енергією. Іншими словами, на відміну від нерелятивістської теорії, неможливо побудувати оператор народження або знищення, який би діяв на координатну, або імпульсну залежність елементів фоківського стану і при цьому не змінював би їх залежності від часу. Відтак, розглядаючи оператори що змінюють числа заповнення одночастинкових станів ми неодмінно повинні розглядати одночастинкові енергії. Лагранжіан взаємодії побудований таким чином, що в процесі зміни чисел заповнення одночастинкових станів зберігатиметься сума одночастинкових енергій. Відтак, ця властивість переходить і до оператору часової еволюції і до оператору розсіяння. В той же час, коли ми розглядаємо зв'язану систему кварків і глюонів, таку як адрон то маємо лише власне значення енергії всієї системи, а одночастинкові енергії навіть неможливо ввести. Відповідно і в експерименті маємо закон збереження енергії-імпульсу саме для адронів, а не для кварків і глюонів. 

Оскільки маємо лише енергію системи в цілому то і зсув в часі повинен розглядатися лише для системи в цілому, а відтак, і час повинен бути один і той самий для всієї системи, а не для кожної частинки окремо. Наведені міркування пояснюють мотивацію розгляду не одночастинкових польових операторів на просторі Мінковського, а  польових операторів, які змінюють числа заповнення багаточастинкових станів (адронів) і які задані на підмножині одночасності. В роботі \cite{Chudak:2016} такі багаточастинкові поля розглядалися. Однак методи отримання динамічних рівнянь для багаточастинкових полів, на наш погляд, потребують суттєвого вдосконалення. Зокрема такі рівняння формулювалися за допомогою одночастинкових рівнянь. Спочатку вони розглядалися на тензорному добутку декількох просторів Мінковського, з послідуючим переходом на підмножину одночасності. Окрім того, для побудови динамічних рівнянь для калібрувальних полів такий метод вимагав формулювання додаткових припущень для замикання рівнянь. В цій роботі ми пропонуємо метод побудови динамічних рівнянь багаточастинкових полів без звертання до одночастинкових рівнянь і відразу на підмножині одночастності. Цей метод полягає в тому що на підмножині одночасності ми вводимо скалярний добуток. Далі застосовується звичайний польовий метод, що базується на принципі найменшої дії і, відповідно рівняннях Ларнажа-Ейлера для полів, заданих на підмножині одночасності. Рівняння для калібрувальних полів отримуються без припущень роботи \cite{Chudak:2016}. Вони дещо відрізняються від рівнянь, що розглядалися в цій роботі а також в \cite{Mercotan2017}. Однак ці нові рівняння відтворюють ті ж фізичні результати, що розглядалися в цих роботах. 

\section{Скалярний добуток на підмножині одночасності}
Розглянемо спочатку двочастинкову підмножину одночасності. При цьому розглядаючи двочастинкову систему ми насамперед матимемо на увазі мезон, як частинку, що складається з конституентних кварка і антикварка. Час і координати події, що може відбутися із першою частинкою позначатимемо $\left( x_{\left( 1 \right)}^{0},x_{\left( 1 \right)}^{1},x_{\left( 1 \right)}^{2},x_{\left( 1 \right)}^{3} \right) $, а з другою $ \left( x_{\left( 2 \right)}^{0},x_{\left( 2 \right)}^{1},x_{\left( 2 \right)}^{2},x_{\left( 2 \right)}^{3} \right) .$ Тут як зазвичай індекс 0 позначає часову координату події, а 1,2,3 - просторові координати. Нижні індекси  дужках ідентифікують першу і другу частинки. Дужки використовуються щоб відрізнити ці індекси від коваріантних координат події. Верхні індекси використовуються для позначення контраваріантних координат. Декартовий добуток просторів Мінковського для двох частинок є восьмивимірним лінійним простором, елементи якого можна розглядати як стовпці
\begin{equation}
\begin{split}
{{z}^{a}}=\left( \begin{matrix}
x_{\left( 1 \right)}^{0}  \\
x_{\left( 1 \right)}^{1}  \\
x_{\left( 1 \right)}^{2}  \\
x_{\left( 1 \right)}^{3}  \\
x_{\left( 2 \right)}^{0}  \\
x_{\left( 2 \right)}^{1}  \\
x_{\left( 2 \right)}^{2}  \\
x_{\left( 2 \right)}^{3}  \\
\end{matrix} \right).
\end{split}
\label{eq:Stovpec}
\end{equation}
 Скалярний добуток на цьому восьмивимірному просторі введемо за допомогою співвідношення 
 \begin{equation}\label{Scalzrnij_dobutok}
 \left\langle  z | z \right\rangle =\frac{1}{2}\left( g_{ab}^{Minc}x_{\left( 1 \right)}^{a}x_{\left( 1 \right)}^{b}+g_{ab}^{Minc}x_{\left( 2 \right)}^{a}x_{\left( 2 \right)}^{b} \right).
 \end{equation}
Тут $ g_{ab}^{Minc}$ позначає тензор Мінковського. По індексах $a$ і $b$ що повторюються мається на увазі підсумування і при цьому кожен з цих індексів пробігатиме значення 0,1,2,3. Далі зручно ввести координати Якобі
\begin{equation}\label{Coordinati_Jacobi}
{{X}^{a}}=\frac{1}{2}\left( x_{\left( 1 \right)}^{a}+x_{\left( 2 \right)}^{a} \right),{{y}^{a}}=x_{\left( 2 \right)}^{a}-x_{\left( 1 \right)}^{a}.
\end{equation}
З урахуванням (\ref{Coordinati_Jacobi}) вираз для скалярного добутку (\ref{Scalzrnij_dobutok}) прийме вид
\begin{equation}\label{Scalarnij_dobutoc_Jacobi}
\left\langle  z | z \right\rangle =g_{ab}^{Minc}\left( {{X}^{a}}{{X}^{b}}+\frac{1}{4}{{y}^{a}}{{y}^{b}} \right)
\end{equation}
Підмножина одночасності виділятиметься в координатах (\ref{Coordinati_Jacobi}) умовою 
\begin{equation}\label{Umova_odnochasnosti}
{{y}^{0}}=0.
\end{equation}
Координати точки на підмножині одночасності позначатимемо семикомпонентним стовпцем
\begin{equation}
\begin{split}
{{q}^{a}}=\left( \begin{matrix}
{{X}^{0}}  \\
{{X}^{1}}  \\
{{X}^{2}}  \\
{{X}^{3}}  \\
{{y}^{1}}  \\
{{y}^{2}}  \\
{{y}^{3}}  \\
\end{matrix} \right).
\end{split}
\label{eq:Stovpec_sim_odnochasnist}
\end{equation}
Скалярний добуток на підмножині одночасності визначимо так щоб він збігався з добутком (\ref{Scalarnij_dobutoc_Jacobi}) з урахуванням умови (\ref{Umova_odnochasnosti}):
\begin{equation}\label{Scal_dobutoc_odnochasnist}
\left\langle  q | q \right\rangle ={{g}_{ab}}{{q}^{a}}{{q}^{b}},
\end{equation}
де метричний тензор що входить в цей вираз має вид 
\begin{equation}
\begin{split}
{{g}^{ab}}=\left( \begin{matrix}
1 & 0 & 0 & 0 & 0 & 0 & 0  \\
0 & -1 & 0 & 0 & 0 & 0 & 0  \\
0 & 0 & -1 & 0 & 0 & 0 & 0  \\
0 & 0 & 0 & -1 & 0 & 0 & 0  \\
0 & 0 & 0 & 0 & -4 & 0 & 0  \\
0 & 0 & 0 & 0 & 0 & -4 & 0  \\
0 & 0 & 0 & 0 & 0 & 0 & -4  \\
\end{matrix} \right).
\end{split}
\label{eq:Metricnij_tenzor_odnochasnist}
\end{equation}
Багаточастинкове поле описуватиметься сукупністю польових функцій ${{\Psi }_{a}}\left( q \right)={{\Psi }_{a}}\left( X,\vec{y} \right).$ Тут через $X$ позначена сукупність координат ${{X}^{0}},{{X}^{1}},{{X}^{2}},{{X}^{3}},$ а через $\vec{y}$ - сукупність внутрішніх змінних ${{y}^{1}},{{y}^{2}},{{y}^{3}}.$ Індекс $a$ нумерує різні компоненти поля і множина його значень визначається представленням групи перетворень, які описують перехід від польових функцій, визначених відносно однієї системи відліку до польових функцій, визначених відносно іншої системи відліку. Розглянемо цю групу перетворень. При цьому враховуватимемо, що як вже зазначалося, координати на підмножинах одночасності в різних системах відліку не пов'язуються між собою. Однак закон перетворення польових функцій при переході від однієї інерційної системи відліку до іншої отримується з інших міркувань. Оскільки квантування багаточастинкових полів буде розглядатися в представленні взаємодії відносно взаємодії саме між багаточастинковими полями (але не між частинками, що утворюють зв'язані стани народження і знищення яких описують квантовані багаточастинкові поля), ми можемо розглядати випадок, коли залежність від внутрішніх змінних $\vec{y}$ відділяється від залежності від координат центру мас $X$ 
\begin{equation}\label{rozdslenna_zminnih}
{{\Psi }_{a}}\left( X,\vec{y} \right)={{\phi }_{a}}\left( X \right)\psi \left( {\vec{y}} \right).
\end{equation}
Множник $\psi \left( {\vec{y}} \right)$ можна визначити як власну функцію внутрішнього гамільтоніану системи взаємодіючих частинок, що утворюють зв'язаний стан. Цей внутрішній гамільтоніан можна визначити як такий оператор, квадрат котрого дорівнює величині ${{P}^{2}}\equiv g_{ab}^{Minc}{{\hat{P}}^{a}}{{\hat{P}}^{b}}$, де ${{\hat{P}}^{a}}-$ генератори просторово-часових зсувів. Відповідно, внутрішній гамільтоніан ${{\hat{H}}^{\text{inernal}}}$  визначатимемо як оператор, власні функції якого збігаються з власними функціями ${{P}^{2}}$, а власні значення є квадратні корені з власних значень  ${{P}^{2}}.$ Якщо ми розглянемо генератор буста ${{\hat{M}}_{0d}},d=1,2,3$ уздовж довільної вісі, то внаслідок комутаційних співвідношень між генераторами групи Пуанкаре такий генератор комутує з оператором ${{P}^{2}}.$ Але за наведеного означення він комутує і з ${{\hat{H}}^{\text{inernal}}}.$ Це означає, що якщо розглянути власний для внутрішнього гамільтоніану стан, такий що відповідає невиродженому власному значенню, то він обов'язково буде власним і для генератора ${{\hat{M}}_{0d}}.$ Якщо розглянути лінійний простір власних станів внутрішнього гамільтоніану, які відповідають деякому виродженому власному значенню, то обравши деякий базис можемо побудувати матрицю генератора ${{\hat{M}}_{0d}}$ на цьому просторі. Діагоналізувавши цю матрицю, отримаємо новий базис, який складатиметься із станів, власних і для внутрішнього гамільтоніану і для генератора ${{\hat{M}}_{0d}}.$ Розглянемо такий стан $\psi \left( {\vec{y}} \right).$ Позначимо $\hat{U}\left( -t \right)$ унітарний оператор, який описує перетворення станів при часовій інверсії. Оскільки внутрішній стан не залежить від часу, він не змінюватиметься при часовій інверсії. Тому якщо перетворення часової інверсії реалізувати як перетворення станів при незмінних операторах, то власне значення оператора  ${{\hat{M}}_{0d}},$ яке відповідає власному стану  $\psi \left( {\vec{y}} \right)$ не зміниться. Якщо ж перетворювати оператори, не змінюючи стани, то оператор, а разом із ним і власне значення повинні змінити знак. Звідси маємо що таке власне значення дорівнює нулю. Це означає що власні стани внутрішнього гамільтоніану не змінюються при переході від однієї інерційної системи відліку до іншої.
Таким чином, на множині значень функції $\psi \left( {\vec{y}} \right)$ буст представлятиметься одиничним оператором. Окрім того, оскільки внутрішні координати $\vec{y}$ в різних системах відліку у випадку буста не пов'язуються між собою ніяким чином, ми можемо лише порівнювати значення функцій $\psi \left( {\vec{y}} \right)$ в різних системах відліку лише при одних і тих самих значеннях аргументів, а не в однієї просторово-часовій точці, як це відбивається у випадку \mbox{<<звичайних>>} одночастинкових полів. Тому і на області визначення функцій  $\psi \left( {\vec{y}} \right),$ буст також предствлятиметься одиничним оператором. Щодо підгрупи обертів при переході з однієї системи відліку до іншої, то в цьому випадку ми не маємо зазначених вище проблем з одночасністю, тому і на області визначення і на області значень функції $\psi \left( {\vec{y}} \right)$ оберти представляються звичайним чином. Оскільки бусти представляються одиничним оператором, то на підпросторі внутрішніх змінних перехід від однієї інерційної системи відліку до іншої представляється групою обертів.  

Функції ${{\phi }_{a}}\left( X \right)$ з виразу (\ref{rozdslenna_zminnih}) формально співпадають із \mbox{<<звичайними>>} одночастинковими польовими функціями. Після квантування вони заміняться на відповідні польові оператори ${{\hat{\phi }}_{a}}\left( X \right)$. Якщо позначити $\hat{U}\left( {\hat{\Lambda }} \right)$ унітарний оператор, що представляє перетворення Лоренца $\hat{\Lambda }$ на тому просторі станів, на якому визначені оператори ${{\hat{\phi }}_{a}}\left( X \right),$ то при переході до нової системи відліку, виходячи з фомальної аналогії з перетвореннм одночастинкових польових операторів, матимемо \cite{Bogolyubov_rus}
\begin{equation}
\begin{split}
  & {{{{\hat{\phi }}'}}_{a}}\left( {{X}'} \right)={{{\hat{U}}}^{\dagger }}\left( \Lambda  \right){{{\hat{\phi }}}_{a}}\left( X \right)\hat{U}\left( \Lambda  \right)= \\ 
& ={{{\hat{D}}}_{ab}}\left( \Lambda  \right){{{\hat{\phi }}}_{b}}\left( X={{{\hat{\Lambda }}}^{-1}}{X}' \right). \\ 
\end{split}
\label{eq:Peretvorenna_fi_a}
\end{equation}
Тут, як зазвичай, ${{\hat{U}}^{\dagger }}\left( \Lambda  \right)$ позначає ермітово спряжений оператор, а ${{\hat{\Lambda }}^{-1}}-$обернену матрицю. Окрім того ${{\hat{D}}_{ab}}\left( \Lambda  \right)$ позначено елементи матриць, що реалізують відповідне представлення груи Лоренца на множині значень польових функцій ${{\phi }_{a}}\left( X \right).$ Наприклад для тричастинкового біспінорного поля \cite{Chudak:2016} матриці  ${{\hat{D}}_{ab}}\left( \Lambda  \right)$ реалізують біспінорне представлення групи Лоренца. По індексах, що повторюються мається на увазі підсумування. Як видно з (\ref{eq:Peretvorenna_fi_a}) сукупність чотирьох величин 
${{X}^{0}},{{X}^{1}},{{X}^{2}},{{X}^{3}}$  потрібно перетворювати при переході з однієї системи відліку до іншої як контраваріантний чотиривектор. Проте звернемо увагу на те, що такий закон перетворення виникає не внаслідок зв'язку між координатами в різних системах відліку (якого, як розглядалося раніше не існує), а внаслідок формальної аналогії між функціями ${{\phi }_{a}}\left( X \right)$ і одночастинковими польовими функціями. З фізичної точки зору такий закон перетворення означає що ми збираємось розглядати адрон, чи іншу складену частинку як точкову при аналізі її взаємодії із приладом, що вимірює положення адрону в просторі, але як ми покажемо далі, ми враховуватимемо його неточковість при аналізі взаємодії з калібрувальним полем, що фізично відповідатиме тому що з калібрувальним полем взаємодіє не власне адрон, а кварки, що його складають. 

Отже на підмножині одночасності діє група матриць 
\begin{equation}
\begin{split}
\hat{G}=\left( \begin{matrix}
\Lambda _{0}^{0} & \Lambda _{1}^{0} & \Lambda _{2}^{0} & \Lambda _{3}^{0} & 0 & 0 & 0  \\
\Lambda _{0}^{1} & \Lambda _{1}^{1} & \Lambda _{2}^{1} & \Lambda _{3}^{1} & 0 & 0 & 0  \\
\Lambda _{0}^{2} & \Lambda _{1}^{2} & \Lambda _{2}^{2} & \Lambda _{3}^{2} & 0 & 0 & 0  \\
\Lambda _{0}^{3} & \Lambda _{1}^{3} & \Lambda _{2}^{3} & \Lambda _{3}^{3} & 0 & 0 & 0  \\
0 & 0 & 0 & 0 & R_{1}^{1} & R_{2}^{1} & R_{3}^{1}  \\
0 & 0 & 0 & 0 & R_{1}^{2} & R_{2}^{2} & R_{3}^{2}  \\
0 & 0 & 0 & 0 & R_{1}^{3} & R_{2}^{3} & R_{3}^{3}  \\
\end{matrix} \right).
\end{split}
\label{eq:Gruppa_matric_G}
\end{equation}
Елементи матриці $G_{b}^{a}$ нумеруються індексами, кожен з яких приймає значення від 0 до 6. Числа $\Lambda _{b}^{a},a,b=0,1,2,3$ є елементами матриці перетворення Лоренца, а $R_{b}^{a},a,b=1,2,3-$елементи матриці обертання. При цьому якщо 
\begin{equation}
\begin{split}
  & \hat{\Lambda }=\left( \begin{matrix}
1 & 0 & 0 & 0  \\
0 & {{R}^{\left( 1 \right)}}_{1}^{1} & {{R}^{\left( 1 \right)}}_{2}^{1} & {{R}^{\left( 1 \right)}}_{3}^{1}  \\
0 & {{R}^{\left( 1 \right)}}_{1}^{2} & {{R}^{\left( 1 \right)}}_{2}^{2} & {{R}^{\left( 1 \right)}}_{3}^{2}  \\
0 & {{R}^{\left( 1 \right)}}_{1}^{3} & {{R}^{\left( 1 \right)}}_{2}^{3} & {{R}^{\left( 1 \right)}}_{3}^{3}  \\
\end{matrix} \right){{{\hat{\Lambda }}}_{0}}\times  \\ 
& \times \left( \begin{matrix}
1 & 0 & 0 & 0  \\
0 & {{R}^{\left( 2 \right)}}_{1}^{1} & {{R}^{\left( 2 \right)}}_{2}^{1} & {{R}^{\left( 2 \right)}}_{3}^{1}  \\
0 & {{R}^{\left( 2 \right)}}_{1}^{2} & {{R}^{\left( 2 \right)}}_{2}^{2} & {{R}^{\left( 2 \right)}}_{3}^{2}  \\
0 & {{R}^{\left( 2 \right)}}_{1}^{3} & {{R}^{\left( 2 \right)}}_{2}^{3} & {{R}^{\left( 2 \right)}}_{3}^{3}  \\
\end{matrix} \right), \\
\end{split}
\label{eq:Dva_oberti_i_bust}
\end{equation}
де ${{\hat{R}}^{\left( 1 \right)}}$ і ${{\hat{R}}^{\left( 2 \right)}}$- матриці тривимірних обертів, то матриця $\hat{R},$
яка входить до (\ref{eq:Gruppa_matric_G}) дорівнює 
\begin{equation}
\begin{split}
& \left( \begin{matrix}
R_{1}^{1} & R_{2}^{1} & R_{3}^{1}  \\
R_{1}^{2} & R_{2}^{2} & R_{3}^{2}  \\
R_{1}^{3} & R_{2}^{3} & R_{3}^{3}  \\
\end{matrix} \right)=\left( \begin{matrix}
{{R}^{\left( 1 \right)}}_{1}^{1} & {{R}^{\left( 1 \right)}}_{2}^{1} & {{R}^{\left( 1 \right)}}_{3}^{1}  \\
{{R}^{\left( 1 \right)}}_{1}^{2} & {{R}^{\left( 1 \right)}}_{2}^{2} & {{R}^{\left( 1 \right)}}_{3}^{2}  \\
{{R}^{\left( 1 \right)}}_{1}^{3} & {{R}^{\left( 1 \right)}}_{2}^{3} & {{R}^{\left( 1 \right)}}_{3}^{3}  \\
\end{matrix} \right)\times  \\ 
& \times \left( \begin{matrix}
{{R}^{\left( 2 \right)}}_{1}^{1} & {{R}^{\left( 2 \right)}}_{2}^{1} & {{R}^{\left( 2 \right)}}_{3}^{1}  \\
{{R}^{\left( 2 \right)}}_{1}^{2} & {{R}^{\left( 2 \right)}}_{2}^{2} & {{R}^{\left( 2 \right)}}_{3}^{2}  \\
{{R}^{\left( 2 \right)}}_{1}^{3} & {{R}^{\left( 2 \right)}}_{2}^{3} & {{R}^{\left( 2 \right)}}_{3}^{3}  \\
\end{matrix} \right). \\ 
\end{split}
\label{eq:MatricaR}
\end{equation} 

Скалярний добуток (\ref{Scal_dobutoc_odnochasnist}) з метричним тензором (\ref{eq:Metricnij_tenzor_odnochasnist}) є інваріантним відносно перетворень групи (\ref{eq:Gruppa_matric_G}). 

Отже нашою подальшою метою буде побудова квантової теорії поля не на просторі Мінковського з групою Лоренца, а на розглянутій вище підмножині одночасності з групою (\ref{eq:Gruppa_matric_G}). Як вже зазначалося, симетрія динамічних рівнянь теорії поля відносно цієї групи є формальним виразом принципу відносності. З іншого боку, якщо простір Мінковського замінити на підмножину одночасності, а групу Лоренца - на групу (\ref{eq:Gruppa_matric_G}), то далі таку теорію можна будувати з використанням міркувань, аналогічних тим, що використовуються при побудові \mbox{<<звичайної>>} одночастинкової теорії поля. 

\section{Лагранжіан двочастинкового мезонного поля}
Двочастинкове мезонне поле, яке після квантування описуватиме процеси народження і знищення зв'язаних станів кварка і антикварка позначатимемо $ {{\psi }_{{{c}_{1}}{{c}_{2}},{{f}_{1}},{{f}_{2}}}}\left( q \right). $. Тут $ q $ позначає сукупність семи величин (\ref{eq:Stovpec_sim_odnochasnist}). Індекси із субіндексом 1 будемо відносити до антикварка, а 2-до кварка. Так $ {c}_{1}$ позначатиме колір антикварка, а $ {c}_{2}$- колір кварка, ${f}_{1}- $ позначатиме аромат антикварка, а  ${f}_{2}- $ аромат кварка. Відповідно, поле $ {{\psi }_{{{c}_{1}}{{c}_{2}},{{f}_{1}},{{f}_{2}}}}\left( q \right) $ приймає значення на якому реалізуються змішані тензорні представлення груп $S{{U}_{c}}\left( 3 \right)$ і $S{{U}_{f}}\left( 3 \right)$:
\begin{equation}
 {{{\psi }'}_{{{c}_{1}}{{c}_{2}},{{f}_{1}},{{f}_{2}}}}\left( q \right)=u_{{{c}_{1}}{{c}_{3}}}^{\left( c \right)\dagger }u_{{{c}_{2}}{{c}_{4}}}^{\left( c \right)}u_{{{f}_{1}}{{f}_{3}}}^{\left( f \right)\dagger }u_{{{f}_{2}}{{f}_{4}}}^{\left( f \right)}{{\psi }_{{{c}_{3}}{{c}_{4}},{{f}_{3}},{{f}_{4}}}}\left( q \right).
\label{Tenzorni_predstavlenna}
\end{equation}
Тут $ u_{{{c}_{2}}{{c}_{4}}}^{\left( c \right)} $ позначені елементи довільної матриці групи $S{{U}_{c}}\left( 3 \right)$, а $ u_{{{f}_{2}}{{f}_{4}}}^{\left( f \right)}- $ елементи незалежної від неї матриці групи $S{{U}_{f}}\left( 3 \right)$. Знак $\dagger $ як зазвичай використовується для позначення елементів ермітово спряженої матриці. По індексах що повторюються як зазвичай мається на увазі підсумування. Динамічні рівняння для поля $ {{\psi }_{{{c}_{1}}{{c}_{2}},{{f}_{1}},{{f}_{2}}}}\left( q \right) $ повинні бути симетричними відносно перетворень (\ref{Tenzorni_predstavlenna}).

Окрім того динамічні рівняння повинні бути симетричними  відносно групи (\ref{eq:Gruppa_matric_G}). Найпростіший лагранжіан, який породжує такі рівняння може бути записаний в виді:
\begin{equation}
\begin{split}
  & {{L}^{\left( 0 \right)}}={{g}^{ab}}\frac{\partial \psi _{{{c}_{1}}{{c}_{2}},{{f}_{1}},{{f}_{2}}}^{*}\left( q \right)}{\partial {{q}^{a}}}\frac{\partial {{\psi }_{{{c}_{1}}{{c}_{2}},{{f}_{1}},{{f}_{2}}}}\left( q \right)}{\partial {{q}^{b}}}- \\ 
& -M_{\mu }^{2}\psi _{{{c}_{1}}{{c}_{2}},{{f}_{1}},{{f}_{2}}}^{*}\left( q \right){{\psi }_{{{c}_{1}}{{c}_{2}},{{f}_{1}},{{f}_{2}}}}\left( q \right). \\ 
\end{split}
\label{eq:L_nol_mezon}
\end{equation}
Тут $ {{g}^{ab}}- $компоненти тензора (\ref{eq:Metricnij_tenzor_odnochasnist}), а величина ${{M}_{\mu }}$ розглядатиметься як \mbox{<<затравочна>>} маса мезона. \mbox{<<Справжня>>} маса мезона розглядатиметься далі.

Оскільки поле $ {{\psi }_{{{c}_{1}}{{c}_{2}},{{f}_{1}},{{f}_{2}}}}\left( q \right) $ повинне описувати динаміку зв'язаних станів кварка і антикварка, лагранжіан (\ref{eq:L_nol_mezon}) вочевидь є недостатнім, бо він не враховує взаємодію між кварком і антикварком, яка забезпечує існування зв'язаного стану. Таку взаємодію можна ввести як зазвичай, вимагаючи симетрії лагранжіану відносно локальних перетворень внутрішньої симетрії типу (\ref{Tenzorni_predstavlenna}). Оскільки існування мезону як зв'язаного стану кварка і антикварка пов'язано саме із сильною взаємодією, то ми зосередимося на симетрії відносно локальних $S{{U}_{c}}\left( 3 \right)-$перетворень.
Досягти цієї симетрії можна також звичайним способом - замінюючи \mbox{<<звичайні>>} похідні в лагранжіані (\ref{eq:L_nol_mezon}) на \mbox{<<подовжені>>} похідні і вводячи відповідні компенсуючі поля: 
\begin{equation}
\begin{split}
   & {{L}_{\mu }}={{g}^{ab}}\left( {\partial \psi _{{{c}_{1}}{{c}_{2}},{{f}_{1}},{{f}_{2}}}^{*}\left( q \right)}/{\partial {{q}^{a}}}\;- \right. \\ 
 & -igA_{a,{{g}_{1}}}^{\left( 1 \right)}\left( q \right)\psi _{{{c}_{21}}{{c}_{2}},{{f}_{1}},{{f}_{2}}}^{*}\left( q \right)\lambda _{{{c}_{1}}{{c}_{21}}}^{{{g}_{1}}}+ \\ 
 & \left. +igA_{a,{{g}_{1}}}^{\left( 2 \right)}\left( q \right)\lambda _{{{c}_{22}}{{c}_{2}}}^{{{g}_{1}}}\psi _{{{c}_{1}}{{c}_{22}},{{f}_{1}},{{f}_{2}}}^{*}\left( q \right) \right)\times  \\ 
 & \times \left( {\partial {{\psi }_{{{c}_{1}}{{c}_{2}},{{f}_{1}},{{f}_{2}}}}\left( q \right)}/{\partial {{q}^{b}}}\; \right.+ \\ 
 & +igA_{b,{{g}_{11}}}^{\left( 1 \right)}\left( q \right){{\psi }_{{{c}_{31}}{{c}_{2}},{{f}_{1}},{{f}_{2}}}}\left( q \right)\lambda _{{{c}_{31}}{{c}_{1}}}^{{{g}_{11}}}- \\ 
 & \left. -igA_{b,{{g}_{11}}}^{\left( 2 \right)}\left( q \right)\lambda _{{{c}_{2}}{{c}_{32}}}^{{{g}_{11}}}{{\psi }_{{{c}_{1}}{{c}_{32}},{{f}_{1}},{{f}_{2}}}}\left( q \right) \right)- \\ 
 & -M_{\mu }^{2}\psi _{{{c}_{1}}{{c}_{2}},{{f}_{1}},{{f}_{2}}}^{*}\left( q \right){{\psi }_{{{c}_{1}}{{c}_{2}},{{f}_{1}},{{f}_{2}}}}\left( q \right) .\\ 
\end{split}
\label{eq:L_mu}
\end{equation} 
Тут $ \lambda _{{{c}_{1}}{{c}_{2}}}^{{{g}_{1}}},{{g}_{1}}=1,2,\ldots 8,{{c}_{1}},{{c}_{2}}=1,2,3- $елементи матриць Гелл-Мана, $ A_{a,{{g}_{1}}}^{\left( 1 \right)}\left( q \right) $ і $ A_{a,{{g}_{1}}}^{\left( 2 \right)}\left( q \right)- $ компенсуючі поля, $g$- константа сильної взаємодії. Задля того щоб лагранжіан (\ref{eq:L_mu}) був інваріантним відносно перетворень  (\ref{Tenzorni_predstavlenna}) в яких матриці $ u_{{{c}_{1}}{{c}_{3}}}^{\left( c \right)\dagger } $ і $ u_{{{c}_{2}}{{c}_{4}}}^{\left( c \right)} $ є функціями від координат 
\begin{equation}
\begin{split}
  & u_{{{c}_{2}}{{c}_{4}}}^{\left( c \right)}\left( q \right)={{\left( \exp \left( i{{\theta }_{{{g}_{1}}}}\left( q \right){{{\hat{\lambda }}}^{{{g}_{1}}}} \right) \right)}_{{{c}_{2}}{{c}_{4}}}}, \\ 
& u_{{{c}_{1}}{{c}_{3}}}^{\left( c \right)\dagger }\left( q\right)={{\left( \exp \left( -i{{\theta }_{{{g}_{1}}}}\left( q \right){{{\hat{\lambda }}}^{{{g}_{1}}}} \right) \right)}_{{{c}_{1}}{{c}_{3}}}}, \\ 
\end{split}
\label{eq:Localni_SU3c}
\end{equation}
де $ {{\theta }_{{{g}_{1}}}}\left( x \right) -$ параметри локального $S{{U}_{c}}\left( 3 \right)-$ перетворення які є довільними функціями від координат на підмножині одночасності (\ref{eq:Stovpec_sim_odnochasnist}), потрібно разом з  (\ref{Tenzorni_predstavlenna}) перетворювати і компенсуючі поля $ A_{a,{{g}_{1}}}^{\left( j \right)}\left( q \right),j=1,2 $ за законом:
\begin{equation}\label{Zacon_peretvorenna_compensujuchix_poliv}
\begin{split}
&{A'}_{a,{{g}_{2}}}^{\left( j \right)}\left( q \right)={{\left( {{D}^{-1}}\left( \theta \left( q \right) \right) \right)}_{{{g}_{2}}{{g}_{1}}}}A_{a,{{g}_{1}}}^{\left( j \right)}\left( q \right)+\\
&+\frac{\partial {{\theta }_{{{g}_{1}}}}\left( q \right)}{\partial {{q}^{a}}}.
\end{split}
\end{equation}
Тут $ {{\left( {{D}^{-1}}\left( \theta \left( q \right) \right) \right)}_{{{g}_{2}}{{g}_{1}}}} $ позначені елементи матриць, обернених до матриць приєднаного представлення групи  $S{{U}_{c}}\left( 3 \right)$. Позначення $ \theta \left( q \right) $ без індексу, як зазвичай, використовується для всієї сукупності восьми параметрів  $S{{U}_{c}}\left( 3 \right)-$ перетворення. Як видно з (\ref{Zacon_peretvorenna_compensujuchix_poliv}) закон перетворення обох компенсуючих полів є однаковим. Динамічні рівняння, яким вони задовольняють також будуть однаковими. Проте наявність серед аргументів цих полів внутрішніх змінних ${{y}^{1}},{{y}^{2}},{{y}^{3}}$  (\ref{eq:Stovpec_sim_odnochasnist}), як буде видно з подальшого, призведе до суттєвої ролі граничних умов, які накладаються на ці поля. Такі умови можуть бути різними для різних полів, і тому матимемо різні розв'язки одних і тих самих динамічних рівнянь з одним і тим самим законом перетворення при локальному $S{{U}_{c}}\left( 3 \right)-$ перетворенні. Різні граничні умови саме й надають можливість розглядати два різних компенсуючих поля $ A_{a,{{g}_{1}}}^{\left( 1 \right)}\left( q \right)$ і $ A_{a,{{g}_{1}}}^{\left( 2 \right)}\left( q \right).$ Далі замість цих полів,  зручніше буде розглядати їх лінійні комбінації, аналогічні змінним Якобі
\begin{equation}
\begin{split}
  & A_{a,{{g}_{1}}}^{\left( + \right)}\left( q \right)=\frac{1}{2}\left( A_{a,{{g}_{1}}}^{\left( 1 \right)}\left( q \right)+A_{a,{{g}_{1}}}^{\left( 2 \right)}\left( q \right) \right), \\ 
& A_{a,{{g}_{1}}}^{\left( - \right)}\left( q \right)=A_{a,{{g}_{1}}}^{\left( 2 \right)}\left( q \right)-A_{a,{{g}_{1}}}^{\left( 1 \right)}\left( q \right). \\ 
\end{split}
\label{eq:Jacobi_dla_A}
\end{equation}
Обернені до (\ref{eq:Jacobi_dla_A}) співвідношення мають вид:
\begin{equation}
\begin{split}
  & A_{a,{{g}_{1}}}^{\left( 1 \right)}\left( q \right)=A_{a,{{g}_{1}}}^{\left( + \right)}\left( q \right)-\frac{1}{2}A_{a,{{g}_{1}}}^{\left( - \right)}\left( q \right), \\ 
& A_{a,{{g}_{1}}}^{\left( 2 \right)}\left( q \right)=A_{a,{{g}_{1}}}^{\left( + \right)}\left( q \right)+\frac{1}{2}A_{a,{{g}_{1}}}^{\left( - \right)}\left( q \right). \\ 
\end{split}
\label{eq:Oberneni_Jacobi_dla_A}
\end{equation}
Оскільки на множині значень польових функцій $ {{\psi }_{{{c}_{1}}{{c}_{2}},{{f}_{1}},{{f}_{2}}}}\left( q \right) $ задане представлення локальної групи $S{{U}_{c}}\left( 3 \right)$ цю область значень можна розкласти в пряму суму інваріантних підпросторів відносно перетворень цього представлення. Внаслідок того що адрон є безкольоровим нас цікавитиме поле, яке має ненульову проекцію тільки на підпростір на якому реалізується скалярне незвідне представлення і який \mbox{<<натягнутий>>} на тензор $ {{\delta }_{{{c}_{1}}{{c}_{2}}}}. $
Це означає що поле $ {{\psi }_{{{c}_{1}}{{c}_{2}},{{f}_{1}},{{f}_{2}}}}\left( q \right) $ може бути представлене в виді 
\begin{equation}\label{Proekcija_na_bezcolorovij_psdprostir}
{{\psi }_{{{c}_{1}}{{c}_{2}},{{f}_{1}},{{f}_{2}}}}\left( q \right)={{\delta }_{{{c}_{1}}{{c}_{2}}}}{{\psi }_{{{f}_{1}},{{f}_{2}}}}\left( q \right),
\end{equation}
де $ {{\psi }_{{{f}_{1}},{{f}_{2}}}}\left( q \right)- $ нові польові функції, для яких далі будуть розглядатися динамічні рівняння, і які після квантування будуть описувати процеси народження і знищення мезонів. Ці динамічні рівняння можна отримати з лагранжіану, що утвориться, якщо в (\ref{eq:L_mu}) підставити (\ref{Proekcija_na_bezcolorovij_psdprostir}) і врахувати позначення (\ref{eq:Jacobi_dla_A}). Після цих перетворень лагранжіан  (\ref{eq:L_mu}) прийме вид
\begin{equation}
\begin{split}
  & {{L}_{\mu }}=3{{g}^{ab}}\left( {\partial \psi _{{{f}_{1}},{{f}_{2}}}^{*}\left( q \right)}/{\partial {{q}^{a}}}\; \right)\left( {\partial {{\psi }_{{{f}_{1}},{{f}_{2}}}}\left( q \right)}/{\partial {{q}^{b}}}\; \right)+ \\ 
& +2{{g}^{2}}{{g}^{ab}}A_{a,{{g}_{1}}}^{\left( - \right)}\left( q \right)A_{b,{{g}_{1}}}^{\left( - \right)}\left( q \right)\psi _{{{f}_{1}},{{f}_{2}}}^{*}\left( q \right){{\psi }_{{{f}_{1}},{{f}_{2}}}}\left( q \right)- \\ 
& -3M_{\mu }^{2}\psi _{{{f}_{1}},{{f}_{2}}}^{*}\left( q \right){{\psi }_{{{f}_{1}},{{f}_{2}}}}\left( q \right) \\
\end{split}
\label{eq:L_mu_bezcvet}
\end{equation}

\section{Рівняння зв'язку для компонент калібрувального поля}
Оскільки мезон складається з кварка і антикварка, двочастинквое мезонне поле можна розглядати як таке що приймає значення на тензорному добутку лінійного простору діраківськи спряжених біспінорів і лінійного простору біспінорів. Тому, як вихідне двочастинкове поле можна було б розглянути таке поле яке окрім пари кольорових і пари ароматових індексів має ще два біспінорних індекси. Таке поле позначатимемо як   
\begin{equation}\label{z_dvoma_bispinornimi_indeksami}
{{\psi }_{{{s}_{1}},{{s}_{2}},{{c}_{1}}{{c}_{2}},{{f}_{1}},{{f}_{2}}}}\left( q \right)={{\psi }_{{{s}_{1}},{{s}_{2}},{{c}_{1}}{{c}_{2}},{{f}_{1}},{{f}_{2}}}}\left( X,\vec{y} \right).
\end{equation}
В цих позначеннях ми відобразили також те, що в цьому розділі, зручно буде відокремити серед координат на підмножині одночасності (\ref{eq:Stovpec_sim_odnochasnist}) сукупність координат центру мас $X\equiv \left( {{X}^{0}},{{X}^{1}},{{X}^{2}},{{X}^{3}} \right)$ і внутрішніх координат $\vec{y}\equiv \left( {{y}^{1}},{{y}^{2}},{{y}^{3}} \right).$ При переході від однієї інерційної системи відліку до іншої, яке описується перетворенням Лоренца $\hat{\Lambda }$ польові функції (\ref{z_dvoma_bispinornimi_indeksami}) перетворюватимуться за законом
\begin{equation}
\begin{split}
  & {{{{\psi }'}}_{{{s}_{1}},{{s}_{2}},{{c}_{1}}{{c}_{2}},{{f}_{1}},{{f}_{2}}}}\left( {X}',\vec{y} \right)= \\ 
& =D_{{{s}_{3}},{{s}_{1}}}^{-1}\left( {\hat{\Lambda }} \right){{D}_{{{s}_{2}}{{s}_{4}}}}\left( {\hat{\Lambda }} \right)\times  \\ 
& \times {{\psi }_{{{s}_{3}},{{s}_{4}},{{c}_{1}}{{c}_{2}},{{f}_{1}},{{f}_{2}}}}\left( X={{{\hat{\Lambda }}}^{-1}}\left( {{X}'} \right),\vec{y} \right), \\ 
\end{split}
\label{eq:dvajdi_bispinornij_zacon}
\end{equation}
де $ {{D}_{{{s}_{4}}{{s}_{2}}}}\left( {\hat{\Lambda }} \right) $ позначає елементи матриці біспінорного представлення групи Лоренца , а $ D_{{{s}_{1}},{{s}_{3}}}^{-1}\left( {\hat{\Lambda }} \right)- $елементи оберненої матриці. 

\begin{sloppypar}
Поле $ {{\psi }_{{{s}_{1}},{{s}_{2}},{{c}_{1}}{{c}_{2}},{{f}_{1}},{{f}_{2}}}}\left( X,\vec{y} \right) $ може бути розкладене по базису з 16 матриць, які утворюються за допомогою лінійно незалежних добутків матриць Дірака \cite{Bogolyubov_rus}. Оскільки мезони є псевдоскалярними частинками, вони повинні описуватися проекцією поля 
$ {{\psi }_{{{s}_{1}},{{s}_{2}},{{c}_{1}}{{c}_{2}},{{f}_{1}},{{f}_{2}}}}\left( X,\vec{y} \right) $ на підпростір, \mbox{<<натягнутий>>} на матрицю $\gamma _{{{s}_{1}}{{s}_{2}}}^{5}.$ Ця проекція вже не залежить від біспінорних індексів і може розглядатися як поле $ {{\psi }_{{{c}_{1}}{{c}_{2}},{{f}_{1}},{{f}_{2}}}}\left( q \right)={{\psi }_{{{c}_{1}}{{c}_{2}},{{f}_{1}},{{f}_{2}}}}\left( X,\vec{y} \right) ,$ розглянуте в попередньому розділі. Якщо $ {{\psi }_{{{s}_{1}},{{s}_{2}},{{c}_{1}}{{c}_{2}},{{f}_{1}},{{f}_{2}}}}\left( X,\vec{y} \right) $ представити в виді 
\end{sloppypar}

\begin{equation}\label{Proekcija_na_gama5}
{{\psi }_{{{s}_{1}},{{s}_{2}},{{c}_{1}}{{c}_{2}},{{f}_{1}},{{f}_{2}}}}\left( X,\vec{y} \right)=\frac{1}{4}{{\psi }_{{{c}_{1}}{{c}_{2}},{{f}_{1}},{{f}_{2}}}}\left( X,\vec{y} \right)\gamma _{{{s}_{1}}{{s}_{2}}}^{5}+\ldots,
\end{equation}
де \mbox{<<трикрапки>>} позначають проекції на решту 15 лінійно незалежних матриць алгебри, утворюючими якої є матриці Дірака, то матимемо:
\begin{equation}\label{Spur}
{{\psi }_{{{c}_{1}}{{c}_{2}},{{f}_{1}},{{f}_{2}}}}\left( X,\vec{y} \right)={{\psi }_{{{s}_{1}},{{s}_{2}},{{c}_{1}}{{c}_{2}},{{f}_{1}},{{f}_{2}}}}\left( X,\vec{y} \right)\gamma _{{{s}_{1}}{{s}_{2}}}^{5}.
\end{equation} 
Тут по біспінорним індексам ${{s}_{1}},{{s}_{2}},$ які повторюються мається на увазі підсумування.

В усіх попередніх міркуваннях ми покладали, що перші індекси відносяться до антикварка, а другі - до кварка, з яких побудовано мезон. Тому, якщо ми обміняємо місцями індекси у розглянутих полів, то кварк певного типу заміниться на його антикварк і навпаки. Така заміна призведе до заміни мезону на антимезон. Ми хочемо з'ясувати яким чином виглядатиме в такому випадку операція зарядового спряження для мезону. Тобто ми хочемо з'ясувати як пов'язані компоненти полів $ {{\psi }_{{{c}_{1}}{{c}_{2}},{{f}_{1}}{{f}_{2}}}}\left( X,\vec{y} \right) $ і $ {{\psi }_{{{c}_{2}}{{c}_{1}},{{f}_{2}}{{f}_{1}}}}\left( X,-\vec{y} \right)$. Знак \mbox{<<мінус>>} перед внутрішніми змінними виникає внаслідок того, що як видно з (\ref{Coordinati_Jacobi}) обмін частинок, для яких вводяться координати Якобі, не змінює координат центру мас, і змінює знак у внутрішніх змінних. Найпростішим тензором, який має трансформаційні властивості (\ref{Tenzorni_predstavlenna}) і (\ref{eq:dvajdi_bispinornij_zacon}) є 
\begin{equation}
\begin{split}
&{{\psi }_{{{s}_{1}}{{s}_{2}},{{c}_{1}}{{c}_{2}},{{f}_{1}}{{f}_{2}}}}\left( {{x}_{\left( 1 \right)}},{{x}_{\left( 2 \right)}} \right)=\\
&={{\bar{\psi }}_{{{s}_{1}},{{c}_{1}},{{f}_{1}}}}\left( {{x}_{\left( 1 \right)}} \right){{\psi }_{{{s}_{2}},{{c}_{2}},{{f}_{2}}}}\left( {{x}_{\left( 2 \right)}} \right).
\end{split}
\label{eq:Naiprostihij_tenzor}
\end{equation}
Виділяючи з нього псевдоскалярний внесок, отримаємо
\begin{equation}
\begin{split}
  & {{\psi }_{{{c}_{1}}{{c}_{2}},{{f}_{1}}{{f}_{2}}}}\left( {{x}_{\left( 1 \right)}},{{x}_{\left( 2 \right)}} \right)= \\ 
& =\frac{1}{4}{{{\bar{\psi }}}_{{{s}_{1}},{{c}_{1}},{{f}_{1}}}}\left( {{x}_{\left( 1 \right)}} \right){{\psi }_{{{s}_{2}},{{c}_{2}},{{f}_{2}}}}\left( {{x}_{\left( 2 \right)}} \right)\gamma _{{{s}_{1}}{{s}_{2}}}^{5}. \\
\end{split}
\label{eq: psevdoskalarnij_vnesok}
\end{equation}
Тоді після обміну частинок місцями, отримаємо
\begin{equation}
\begin{split}
& {{\psi }_{{{c}_{2}}{{c}_{1}},{{f}_{2}}{{f}_{1}}}}\left( {{x}_{\left( 2 \right)}},{{x}_{\left( 1 \right)}} \right)= \\ 
& =\frac{1}{4}{{{\bar{\psi }}}_{{{s}_{2}},{{c}_{2}},{{f}_{2}}}}\left( {{x}_{\left( 2 \right)}} \right){{\psi }_{{{s}_{1}},{{c}_{1}},{{f}_{1}}}}\left( {{x}_{\left( 1 \right)}} \right)\gamma _{{{s}_{2}}{{s}_{1}}}^{5}= \\ 
& =\frac{1}{4}\psi _{{{s}_{3}},{{c}_{2}},{{f}_{2}}}^{*}\left( {{x}_{\left( 2 \right)}} \right)\gamma _{{{s}_{3}}{{s}_{2}}}^{0}{{\psi }_{{{s}_{1}},{{c}_{1}},{{f}_{1}}}}\left( {{x}_{\left( 1 \right)}} \right)\gamma _{{{s}_{2}}{{s}_{1}}}^{5}= \\ 
& ={{\left( \frac{1}{4}{{\psi }_{{{s}_{3}},{{c}_{2}},{{f}_{2}}}}\left( {{x}_{\left( 2 \right)}} \right)\gamma _{{{s}_{3}}{{s}_{2}}}^{0}\psi _{{{s}_{1}},{{c}_{1}},{{f}_{1}}}^{*}\left( {{x}_{\left( 1 \right)}} \right)\gamma _{{{s}_{2}}{{s}_{1}}}^{5} \right)}^{*}}= \\ 
& ={{\left( \frac{1}{4}{{\psi }_{{{s}_{3}},{{c}_{2}},{{f}_{2}}}}\left( {{x}_{\left( 2 \right)}} \right)\psi _{{{s}_{1}},{{c}_{1}},{{f}_{1}}}^{*}\left( {{x}_{\left( 1 \right)}} \right)\gamma _{{{s}_{3}}{{s}_{2}}}^{0}\gamma _{{{s}_{2}}{{s}_{1}}}^{5} \right)}^{*}}= \\ 
& =-{{\left( \frac{1}{4}\psi _{{{s}_{1}},{{c}_{1}},{{f}_{1}}}^{*}\left( {{x}_{\left( 1 \right)}} \right)\gamma _{{{s}_{2}}{{s}_{1}}}^{0}{{\psi }_{{{s}_{3}},{{c}_{2}},{{f}_{2}}}}\left( {{x}_{\left( 2 \right)}} \right)\gamma _{{{s}_{3}}{{s}_{2}}}^{5} \right)}^{*}}= \\ 
& =-{{\left( \frac{1}{4}\psi _{{{s}_{1}},{{c}_{1}},{{f}_{1}}}^{*}\left( {{x}_{\left( 1 \right)}} \right)\gamma _{{{s}_{1}}{{s}_{2}}}^{0}{{\psi }_{{{s}_{3}},{{c}_{2}},{{f}_{2}}}}\left( {{x}_{\left( 2 \right)}} \right)\gamma _{{{s}_{3}}{{s}_{2}}}^{5} \right)}^{*}}= \\ 
& =-{{\left( \frac{1}{4}{{{\bar{\psi }}}_{{{s}_{2}},{{c}_{1}},{{f}_{1}}}}\left( {{x}_{\left( 1 \right)}} \right){{\psi }_{{{s}_{3}},{{c}_{2}},{{f}_{2}}}}\left( {{x}_{\left( 2 \right)}} \right)\gamma _{{{s}_{3}}{{s}_{2}}}^{5} \right)}^{*}}= \\ 
& =-{{\left( {{\psi }_{{{c}_{1}}{{c}_{2}},{{f}_{1}}{{f}_{2}}}}\left( {{x}_{\left( 1 \right)}},{{x}_{\left( 2 \right)}} \right) \right)}^{*}}. \\ 
\end{split}
\label{eq:Pisla_obminu}
\end{equation}
На підмножині одночасності в координатах Якобі, матимемо:
\begin{equation}\label{Rivnanna_zvazku}
{{\psi }_{{{c}_{1}}{{c}_{2}},{{f}_{1}},{{f}_{2}}}}\left( X,\vec{y} \right)=-{{\left( {{\psi }_{{{c}_{2}}{{c}_{1}},{{f}_{2}},{{f}_{1}}}}\left( X,-\vec{y} \right) \right)}^{*}}
\end{equation}
Оскільки тензор (\ref{z_dvoma_bispinornimi_indeksami}) має такі ж трансформаційні властивості що й (\ref{eq:Naiprostihij_tenzor}) для його проекції на підпростір \mbox{<<натягнутий>>} на ${{\hat{\gamma }}^{5}}$ приймемо такий самий вид операції зарядового спряження (\ref{Rivnanna_zvazku}). Дійсно, \mbox{<<розщеплений>>} вид тензору (\ref{eq:Naiprostihij_tenzor}), говорить про те, що він не враховує взаємодію між кварками. Ця взаємодія вводиться шляхом вимоги локальної $S{{U}_{c}}\left( 3 \right)$ симетрії, яка як окремий випадок включає глобальну симетрію. Тому симетрія  (\ref{z_dvoma_bispinornimi_indeksami}) не нижча за симетрію (\ref{eq:Naiprostihij_tenzor}). Оскільки при отриманні (\ref{Rivnanna_zvazku}) ми користувалися лише симетрійними властивостями, то ми й приймаємо аналогічне співвідношення для тензора, побудованого з урахуванням взаємодії. Оскільки внутрішній стан кварків в мезоні повинен бути основним (бо збуджений стан представлятиме іншу частинку), ми будемо його вважати сферично симетричним. Тоді
\begin{equation}\label{Sfericheskaja_simmetria}
{{\psi }_{{{c}_{1}}{{c}_{2}},{{f}_{1}},{{f}_{2}}}}\left( {{X}^{b}},-\vec{y} \right)={{\psi }_{{{c}_{1}}{{c}_{2}},{{f}_{1}},{{f}_{2}}}}\left( {{X}^{b}},\vec{y} \right).
\end{equation}

Рівняння зв'язку  (\ref{Rivnanna_zvazku}) породжує певні зв'язки між компонентами калібрувальних полів $ A_{a,{{g}_{1}}}^{\left( 1 \right)}\left( q \right) $ і $ A_{a,{{g}_{1}}}^{\left( 2 \right)}\left( q \right). $ Розглянемо ці зв'язки.

 Якщо паралельно зсунути тензор $ {{\psi }_{{{c}_{1}}{{c}_{2}},{{f}_{1}}{{f}_{2}}}}\left( q \right) $ уздовж підмножини одночасності, його зміна описуватиметься співвідношенням: 
\begin{equation}
\begin{split}
  & \delta {{\psi }_{{{c}_{1}}{{c}_{2}},{{f}_{1}}{{f}_{2}}}}\left( q \right)=\left( igA_{b,{{g}_{11}}}^{\left( 1 \right)}\left( q \right){{\psi }_{{{c}_{21}}{{c}_{2}},{{f}_{1}}{{f}_{2}}}}\left( q \right)\lambda _{{{c}_{21}}{{c}_{1}}}^{{{g}_{11}}} \right.- \\ 
& \left. -igA_{b,{{g}_{11}}}^{\left( 2 \right)}\left( q \right)\lambda _{{{c}_{2}}{{c}_{22}}}^{{{g}_{11}}}{{\psi }_{{{c}_{1}}{{c}_{22}},{{f}_{1}}{{f}_{2}}}}\left( q \right) \right)d{{q}^{b}}. \\ 
\end{split}
\label{eq: parallelnij_zsuv}
\end{equation}
Враховуючи (\ref{Proekcija_na_bezcolorovij_psdprostir}) отримаємо:
\begin{equation}
\begin{split}
  & \delta \left( {{\psi }_{{{f}_{1}},{{f}_{2}}}}\left( q \right){{\delta }_{{{c}_{1}}{{c}_{2}}}} \right)= \\ 
& =ig\left( A_{b,{{g}_{1}}}^{\left( 1 \right)}\left( q \right)-A_{b,{{g}_{1}}}^{\left( 2 \right)}\left( q \right) \right)\lambda _{{{c}_{2}}{{c}_{1}}}^{{{g}_{1}}}{{\psi }_{{{f}_{1}}{{f}_{2}}}}\left( q \right)d{{q}^{b}}. \\ 
\end{split}
\label{eq:zsuv_bezcvet}
\end{equation}
Далі в цьому виразі нам потрібно відділити внески від зсуву по напрямках координат центру мас і внутрішніх координат
\begin{equation}
\begin{split}
  & \delta \left( {{\psi }_{{{f}_{1}},{{f}_{2}}}}\left( X,\vec{y} \right){{\delta }_{{{c}_{1}}{{c}_{2}}}} \right)= \\ 
& =-\sum\limits_{b=0}^{3}{igA_{b,{{g}_{1}}}^{\left( - \right)}\left( X,\vec{y} \right)\lambda _{{{c}_{2}}{{c}_{1}}}^{{{g}_{1}}}{{\psi }_{{{f}_{1}}{{f}_{2}}}}\left( X,\vec{y} \right)d{{X}^{b}}}- \\ 
& -\sum\limits_{b=1}^{3}{igA_{3+b,{{g}_{1}}}^{\left( - \right)}\left( X,\vec{y} \right)\lambda _{{{c}_{2}}{{c}_{1}}}^{{{g}_{1}}}{{\psi }_{{{f}_{1}}{{f}_{2}}}}\left( X,\vec{y} \right)d{{y}^{b}}}. \\ 
\end{split}
\label{eq:zsuv_bezcvet_Jacobi}
\end{equation}
Тут для скорочення запису ми скористалися введеним раніше позначенням (\ref{eq:Jacobi_dla_A}). Виходячи з 
(\ref{eq:zsuv_bezcvet_Jacobi}) обмінюючи кварк і антикварк місцями і беручи комплексне спряження, отримаємо
\begin{equation}
\begin{split}
  & \delta {{\left( {{\psi }_{{{f}_{2}},{{f}_{1}}}}\left( X,-\vec{y} \right){{\delta }_{{{c}_{2}}{{c}_{1}}}} \right)}^{*}}= \\ 
& =\sum\limits_{b=0}^{3}{igA_{b,{{g}_{1}}}^{\left( - \right)}\left( X,-\vec{y} \right)\lambda _{{{c}_{2}}{{c}_{1}}}^{{{g}_{1}}}\psi _{{{f}_{2}}{{f}_{1}}}^{*}\left( X,-\vec{y} \right)d{{X}^{b}}}+ \\ 
& +\sum\limits_{b=1}^{3}{igA_{3+b,{{g}_{1}}}^{\left( - \right)}\left( X,-\vec{y} \right)\lambda _{{{c}_{2}}{{c}_{1}}}^{{{g}_{1}}}\psi _{{{f}_{2}}{{f}_{1}}}^{*}\left( X,-\vec{y} \right)d\left( -{{y}^{b}} \right)}. \\ 
\end{split}
\label{eq: zsuv_iz_zminoju_quarqa_na_antiquarq}
\end{equation}
Враховуючи (\ref{Rivnanna_zvazku}) і (\ref{Proekcija_na_bezcolorovij_psdprostir}) отримуємо, 
що
\begin{equation}\label{zvazoc_tilcu_dla_flavoru}
\psi _{{{f}_{2}},{{f}_{1}}}^{*}\left( X,-\vec{y} \right)=-{{\psi }_{{{f}_{1}},{{f}_{2}}}}\left( X,\vec{y} \right)
\end{equation}
Окрім того, оскільки внутрішній стан мезону, як це обговорювалося з приводу (\ref{Sfericheskaja_simmetria}), є сферично симетричний і залежить від $\left| {\vec{y}} \right|$, то й поле $ A_{b,{{g}_{1}}}^{\left( - \right)}\left( X,\vec{y} \right) $, створене кварком і антикварком в цьому стані, також повинне залежати від $\left| {\vec{y}} \right|.$
 Відтак, маємо
 \begin{equation}\label{A_ot_minus_y_ravno_A_ot_y}
 A_{b,{{g}_{1}}}^{\left( - \right)}\left( X,-\vec{y} \right)=A_{b,{{g}_{1}}}^{\left( - \right)}\left( X,\vec{y} \right).
 \end{equation} 
Враховуючи (\ref{zvazoc_tilcu_dla_flavoru}) і (\ref{A_ot_minus_y_ravno_A_ot_y}), і виносячи \mbox{<<мінус>>} з під знаку диференціалу в (\ref{eq: zsuv_iz_zminoju_quarqa_na_antiquarq}), це співвідношення перепишемо в виді
 \begin{equation}
 \begin{split}
   & \delta {{\left( {{\psi }_{{{f}_{2}},{{f}_{1}}}}\left( X,-\vec{y} \right){{\delta }_{{{c}_{2}}{{c}_{1}}}} \right)}^{*}}= \\ 
 & =-\sum\limits_{b=0}^{3}{igA_{b,{{g}_{1}}}^{\left( - \right)}\left( X,\vec{y} \right)\lambda _{{{c}_{2}}{{c}_{1}}}^{{{g}_{1}}}{{\psi }_{{{f}_{1}}{{f}_{2}}}}\left( X,\vec{y} \right)d{{X}^{b}}}+ \\ 
 & +\sum\limits_{b=1}^{3}{igA_{3+b,{{g}_{1}}}^{\left( - \right)}\left( X,\vec{y} \right)\lambda _{{{c}_{2}}{{c}_{1}}}^{{{g}_{1}}}{{\psi }_{{{f}_{1}}{{f}_{2}}}}\left( X,\vec{y} \right)d{{y}^{b}}} \\ 
  \end{split}
 \label{eq: delta_psi_so_zvezdochcouj}
 \end{equation}
 
 Отже, враховуючи (\ref{eq:zsuv_bezcvet_Jacobi}) маємо:
 \begin{equation}
 \begin{split}
   & \delta {{\left( -{{\psi }_{{{f}_{2}},{{f}_{1}}}}\left( X,-\vec{y} \right){{\delta }_{{{c}_{2}}{{c}_{1}}}} \right)}^{*}}= \\ 
 & =\sum\limits_{b=0}^{3}{igA_{b,{{g}_{1}}}^{\left( - \right)}\left( X,\vec{y} \right)\lambda _{{{c}_{2}}{{c}_{1}}}^{{{g}_{1}}}{{\psi }_{{{f}_{1}}{{f}_{2}}}}\left( X,\vec{y} \right)d{{X}^{b}}}- \\ 
 & -\sum\limits_{b=1}^{3}{igA_{3+b,{{g}_{1}}}^{\left( - \right)}\left( X,\vec{y} \right)\lambda _{{{c}_{2}}{{c}_{1}}}^{{{g}_{1}}}{{\psi }_{{{f}_{1}}{{f}_{2}}}}\left( X,\vec{y} \right)d{{y}^{b}}}, \\ 
 & \delta \left( {{\psi }_{{{f}_{1}},{{f}_{2}}}}\left( X,\vec{y} \right){{\delta }_{{{c}_{1}}{{c}_{2}}}} \right)= \\ 
 & =-\sum\limits_{b=0}^{3}{igA_{b,{{g}_{1}}}^{\left( - \right)}\left( X,\vec{y} \right)\lambda _{{{c}_{2}}{{c}_{1}}}^{{{g}_{1}}}{{\psi }_{{{f}_{1}}{{f}_{2}}}}\left( X,\vec{y} \right)d{{X}^{b}}}- \\ 
 & -\sum\limits_{b=1}^{3}{igA_{3+b,{{g}_{1}}}^{\left( - \right)}\left( X,\vec{y} \right)\lambda _{{{c}_{2}}{{c}_{1}}}^{{{g}_{1}}}{{\psi }_{{{f}_{1}}{{f}_{2}}}}\left( X,\vec{y} \right)d{{y}^{b}}}. \\ 
 \end{split}
 \label{eq:Porivnanna_zsuviv}
 \end{equation}
 Виходячи з (\ref{Rivnanna_zvazku}) ці величини ми повинні прирівняти. Таке прирівнювання, виходячи з (\ref{eq:Porivnanna_zsuviv}) призведе до висновку:
 \begin{equation}
 \begin{split}
   & A_{b,{{g}_{1}}}^{\left( - \right)}\left( X,\vec{y} \right)=0, \\ 
 & b=0,1,2,3. \\ 
  \end{split}
 \label{eq:A_vnechn_ravno_nulu}
 \end{equation}
 Фізичний сенс цього результату полягає в тому що якщо ми безкольоровий адрон піддаємо паралельному переносу як ціле, то внаслідок його безкольоровості поле не виникає. Якщо ж ми будемо його паралельно переносити у напрямках, що відповідають зміні відстані між кварком і антикварком всередині адрону, то такий перенос породжуватиме внутрішнє поле адрону.
 
 Отже (\ref{eq:L_mu_bezcvet}) може бути записане в виді
 \begin{equation}
 \begin{split}
 & {{L}_{\mu }}=3{{g}^{ab}}\left( {\partial \psi _{{{f}_{1}},{{f}_{2}}}^{*}\left( q \right)}/{\partial {{q}^{a}}}\; \right)\left( {\partial {{\psi }_{{{f}_{1}},{{f}_{2}}}}\left( q \right)}/{\partial {{q}^{b}}}\; \right)+ \\ 
 & -8{{g}^{2}}\left( \sum\limits_{b=4}^{6}{{{\left( A_{a,{{g}_{1}}}^{\left( - \right)}\left( q \right) \right)}^{2}}} \right)\psi _{{{f}_{1}},{{f}_{2}}}^{*}\left( q \right){{\psi }_{{{f}_{1}},{{f}_{2}}}}\left( q \right)- \\ 
 & -3M_{\mu }^{2}\psi _{{{f}_{1}},{{f}_{2}}}^{*}\left( q \right){{\psi }_{{{f}_{1}},{{f}_{2}}}}\left( q \right). \\ 
  \end{split}
 \label{eq:L_mu_iz_vrachuvannam_zvazkiv}
 \end{equation}
 Тут окрім (\ref{eq:L_mu_bezcvet}) і (\ref{eq:A_vnechn_ravno_nulu}) ми ще врахували вид метричного тензору (\ref{eq:Metricnij_tenzor_odnochasnist}). 
 
 Звернемо увагу на те, що неоднородні доданки в законах перетворення (\ref{Zacon_peretvorenna_compensujuchix_poliv}) для полів $ A_{a,{{g}_{1}}}^{\left( 1 \right)}\left( q \right) $ і $ A_{a,{{g}_{1}}}^{\left( 2 \right)}\left( q \right) $ є однаковими. Тому, враховуючи визначення (\ref{eq:Jacobi_dla_A}), бачимо, що в законі перетворення  $ A_{a,{{g}_{1}}}^{\left( - \right)}\left( q \right) $ ці неоднорідні доданки компенсуються і поле  $ A_{a,{{g}_{1}}}^{\left( - \right)}\left( q \right) $ перетворюються за \mbox{<<чистим>>} представленням локальної групи $S{{U}_{c}}\left( 3 \right).$ Внаслідок цього величини $ {{g}^{ab}}A_{a,{{g}_{1}}}^{\left( - \right)}\left( q \right)A_{b,{{g}_{1}}}^{\left( - \right)}\left( q \right) $ і $ \left( \sum\limits_{b=4}^{7}{{{\left( A_{a,{{g}_{1}}}^{\left( - \right)}\left( q \right) \right)}^{2}}} \right) $, які входять в (\ref{eq:L_mu_bezcvet}) і в (\ref{eq:L_mu_iz_vrachuvannam_zvazkiv}) є локально калібрувально-інварінтними. Окрім того безкольоровий тензор $ {{\psi }_{{{c}_{1}}{{c}_{2}},{{f}_{1}},{{f}_{2}}}}\left( q \right)={{\delta }_{{{c}_{1}}{{c}_{2}}}}{{\psi }_{{{f}_{1}},{{f}_{2}}}}\left( q \right) $ є також локально калібрувально-інваріантним. Звернемо увагу на те, що в цьому полягає відмінність від роботи \cite{Chudak:2016}, в якій не вдавалося описати безкольоровість адрону за допомогою локального інваріанту, а лише з використанням множини калібрувально еквівалентних між собою польових конфігурацій. Оскільки $ {{\psi }_{{{c}_{1}}{{c}_{2}},{{f}_{1}},{{f}_{2}}}}\left( q \right)={{\delta }_{{{c}_{1}}{{c}_{2}}}}{{\psi }_{{{f}_{1}},{{f}_{2}}}}\left( q \right) $ є локальним інваріантом, величина $ \delta \left( {{\psi }_{{{f}_{1}},{{f}_{2}}}}\left( X,\vec{y} \right){{\delta }_{{{c}_{1}}{{c}_{2}}}} \right), $ що визначається в (\ref{eq:zsuv_bezcvet_Jacobi}) і з урахуванням (\ref{eq:A_vnechn_ravno_nulu}) приймає вид
 \begin{equation}
 \begin{split}
   & \delta \left( {{\psi }_{{{f}_{1}},{{f}_{2}}}}\left( X,\vec{y} \right){{\delta }_{{{c}_{1}}{{c}_{2}}}} \right)= \\ 
 & =-\sum\limits_{b=1}^{3}{igA_{3+b,{{g}_{1}}}^{\left( - \right)}\left( X,\vec{y} \right)\lambda _{{{c}_{2}}{{c}_{1}}}^{{{g}_{1}}}{{\psi }_{{{f}_{1}}{{f}_{2}}}}\left( X,\vec{y} \right)d{{y}^{b}}}, \\ 
  \end{split}
 \label{eq:delta_psi_z_urachuvannam_zvazkiv}
 \end{equation} 

\noindent
\begin{sloppypar}
повинна дорівнювати нулю. Як видно з (\ref{eq:delta_psi_z_urachuvannam_zvazkiv}) це означає, що вектор 
 $\vec{A}_{{{g}_{1}}}^{\left( - \right)} \left( X,\vec{y} \right) = \left( A_{4,{{g}_{1}}}^{\left( - \right)}\left( X,\vec{y} \right),A_{5,{{g}_{1}}}^{\left( - \right)} \left( X,\vec{y} \right),A_{6,{{g}_{1}}}^{\left( - \right)} \left( X,\vec{y} \right) \right) $ повинен бути ортогональний до вектора $\vec{y}.$ В цій роботі ми не розглядатимемо динамічних рівнянь для цього поля, а спробуємо отримати динамічні рівняння відразу для поля
\end{sloppypar}
 \begin{equation}\label{Pole_V_ot_q}
  V\left( q \right)={{g}^{2}}\left( \sum\limits_{b=4}^{6}{{{\left( A_{a,{{g}_{1}}}^{\left( - \right)}\left( q \right) \right)}^{2}}} \right), 
 \end{equation}
що безпосередньо входить до лагранжіану (\ref{eq:L_mu_iz_vrachuvannam_zvazkiv}).

\section{Динамічне рівняння для поля $ V\left( q \right) $.}
Хоча поля $ A_{a,{{g}_{1}}}^{\left( 1 \right)}\left( q \right) $ і $ A_{a,{{g}_{1}}}^{\left( 2 \right)}\left( q \right) $ мають залежність від внутрішніх змінних вони залишаються одночастинковими, бо на відміну від тих полів, з якими вони взаємодіють, ці поля не описують зв'язани стани частинок. Дійсно, двочастинкове поле поавинне перетворюватися за тензорним добутком декількох представлень групи Лоренца, або групи (\ref{eq:Gruppa_matric_G})  і групи внутрішньої симетрії, з можливим виділенням з цього тензорного незвідного представлення, яке реалізується на певному інваріантному піпідпросторі. В той же час $ A_{a,{{g}_{1}}}^{\left( 1 \right)}\left( q \right) $ і $ A_{a,{{g}_{1}}}^{\left( 2 \right)}\left( q \right) $, а аткож їх лінійні комбінації $ A_{a,{{g}_{1}}}^{\left(- \right)}\left( q \right) $ і $ A_{a,{{g}_{1}}}^{\left(+ \right)}\left( q \right) $ (\ref{eq:Jacobi_dla_A}) такої властивості не мають.   Такі поля можна розглядати як такі, що описують взаємодію глюону із багаточастинковим кварковиим станом, звідки й виникає залежність від внутрішніх змінних. Як буде показано далі, така взаємодія призводить до конфайнменту кварків і, відповідно утриманню їх всередині двочастинкового стану. Але, як відомо, для глюонів також властиве явище конфайнменту. Тому природно припустити що взаємодія між зв'язаними станами кварків здійснюється не шляхом обміну глюонами, а шляхом обміну їх зв'язаними станами - глюболами \cite{Olive:2016xmw}. Для того, щоб отримати динамічні рівняння для двоглюонного поля розглянемо найпростіший тензор, який можна утворити з одноглюонних полів
\begin{equation}
\begin{split}
  & {{A}_{ab,{{g}_{1}}{{g}_{2}}}}\left( q \right)={{g}^{2}}\left( A_{a,{{g}_{1}}}^{\left( - \right)}\left( q \right)A_{b,{{g}_{2}}}^{\left( - \right)}\left( q \right) \right), \\ 
& a,b=4,5,6. \\ 
\end{split}
\label{eq:Naiprostichij_tenzor_A}
\end{equation}
Розкладаючи лінійний простір тензорів $ {{A}_{ab,{{g}_{1}}{{g}_{2}}}}\left( q \right) $ відносно групи (\ref{eq:Gruppa_matric_G})  в пряму суму інваріантних підпросторів виділимо доданок, що відповідає проекції на скалярний підпростір
\begin{equation}\label{Proekcsja_na_skalarnij_pidprostir_A}
{{A}_{ab,{{g}_{1}}{{g}_{2}}}}\left( q \right)=-{{A}_{{{g}_{1}}{{g}_{2}}}}\left( q \right){{g}_{ab}}+\ldots 
\end{equation}
Тут знов як \mbox{<<трикрапки>>} позначені внески від проекцій на решту інваріантних підпросторів, які нас не цікавитимуть, бо ми хочемо розглянути скалярний глюбол і тому розглядатимемо окремий випадок коли всі ці проекції дорівнюють нулю. Знак \mbox{<<мінус>>} в цьому означенні є несуттєвим і використаний виключно заради зручності наступних обчислень. Згортаючи обидві частини рівності (\ref{Proekcsja_na_skalarnij_pidprostir_A}) з метричним тензором ${{g}^{ab}}$ і враховуючи співвідношення (\ref{eq:A_vnechn_ravno_nulu}) матимемо:
\begin{equation}\label{Ag1g2}
{{A}_{{{g}_{1}}{{g}_{2}}}}\left( q \right)=\frac{4}{7}{{g}^{2}}\sum\limits_{b=4}^{6}{\left( A_{b,{{g}_{1}}}^{\left( - \right)}\left( q \right)A_{b,{{g}_{2}}}^{\left( - \right)}\left( q \right) \right)}
\end{equation}
Аналогічну процедуру застосуємо для внутрішніх індексів 
\begin{equation}
\begin{split}
  & {{A}_{{{g}_{1}}{{g}_{2}}}}\left( q \right)=A\left( q \right){{\delta }_{{{g}_{1}}{{g}_{2}}}}+\ldots , \\ 
& A\left( q \right)=\frac{1}{14}{{g}^{2}}\sum\limits_{b=4}^{6}{\left( A_{b,{{g}_{1}}}^{\left( - \right)}\left( q \right)A_{b,{{g}_{1}}}^{\left( - \right)}\left( q \right) \right)}= \\ 
& =\frac{1}{14}V\left( q \right), \\ 
\end{split}
\label{eq:Videlenie_scalara_po_cvetu_glubola}
\end{equation}
де $V\left( q \right)$ визначене співвідношенням (\ref{Pole_V_ot_q}) і по індексу ${{g}_{1}},$ який повторюється проводиться підсумування.

Кінетична частина лагранжіану для поля $ {{A}_{{{g}_{1}}{{g}_{2}}}}\left( q \right) $ може бути записана у вигляді
\begin{equation}
\begin{split}
  & L_{G}^{\left( 0 \right)}=\frac{1}{2}{{g}^{ab}}\frac{\partial {{A}_{{{g}_{1}}{{g}_{2}}}}\left( q \right)}{\partial {{q}^{a}}}\frac{\partial {{A}_{{{g}_{1}}{{g}_{2}}}}\left( q \right)}{\partial {{q}^{b}}}- \\ 
& -\frac{1}{2}M_{G}^{2}{{A}_{{{g}_{1}}{{g}_{2}}}}\left( q \right){{A}_{{{g}_{1}}{{g}_{2}}}}\left( q \right). \\ 
\end{split}
\label{eq:Cineticna_chastina_lagranjianu_G}
\end{equation}
Величину $M_{G}$, що входить в (\ref{eq:Cineticna_chastina_lagranjianu_G}) із звичайних міркувань, що використовуються при побудові польових лагранжіанів, називатимемо затравочною масою глюболу. При цьому \mbox{<<справжня>>} маса глюболу розглядатиметься далі. Оскільки поле $ A_{a,{{g}_{1}}}^{\left(- \right)}\left( q \right) $ перетворюється за \mbox{<<чистим>>} представленням локальної групи $S{{U}_{c}}\left( 3 \right)$ матрицями оберненими до матриць приєднаного представлення, доданок що містить $M_{G}$ не порушує локальну калібрувальну інваріантність. Натомість її порушують доданки, які містять похідні. Але така ситуація не відрізняється від такої що має місце при розгляді інших полів і вона може бути виправлена шляхом заміни звичайних похідних на коваріантні. 
\begin{equation}
\begin{split}
  & {{D}_{a}}\left( {{A}_{{{g}_{1}}{{g}_{2}}}}\left( q \right) \right)=\frac{\partial {{A}_{{{g}_{1}}{{g}_{2}}}}\left( q \right)}{\partial {{q}^{a}}}- \\ 
& -gA_{{{g}_{3}},a}^{\left( 1 \right)}\left( q \right)I_{{{g}_{1}},{{g}_{11}}}^{{{g}_{3}}}{{A}_{{{g}_{11}}{{g}_{2}}}}\left( q \right)- \\ 
& -gA_{{{g}_{3}},a}^{\left( 2 \right)}\left( q \right)I_{{{g}_{2}},{{g}_{12}}}^{{{g}_{3}}}{{A}_{{{g}_{1}}{{g}_{12}}}}\left( q \right). \\ 
\end{split}
\label{eq:Kovariantna_pohidna_vid_Ag1g2}
\end{equation}
Тут $ I_{{{g}_{2}},{{g}_{3}}}^{{{g}_{1}}}- $ генератори приєднаного представлення групи $S{{U}_{c}}\left( 3 \right),$ які можна побудувати зі структурних констант цієї групи. Враховуючи (\ref{eq:Videlenie_scalara_po_cvetu_glubola}) і покладаючи нулю проекції на всі інші інваріантні підпростори окрім скалярного, маємо замість (\ref{eq:Kovariantna_pohidna_vid_Ag1g2}):
\begin{equation}
\begin{split}
  & {{D}_{a}}\left( V\left( q \right){{\delta }_{{{g}_{1}}{{g}_{2}}}} \right)=\frac{\partial V\left( q \right)}{\partial {{q}^{a}}}{{\delta }_{{{g}_{1}}{{g}_{2}}}}+ \\ 
& +g\left( A_{{{g}_{3}},a}^{\left( 2 \right)}\left( q \right)-A_{{{g}_{3}},a}^{\left( 1 \right)}\left( q \right) \right)I_{{{g}_{1}},{{g}_{2}}}^{{{g}_{3}}}V\left( q \right), \\ 
\end{split}
\label{eq:Kovariantna_pohidna_na_skalarnomu_pidprostori}
\end{equation}
або з урахуванням (\ref{eq:Jacobi_dla_A})
\begin{equation}
\begin{split}
& {{D}_{a}}\left( V\left( q \right){{\delta }_{{{g}_{1}}{{g}_{2}}}} \right)= \\ 
& =\frac{\partial V\left( q \right)}{\partial {{q}^{a}}}{{\delta }_{{{g}_{1}}{{g}_{2}}}}+gA_{{{g}_{3}},a}^{\left( - \right)}\left( q \right)I_{{{g}_{1}},{{g}_{2}}}^{{{g}_{3}}}V\left( q \right). \\ 
\end{split}
\label{eq:Kovariantna_pohidna_na_skalarnomu_pidprostori_Aminus}
\end{equation}
З урахуванням (\ref{eq:Videlenie_scalara_po_cvetu_glubola}) і заміни звичайних похідних на коваріантні замість лагрнажіану (\ref{eq:Cineticna_chastina_lagranjianu_G}) , отримаємо (тут і далі для зручності, числовий множник 1/14 з виразу (\ref{eq:Videlenie_scalara_po_cvetu_glubola}) ми включили в функцію $V\left( q \right)$ тим самим несуттєво її перевизначивши)
\begin{equation}
\begin{split}
  & {{L}_{G}}=\frac{1}{2}{{g}^{ab}}\left( \frac{\partial V\left( q \right)}{\partial {{q}^{a}}}{{\delta }_{{{g}_{1}}{{g}_{2}}}}+gA_{a,{{g}_{3}}}^{\left( - \right)}\left( q \right)I_{{{g}_{1}},{{g}_{2}}}^{{{g}_{3}}}V\left( q \right) \right)\times  \\ 
& \times \left( \frac{\partial V\left( q \right)}{\partial {{q}^{b}}}{{\delta }_{{{g}_{1}}{{g}_{2}}}}+gA_{b,{{g}_{13}}}^{\left( - \right)}\left( q \right)I_{{{g}_{1}},{{g}_{2}}}^{{{g}_{13}}}V\left( q \right) \right) \\ 
& -4M_{G}^{2}{{\left( V\left( q \right) \right)}^{2}}. \\ 
\end{split}
\label{eq:}
\end{equation}
Після перетворень, замість цього виразу отримаємо
\begin{equation}
\begin{split}
  & {{L}_{G}}=4{{g}^{ab}}\frac{\partial V\left( q \right)}{\partial {{q}^{a}}}\frac{\partial V\left( q \right)}{\partial {{q}^{b}}}+ \\ 
& +{{g}^{2}}{{g}^{ab}}A_{a,{{g}_{3}}}^{\left( - \right)}\left( q \right)A_{b,{{g}_{13}}}^{\left( - \right)}\left( q \right)I_{{{g}_{1}},{{g}_{2}}}^{{{g}_{3}}}I_{{{g}_{1}},{{g}_{2}}}^{{{g}_{13}}}{{\left( V\left( q \right) \right)}^{2}}- \\ 
& -4M_{G}^{2}{{\left( V\left( q \right) \right)}^{2}}. \\ 
\end{split}
\label{eq:Lagranjian_skalarnogo_glubolu}
\end{equation}
Оскільки структурні константи групи  $S{{U}_{c}}\left( 3 \right)$ відомі з комутаційних співвідношень між матрицями Гелл-Мана, то безпосереднім розрахунком встановлюємо що
\begin{equation}\label{zgortka_structurnix_konstant}
I_{{{g}_{1}},{{g}_{2}}}^{{{g}_{3}}}I_{{{g}_{1}},{{g}_{2}}}^{{{g}_{13}}}=12{{\delta }^{{{g}_{3}}{{g}_{13}}}}.
\end{equation}
З урахуванням цього результату і визначення (\ref{Pole_V_ot_q}) функції $ V\left( q \right) $, отримаємо лагранжіан, що містить лише цю польову функцію:
\begin{equation}
\begin{split}
  & {{L}_{V}}=\frac{1}{2}{{g}^{ab}}\frac{\partial V\left( q \right)}{\partial {{q}^{a}}}\frac{\partial V\left( q \right)}{\partial {{q}^{b}}}+ \\ 
& -\frac{3}{2}{{\left( V\left( q \right) \right)}^{3}}-\frac{1}{2}M_{G}^{2}{{\left( V\left( q \right) \right)}^{2}}. \\
\end{split}
\label{eq:Lagranjian_dla_V}
\end{equation}
Тут ми поділили лагранжіан на коефіцієнт 8, щоб надати його кінетичні частині вид, звичайний для дійсного скалярного поля. 

Маючи лагрнжіан для поля $ V\left( q \right) $ можемо отримати динамічне рівняння для цього поля, як рівняння Лагранжа-Ейлера:
\begin{equation}\label{Lagrang_Eiler_obhij_vid}
\frac{\partial {{L}_{V}}}{\partial V}-\frac{\partial }{\partial {{q}^{c}}}\left( \frac{\partial {{L}_{V}}}{\partial \left( {\partial V\left( q \right)}/{\partial {{q}^{c}}}\; \right)} \right)=0.
\end{equation}
З урахуванням (\ref{eq:Lagranjian_dla_V}) маємо
\begin{equation}
\begin{split}
  & \frac{\partial {{L}_{V}}}{\partial V}=-\frac{9}{2}{{\left( V\left( q \right) \right)}^{2}}-M_{G}^{2}V\left( q \right), \\ 
& \frac{\partial {{L}_{V}}}{\partial \left( {\partial V\left( q \right)}/{\partial {{q}^{c}}}\; \right)}= \\ 
& =\frac{1}{2}{{g}^{cb}}\frac{\partial V\left( q \right)}{\partial {{q}^{b}}}+\frac{1}{2}{{g}^{ac}}\frac{\partial V\left( q \right)}{\partial {{q}^{a}}}= \\ 
& =\frac{1}{2}{{g}^{ca}}\frac{\partial V\left( q \right)}{\partial {{q}^{a}}}+\frac{1}{2}{{g}^{ca}}\frac{\partial V\left( q \right)}{\partial {{q}^{a}}}= \\ 
& ={{g}^{ca}}\frac{\partial V\left( q \right)}{\partial {{q}^{a}}}. \\ 
\end{split}
\label{eq:Rozraxunoc_poxidnix_dla_Lagranga_Eilera}
\end{equation}
Тут ми скористалися симетричністю тензора Мінковськго. Підставляючи (\ref{eq:Rozraxunoc_poxidnix_dla_Lagranga_Eilera}) в (\ref{Lagrang_Eiler_obhij_vid}) отримуємо динамічне рівняння для поля $ V\left( q \right) :$
\begin{equation}
\begin{split}
  & -{{g}^{ca}}\frac{{{\partial }^{2}}V\left( q \right)}{\partial {{q}^{c}}\partial {{q}^{a}}}-M_{G}^{2}V\left( q \right)+ \\ 
& -\frac{9}{2}{{\left( V\left( q \right) \right)}^{2}}=0. \\ 
\end{split}
\label{eq:Dinamichne_rivnanna_dla_pola_Votq}
\end{equation}

Отже тепер, лагранжіан $L={{L}_{\mu }}+{{L}_{V}}$, де ${{L}_{\mu }}$ визначається (\ref{eq:L_mu_iz_vrachuvannam_zvazkiv}), а ${{L}_{V}}$ - визначається (\ref{eq:Lagranjian_dla_V}), можна розглядати як лагранжіан для взаємодіючих мезонного поля $ {{\psi }_{{{f}_{1}},{{f}_{2}}}}\left( q \right) $ і скалярного глюбольного поля $ V\left( q \right) $, бо як показно в роботі \cite{Chudak:2016} після квантування оператори народження і знищення, що відповідають цим полям змінюють числа заповнення зв'язаних станів кварка і антикварка, або двох глюонів.

Якщо лагранжіан $ {{L}_{\mu }}$ поділити на коефіцієнт 3, щоб його кінетична частина прийняла вид, звичайний для комплексного скалярного поля, то лагранжіан  $L$ запишеться таким чином
\begin{equation}
\begin{split}
  & L={{g}^{ab}}\left( {\partial \psi _{{{f}_{1}},{{f}_{2}}}^{*}\left( q \right)}/{\partial {{q}^{a}}}\; \right)\left( {\partial {{\psi }_{{{f}_{1}},{{f}_{2}}}}\left( q \right)}/{\partial {{q}^{b}}}\; \right)+ \\ 
& -\frac{8}{3}V\left( q \right)\psi _{{{f}_{1}},{{f}_{2}}}^{*}\left( q \right){{\psi }_{{{f}_{1}},{{f}_{2}}}}\left( q \right)- \\ 
& -M_{\mu }^{2}\psi _{{{f}_{1}},{{f}_{2}}}^{*}\left( q \right){{\psi }_{{{f}_{1}},{{f}_{2}}}}\left( q \right)+ \\ 
& +\frac{1}{2}{{g}^{ab}}\left( {\partial V\left( q \right)}/{\partial {{q}^{a}}}\; \right)\left( {\partial V\left( q \right)}/{\partial {{q}^{b}}}\; \right)+ \\ 
&-\frac{3}{2}{{\left( V\left( q \right) \right)}^{3}}-\frac{1}{2}M_{G}^{2}{{\left( V\left( q \right) \right)}^{2}}. \\ 
\end{split}
\label{eq:Lagrangian_L}
\end{equation}
Виділимо окремо залежності і підсумування по координатах центру мас і по відносних координатах:
\begin{equation}
\begin{split}
  & L={{g}^{ab,Minc}}\left( {\partial \psi _{{{f}_{1}},{{f}_{2}}}^{*}\left( X,\vec{y} \right)}/{\partial {{X}^{a}}}\; \right)\times  \\ 
& \times \left( {\partial {{\psi }_{{{f}_{1}},{{f}_{2}}}}\left( X,\vec{y} \right)}/{\partial {{X}^{b}}}\; \right)- \\ 
& -4\sum\limits_{b=1}^{3}{\left( {\partial \psi _{{{f}_{1}},{{f}_{2}}}^{*}\left( X,\vec{y} \right)}/{\partial {{y}^{b}}}\; \right)}\left( {\partial {{\psi }_{{{f}_{1}},{{f}_{2}}}}\left( X,\vec{y} \right)}/{\partial {{y}^{b}}}\; \right) \\ 
& -\frac{8}{3}V\left( X,\vec{y} \right)\psi _{{{f}_{1}},{{f}_{2}}}^{*}\left( X,\vec{y} \right){{\psi }_{{{f}_{1}},{{f}_{2}}}}\left( X,\vec{y} \right)- \\ 
& -M_{\mu }^{2}\psi _{{{f}_{1}},{{f}_{2}}}^{*}\left( X,\vec{y} \right){{\psi }_{{{f}_{1}},{{f}_{2}}}}\left( X,\vec{y} \right)+ \\ 
& +\frac{1}{2}{{g}^{ab,Minc}}\left( {\partial V\left( X,\vec{y} \right)}/{\partial {{X}^{a}}}\; \right)\left( {\partial V\left( X,\vec{y} \right)}/{\partial {{X}^{b}}}\; \right)- \\ 
& -4\sum\limits_{b=1}^{3}{{{\left( {\partial V\left( X,\vec{y} \right)}/{\partial {{y}^{b}}}\; \right)}^{2}}}+ \\ 
& -\frac{3}{2}{{\left( V\left( X,\vec{y} \right) \right)}^{3}}-\frac{1}{2}M_{G}^{2}{{\left( V\left( X,\vec{y} \right) \right)}^{2}}. \\ 
\end{split}
\label{eq:Polnij_lagranjian_v_peremennih_Jacobi}
\end{equation}
Тут ${g}^{ab,Minc}, $ як і раніше, тензор Мінковського, але тепер він розглядається як тензор з верхніми індексами.
Якщо доданок, що містить похідні від мезонного поля по внутрішніх координатах проінтегрувати по частинах, отримаємо:
\begin{equation}
\begin{split}
& L={{g}^{ab,Minc}}\left( {\partial \psi _{{{f}_{1}},{{f}_{2}}}^{*}\left( X,\vec{y} \right)}/{\partial {{X}^{a}}}\; \right)\times  \\ 
& \times \left( {\partial {{\psi }_{{{f}_{1}},{{f}_{2}}}}\left( X,\vec{y} \right)}/{\partial {{X}^{b}}}\; \right)- \\ 
& -\psi _{{{f}_{1}},{{f}_{2}}}^{*}\left( X,\vec{y} \right)2\hat{H}_{0}^{\text{internal}}\left( {{\psi }_{{{f}_{1}},{{f}_{2}}}}\left( X,\vec{y} \right) \right) \\ 
& -\frac{8}{3}V\left( X,\vec{y} \right)\psi _{{{f}_{1}},{{f}_{2}}}^{*}\left( X,\vec{y} \right){{\psi }_{{{f}_{1}},{{f}_{2}}}}\left( X,\vec{y} \right)- \\ 
& -M_{\mu }^{2}\psi _{{{f}_{1}},{{f}_{2}}}^{*}\left( X,\vec{y} \right){{\psi }_{{{f}_{1}},{{f}_{2}}}}\left( X,\vec{y} \right)+ \\ 
& +\frac{1}{2}{{g}^{ab,Minc}}\left( {\partial V\left( X,\vec{y} \right)}/{\partial {{X}^{a}}}\; \right)\left( {\partial V\left( X,\vec{y} \right)}/{\partial {{X}^{b}}}\; \right)- \\ 
& -4\sum\limits_{b=1}^{3}{{{\left( {\partial V\left( X,\vec{y} \right)}/{\partial {{y}^{b}}}\; \right)}^{2}}}+ \\ 
& -\frac{3}{2}{{\left( V\left( X,\vec{y} \right) \right)}^{3}}-\frac{1}{2}M_{G}^{2}{{\left( V\left( X,\vec{y} \right) \right)}^{2}}, \\ 
\end{split}
\label{eq:Z_nulovim_H_internal}
\end{equation}
де введене позначення
\begin{equation}
\begin{split}
\hat{H}_{0}^{\text{internal}}=-2{{\Delta }_{{\vec{y}}}}=-2\sum\limits_{b=1}^{3}{\frac{{{\partial }^{2}}}{\partial {{\left( {{y}^{b}} \right)}^{2}}}}.
\end{split}
\label{eq:Poznachenna_H0internal}
\end{equation}
Тут позначення $ {{\Delta }_{{\vec{y}}}}\equiv \sum\limits_{b=1}^{3}{\frac{{{\partial }^{2}}}{{{\left( \partial {{y}_{b}} \right)}^{2}}}} $ означає оператор Лапласа по внутрішніх координатах.
Оператор $ \hat{H}_{0}^{\text{internal}} $ збігається з кінетичною частиною внутрішнього гамільтоніану системи двох частинок в нерелятивістській квантовій механіці. Якщо тепер повернутися до динамічного рівняння (\ref{eq:Dinamichne_rivnanna_dla_pola_Votq}) для функції $V\left( q \right)=V\left( X,\vec{y} \right)$, то можемо розглянути його частковий розв'язок ${{V}_{0}}\left( {\vec{y}} \right),$ який залежить тільки від внутрішніх змінних. Функція ${{V}_{0}}\left( {\vec{y}} \right),$ виходячи з (\ref{eq:Dinamichne_rivnanna_dla_pola_Votq}), визначатиметься рівнянням 
 \begin{equation}\label{Rivnanna_dla_potencialu}
 4{{\Delta }_{{\vec{y}}}}{{V}_{0}}\left( {\vec{y}} \right)-M_{G}^{2}{{V}_{0}}\left( {\vec{y}} \right)-\frac{9}{2}{{\left( {{V}_{0}}\left( {\vec{y}} \right) \right)}^{2}}=0.
 \end{equation}
Якщо тепер представити функцію $ V\left( X,\vec{y} \right) $ в виді
\begin{equation}
\begin{split}
  & V\left( X,\vec{y} \right)={{V}_{0}}\left( {\vec{y}} \right)+{{V}_{1}}\left( X,\vec{y} \right), \\ 
& {{V}_{1}}\left( X,\vec{y} \right)\equiv V\left( X,\vec{y} \right)-{{V}_{0}}\left( {\vec{y}} \right), \\ 
\end{split}
\label{eq:Potencial}
\end{equation}
то частина лагранжіану (\ref{eq:Z_nulovim_H_internal}), що містить польові функції мезонного поля запишеться так
\begin{equation}
\begin{split}
  & L={{g}^{ab,Minc}}\left( {\partial \psi _{{{f}_{1}},{{f}_{2}}}^{*}\left( X,\vec{y} \right)}/{\partial {{X}^{a}}}\; \right)\times  \\ 
& \times \left( {\partial {{\psi }_{{{f}_{1}},{{f}_{2}}}}\left( X,\vec{y} \right)}/{\partial {{X}^{b}}}\; \right)- \\ 
& -\psi _{{{f}_{1}},{{f}_{2}}}^{*}\left( X,\vec{y} \right)\Biggl(  M_{\mu }^{2}+2{{M}_{\mu }} \times \\ 
& \left. \times \left( -\frac{1}{\left( {{{M}_{\mu }}}/{2}\; \right)}{{\Delta }_{{\vec{y}}}}+\frac{4}{3{{M}_{\mu }}}{{V}_{0}}\left( {\vec{y}} \right) \right) \Biggr) \right. \left( {{\psi }_{{{f}_{1}},{{f}_{2}}}}\left( X,\vec{y} \right) \right)- \\ 
& -\frac{8}{3}{{V}_{1}}\left( X,\vec{y} \right)\psi _{{{f}_{1}},{{f}_{2}}}^{*}\left( X,\vec{y} \right){{\psi }_{{{f}_{1}},{{f}_{2}}}}\left( X,\vec{y} \right)+\ldots  \\ 
\end{split}
\label{eq:Lagrangian_s_Hinternal}
\end{equation}
Тут як трикрапки позначена решта доданків лагарнжіана (\ref{eq:Z_nulovim_H_internal}), які ми проаналізуємо далі.
Вираз $\left( -{1}/{\left( {{{M}_{\mu }}}/{2}\; \right)}\;{{\Delta }_{{\vec{y}}}}+\left( {4}/{\left( 3{{M}_{\mu }} \right)}\; \right){{V}_{0}}\left( {\vec{y}} \right) \right) $ має вид внутрішнього, нерелятивістського двочастинкового гамільтоніану, в якому роль потенційної енергії взаємодії кварка з антикварком з точністю до множника $ \left( {4}/{\left( 3{{M}_{\mu }} \right)}\; \right) $ грає функція $ {{V}_{0}}\left( {\vec{y}} \right) $ що є розв'язком рівняння (\ref{Rivnanna_dla_potencialu}). Втім, внутрішній гамільтоніан зручніше ввести таким чином:
\begin{equation}
\begin{split}
&{{\hat{H}}^{\text{internal}}}=\\
&={{M}_{\mu }}+\left( -\frac{1}{\left( {{{M}_{\mu }}}/{2}\; \right)}{{\Delta }_{{\vec{y}}}}+\left( {4}/{\left( 3{{M}_{\mu }} \right)}\; \right){{V}_{0}}\left( {\vec{y}} \right) \right).
\end{split}
\label{eq:Hinternal}
\end{equation}
Якщо вважати що внутрішній стан мезону  може бути добре описаний в нерелятивістському наближенні, то власні значення оператору  $\left( -{1}/{\left( {{{M}_{\mu }}}/{2}\; \right)}\;{{\Delta }_{{\vec{y}}}}+\left( {4}/{\left( 3{{M}_{\mu }} \right)}\; \right){{V}_{0}}\left( {\vec{y}} \right) \right)$ повинні бути суттєво меншими ніж затравочна маса мезону $ {{M}_{\mu }}$. У якості цієї затравочної маси можна взяти наприклад сумарну масу кварка і антикварка. Оскільки ми вважаємо, що мезон складається з певного кварка і антикварка, то це означає що за рахунок внутрішніх процесів в мезоні, ані нові конституентні кварки, ані нові конституентні антикварки народжуватися не можуть, що й означає що характерні власні значення оператору $\left( -{1}/{\left( {{{M}_{\mu }}}/{2}\; \right)}\;{{\Delta }_{{\vec{y}}}}+\left( {4}/{\left( 3{{M}_{\mu }} \right)}\; \right){{V}_{0}}\left( {\vec{y}} \right) \right) $ є набагато меншими ніж $ {{M}_{\mu }}$. Але тоді, нехтуючи квадратом оператору $\left( -{1}/{\left( {{{M}_{\mu }}}/{2}\; \right)}\;{{\Delta }_{{\vec{y}}}}+\left( {4}/{\left( 3{{M}_{\mu }} \right)}\; \right){{V}_{0}}\left( {\vec{y}} \right) \right)$ у порівнянні з $ M_{\mu }^{2} $ маємо
\begin{equation}
\begin{split}
  & {{\left( {{{\hat{H}}}^{\text{internal}}} \right)}^{2}}\approx M_{\mu }^{2}+ \\ 
& +2{{M}_{\mu }}\left( -\frac{1}{\left( {{{M}_{\mu }}}/{2}\; \right)}{{\Delta }_{{\vec{y}}}}+\left( {4}/{\left( 3{{M}_{\mu }} \right)}\; \right){{V}_{0}}\left( {\vec{y}} \right) \right) \\ 
\end{split}
\label{eq:Hinternal_kvadrat}
\end{equation}
Тоді замість (\ref{eq:Lagrangian_s_Hinternal}) маємо
\begin{equation}
\begin{split}
  & L={{g}^{ab,Minc}}\left( {\partial \psi _{{{f}_{1}},{{f}_{2}}}^{*}\left( X,\vec{y} \right)}/{\partial {{X}^{a}}}\; \right)\times  \\ 
& \times \left( {\partial {{\psi }_{{{f}_{1}},{{f}_{2}}}}\left( X,\vec{y} \right)}/{\partial {{X}^{b}}}\; \right)- \\ 
& -\psi _{{{f}_{1}},{{f}_{2}}}^{*}\left( X,\vec{y} \right){{\left( {{{\hat{H}}}^{\text{internal}}} \right)}^{2}}\left( {{\psi }_{{{f}_{1}},{{f}_{2}}}}\left( X,\vec{y} \right) \right)- \\ 
& -\frac{8}{3}{{V}_{1}}\left( X,\vec{y} \right)\psi _{{{f}_{1}},{{f}_{2}}}^{*}\left( X,\vec{y} \right){{\psi }_{{{f}_{1}},{{f}_{2}}}}\left( X,\vec{y} \right)+\ldots  \\ 
\end{split}
\label{eq:Lagrangian_s_Hinternal_kvadrat}
\end{equation}
Тоді якщо залежність функції $ {{\psi }_{{{f}_{1}},{{f}_{2}}}}\left( X,\vec{y} \right) $ від внутрішніх змінних $ \vec{y} $ описувати за допомогою власної функції оператора $ {{\hat{H}}^{\text{internal}}} $ (\ref{eq:Hinternal}), а власне значення розглядати як \mbox{<<справжню>>} масу мезону, то як видно з (\ref{eq:Lagrangian_s_Hinternal_kvadrat}) така маса \mbox{<<правильно>>} увійде в лагранжіан і динамічні рівняння.

Отже, аналізуючи властивості розв'язків рівняння (\ref{Rivnanna_dla_potencialu}) ми можемо отримати інформацію щодо потенціалу міжкваркової взаємодії. Перш ніж проаналізувати ці властивості, знерозміримо це рівняння. Задля цього виразимо розмірності всіх полів, що входять в лагранжіан (\ref{eq:Polnij_lagranjian_v_peremennih_Jacobi}) в ступенях характерної довжини задачі, яку позначимо як $l$.
Оскільки доданки $ \left( -4\sum\limits_{b=1}^{3}{{{\left( {\partial V\left( X,\vec{y} \right)}/{\partial {{y}^{b}}}\; \right)}^{2}}} \right) $ і $ \left( -\frac{1}{2}M_{G}^{2}{{\left( V\left( X,\vec{y} \right) \right)}^{2}} \right) $ в (\ref{eq:Polnij_lagranjian_v_peremennih_Jacobi})  повинні мати однакову розмірність, величина $ M_{G}^{2}$ повинна мати розмірність ${{l}^{-2}}$. Відповідно маса $ M_{G},$ як і повинно бути, має розмірність ${{l}^{-1}}.$
Для того, щоб доданок $ \frac{3}{2}{{\left( V\left( X,\vec{y} \right) \right)}^{3}} $ був тієї самої розмірності, що й $ \left( -4\sum\limits_{b=1}^{3}{{{\left( {\partial V\left( X,\vec{y} \right)}/{\partial {{y}^{b}}}\; \right)}^{2}}} \right) $ і $ \left( -\frac{1}{2}M_{G}^{2}{{\left( V\left( X,\vec{y} \right) \right)}^{2}} \right) ,$ поле $ V\left( X,\vec{y} \right) $ повинне бути порядку ${{l}^{-2}}.$ Це також узгоджується із визначенням (\ref{Pole_V_ot_q}). Дійсно, як видно з виразу (\ref{eq:Kovariantna_pohidna_vid_Ag1g2}), поля $ A_{{{g}_{3}},a}^{\left( 1 \right)}\left( q \right) $ і  $ A_{{{g}_{3}},a}^{\left( 2 \right)}\left( q \right) $ повинні мати ту ж розмірність, що й оператор звичайної похідної ${\partial }/{\partial {{q}^{b}}}\;$. Тобто ці поля, а також їх різниця $ A_{{{g}_{3}},a}^{\left( - \right)}\left( q \right) $ (\ref{eq:Jacobi_dla_A}) повинні мати розмірність ${{l}^{-1}},$ оскільки константа зв'язку $g$ повинна бути безрозмірною щоб можна було порівнювати і додавати величини, що містять константу зв'язку в різних ступенях. Отже з цих міркувань, виходячи з означення (\ref{Pole_V_ot_q}) також дістаємо висновку що поле  $ V\left( X,\vec{y} \right) $ має розмірність ${{l}^{-2}}.$ Виходячи з цього, всі доданки в (\ref{eq:Polnij_lagranjian_v_peremennih_Jacobi}) мають розмірність ${{l}^{-6}}.$ Тому мезонне поле $ {{\psi }_{{{f}_{1}},{{f}_{2}}}}\left( X,\vec{y} \right) $ також повинне бути порядку ${{l}^{-2}}.$ Оскільки дія, що відповідає лагранжіана (\ref{eq:Polnij_lagranjian_v_peremennih_Jacobi}) передбачатиме семикратне інтегрування цього лагранжіана по $d{{X}^{0}}d{{X}^{1}}d{{X}^{2}}d{{X}^{3}}d{{y}^{1}}d{{y}^{2}}d{{y}^{3}},$ задля того, щоб дія була безрозмірна, лагранжіан потрібно помножити на константу розмірності ${{l}^{-1}},$ що ніяк не вплине на динамічні рівняння, що породжуються цим лагранжіаном.

 Отже, введемо безрозмірні внутрішні координати $\vec{r}$, безрозмірну масу глюбола ${{m}_{G}}$ і безрозмірну потенційну енергію $ u\left( {\vec{r}} \right) $ наступним чином:
 \begin{equation}
 \begin{split}
   & \vec{y}=l\vec{r},{{M}_{G}}={{l}^{-1}}{{m}_{G}}, \\ 
 & {{V}_{0}}\left( {\vec{y}} \right)={{V}_{0}}\left( l\vec{r} \right)={{l}^{-2}}u\left( {\vec{r}} \right). \\ 
  \end{split}
 \label{eq:Bezrazmer}
 \end{equation}
 Тоді замість рівняння (\ref{Rivnanna_dla_potencialu}) у введених безрозмірних змінних, отримаємо:
 \begin{equation}\label{Rivnanna_dla_potencialu_Bezrazmer}
 4{{\Delta }_{{\vec{r}}}}u\left( {\vec{r}} \right)-m_{G}^{2}u\left( {\vec{r}} \right)-\frac{9}{2}{{\left( u\left( {\vec{r}} \right) \right)}^{2}}=0.
 \end{equation}
 Тут $ {{\Delta }_{{\vec{r}}}}\equiv \sum\limits_{b=1}^{3}{\frac{{{\partial }^{2}}}{\partial {{\left( {{r}^{b}} \right)}^{2}}}}- $оператор Лапласа по безрозмірних змінних $\vec{r}.$
 
 Розглянемо властивості сферично симетричного розв'язку рівняння (\ref{Rivnanna_dla_potencialu_Bezrazmer}). Тобто, замість змінних $\vec{r}\left( {{r}^{1}},{{r}^{2}},{{r}^{3}} \right)$ перейдемо до сферичних координат $r,\theta ,\phi :$
 \begin{equation}
 \begin{split}
 & {{r}^{1}}=r\sin \left( \theta  \right)\cos \left( \phi  \right), \\ 
 & {{r}^{2}}=r\sin \left( \theta  \right)\sin \left( \phi  \right), \\ 
 & {{r}^{3}}=r\cos \left( \theta  \right), \\ 
  \end{split}
 \label{eq:Sfericheskie_koordinati}
 \end{equation} 
 і розглянемо розв'язок рівняння  (\ref{Rivnanna_dla_potencialu_Bezrazmer}) який є функцією лише від змінної \mbox{$r:u=u\left( r \right).$ }З урахуванням виду оператора Лапласа в сферичних координатах, це рівняння перепишеться таким чином:
 \begin{equation}
 \begin{split}
 \frac{4}{{{r}^{2}}}\frac{d}{dr}\left( {{r}^{2}}\frac{du\left( r \right)}{dr} \right)-m_{G}^{2}u\left( r \right)-\frac{9}{2}{{\left( u\left( r \right) \right)}^{2}}=0.
 \end{split}
 \label{eq:Sfer_sym_uravnenie}
 \end{equation}
 Після стандартної заміни 
 \begin{equation}\label{Standartna_zamina}
 u\left( r \right)=\frac{\chi \left( r \right)}{r},
 \end{equation}
 отримаємо:
 \begin{equation}\label{Newton}
 \frac{{{d}^{2}}\chi \left( r \right)}{d{{r}^{2}}}=\frac{9}{8}\frac{\chi \left( r \right)\left( \chi \left( r \right)+\left( {m_{G}^{2}}/{9}\; \right)r \right)}{r}.
 \end{equation}

Для того, щоб проаналізувати властивості розв'язків рівняння (\ref{Newton}) скористаємось аналогією з класичною механікою. Незалежну змінну $r$ розглядатимемо аналогічно часу. Величину $\chi $ називатимемо \mbox{<<координатою>>}, її першу похідну ${d\chi }/{dr}\;-$ \mbox{<<швидкістю>>}, а другу ${{{d}^{2}}\chi }/{d{{r}^{2}}}\;-$ \mbox{<<прискоренням>>}. Залежність \mbox{<<прискорення>>} від \mbox{<<координати>>}, що визначається правою частиною рівняння  (\ref{Newton}) призводить до того, що на координатній площині $\left( r,\chi  \right)$ виділяються три області, показані на Рис.1,всередині кожної з яких \mbox{<<прискорення>>} має постійний знак, показаний на тому ж Рис.1. Тобто, якщо графік залежності $\chi \left( r \right)$ потрапляє в одну з цих трьох виділених областей, то подальший хід цього графіка визначається відповідним знаком \mbox{<<прискорення>>}.

\begin{figure}
	\center{\includegraphics[scale=0.35]{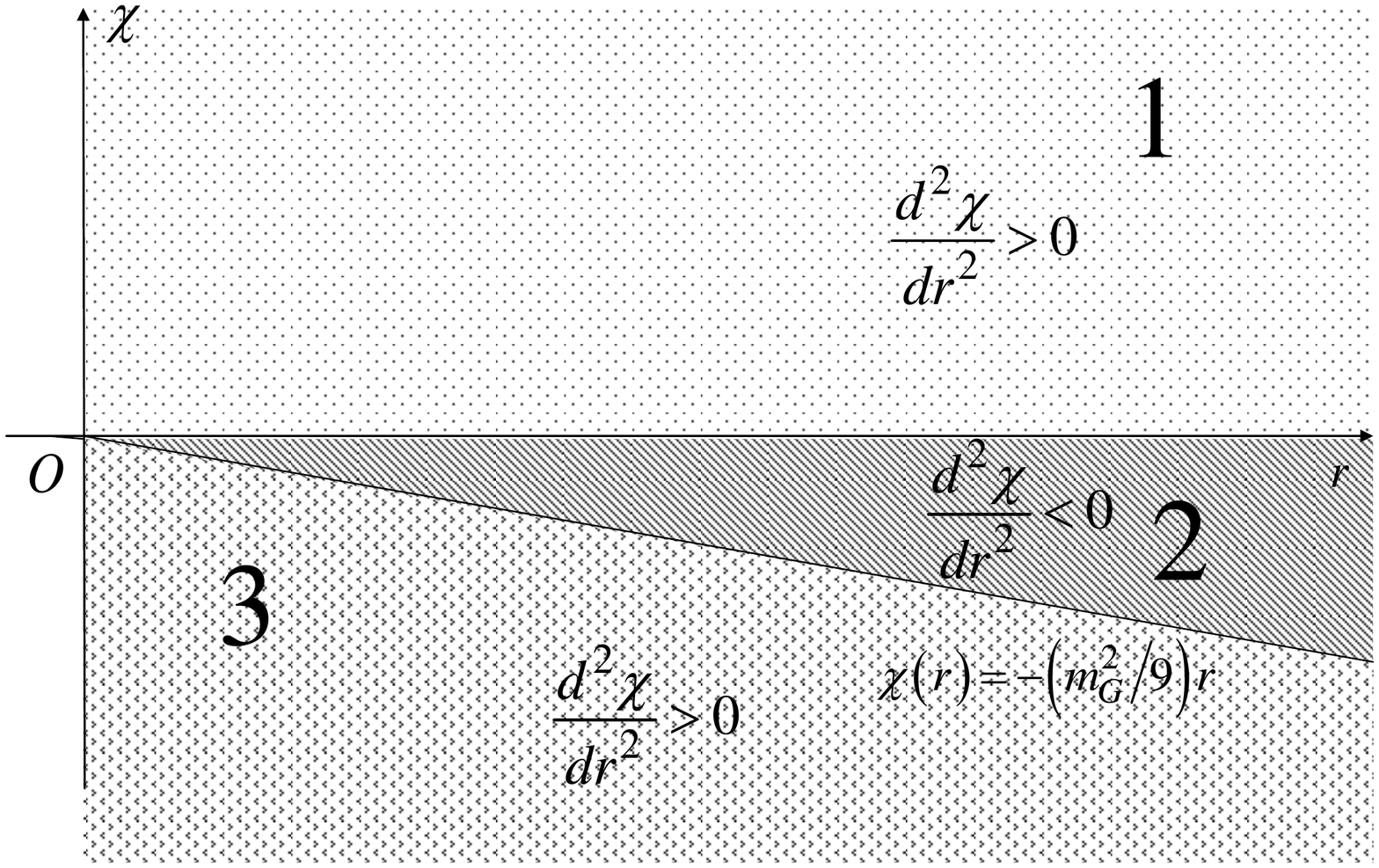}}
	\vskip-1mm\noindent{\footnotesize Рис. 1. Області певного знаку \mbox{<<прискорення>>} для рівняння (\ref{Newton})}%
	\vskip15pt
\end{figure}

Розглянемо граничні умови, які можна накласти на функцію  $\chi \left( r \right).$  З рівняння (\ref{Standartna_zamina}) видно, що якщо ми хочемо отримати скінчену потенційну енергію $u\left( r \right)$ при всіх скінчених значеннях $r$ потрібно покласти:
\begin{equation}\label{Granichna_umova_1}
{{\left. \chi \left( r \right) \right|}_{r=0}}=0.
\end{equation}  
При цьому \mbox{<<початкова швидкість>>} вже не повинна дорівнювати нулю і ми можемо покладати її довільному дійсному числу. Отже покладемо:
\begin{equation}\label{Granichna_umova_2}
{{\left. \frac{d\chi \left( r \right)}{dr} \right|}_{r=0}}=C,C\in \mathbb{R},
\end{equation}
і розглянемо властивості розв'язку рівняння (\ref{Newton}) в залежності від обрання значення $ C. $ Оскільки виходячи з граничних умов (\ref{Granichna_umova_1}) і (\ref{Granichna_umova_2}) при малих $r$ маємо $\chi \left( r \right)=Cr,$ то різні вибори $C$ будуть призводити до потрапляння графіку $\chi \left( r \right)$ при цих малих $  r $ в різні області на Рис.1. При цьому, підставляючи $\chi \left( r \right)=Cr$ в праву частину рівняння  (\ref{Newton}) отримаємо що чисельник в цій правій частині буде пропорційний ${{r}^{2}}$, а знаменник - пропорційний першому ступеню $  r $ . Внаслідок цього ${{\left. \frac{{{d}^{2}}\chi \left( r \right)}{d{{r}^{2}}} \right|}_{r=0}}=0.$
Окрім того, оскільки $\chi \left( r \right)=Cr,$ з то (\ref{Standartna_zamina}) отримаємо 
\begin{equation}
\begin{split}
  & \frac{du\left( r \right)}{dr}=\frac{1}{r}\frac{d\chi \left( r \right)}{dr}- \\ 
& -\frac{1}{{{r}^{2}}}\chi \left( r \right)\xrightarrow{r\to 0}\frac{C}{r}-\frac{Cr}{{{r}^{2}}}=0. \\ 
\end{split}
\label{eq:Asimptotichna_svoboda}
\end{equation}
Тобто модель, що розглядається \textbf{\textit{описує асимптотичну свободу кварків}}. 

Розглянемо розв'язок, що задовольняє граничним умовам (\ref{Granichna_umova_1}) і (\ref{Granichna_umova_2}) при $C>0.$ Тоді при малих $r$ графік $\chi \left( r \right)$ потраплятиме в область 1 на Рис. 1, як показано на Рис. 2.
\begin{figure}
	\center{\includegraphics[scale=0.35]{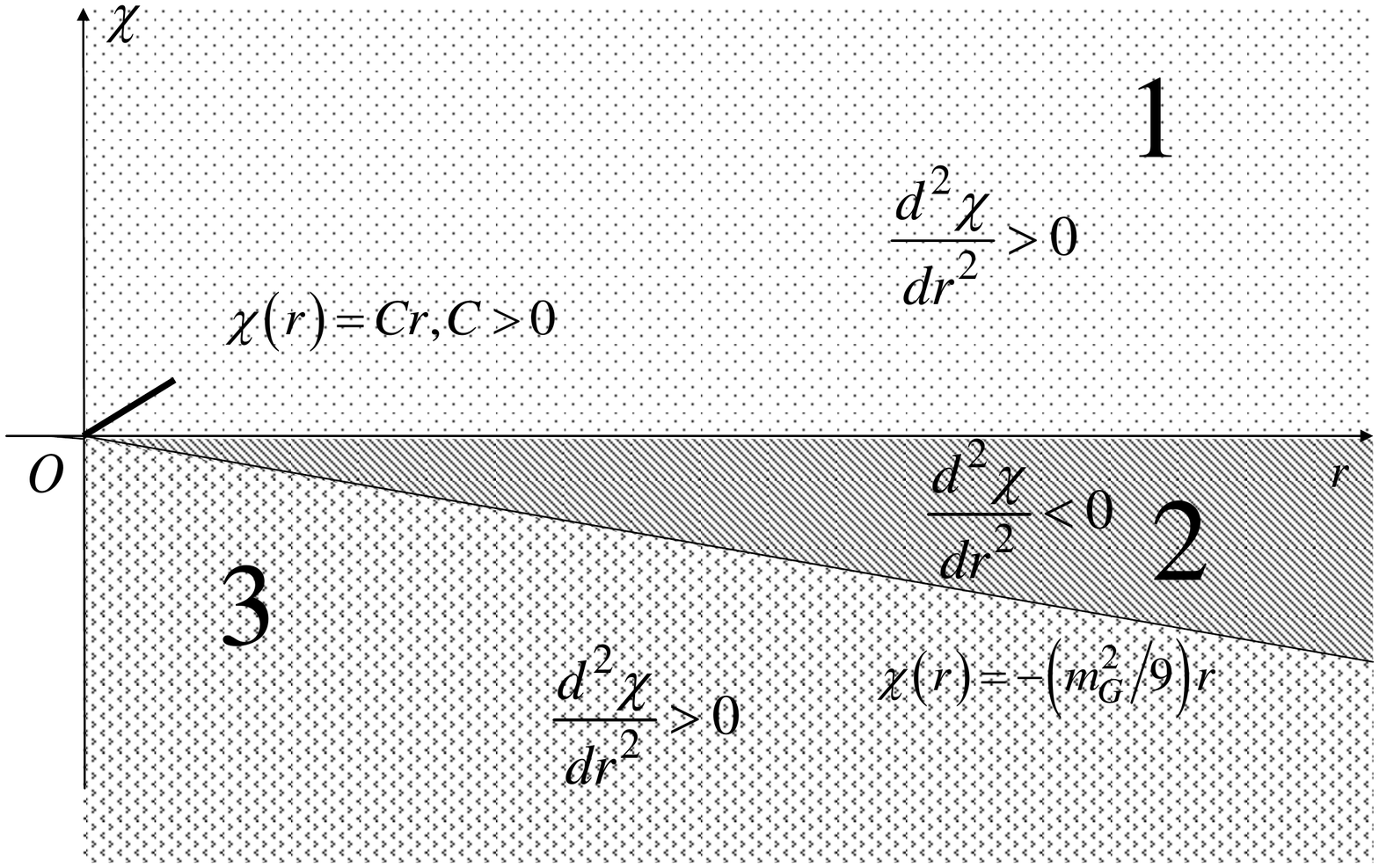}}
	\vskip-1mm\noindent{\footnotesize Рис. 2. Залежність $\chi \left( r \right)$  при малих $r$  для випадку $C>0.$} %
	\vskip15pt
\end{figure}
  Оскільки \mbox{<<швидкість>>} і \mbox{<<прискорення>>} є додатними, то швидкість буде зростати. Це означає що графік  $\chi \left( r \right)$ за таких умов не вийде за межі області 1 на Рис.1, або Рис.2. Тобто швидкість не зможе стати від'ємною, і оскільки вона постійно зростатиме - то не зможе стати рівною нулю. Тому при $r$ що наближатиметься до плюс нескінченості \mbox{<<координата>>} $\chi \left( r \right)$ також наближатиметься до плюс нескінченості. Схематично це показано на Рис.3
 \begin{figure}
 	\center{\includegraphics[scale=0.35]{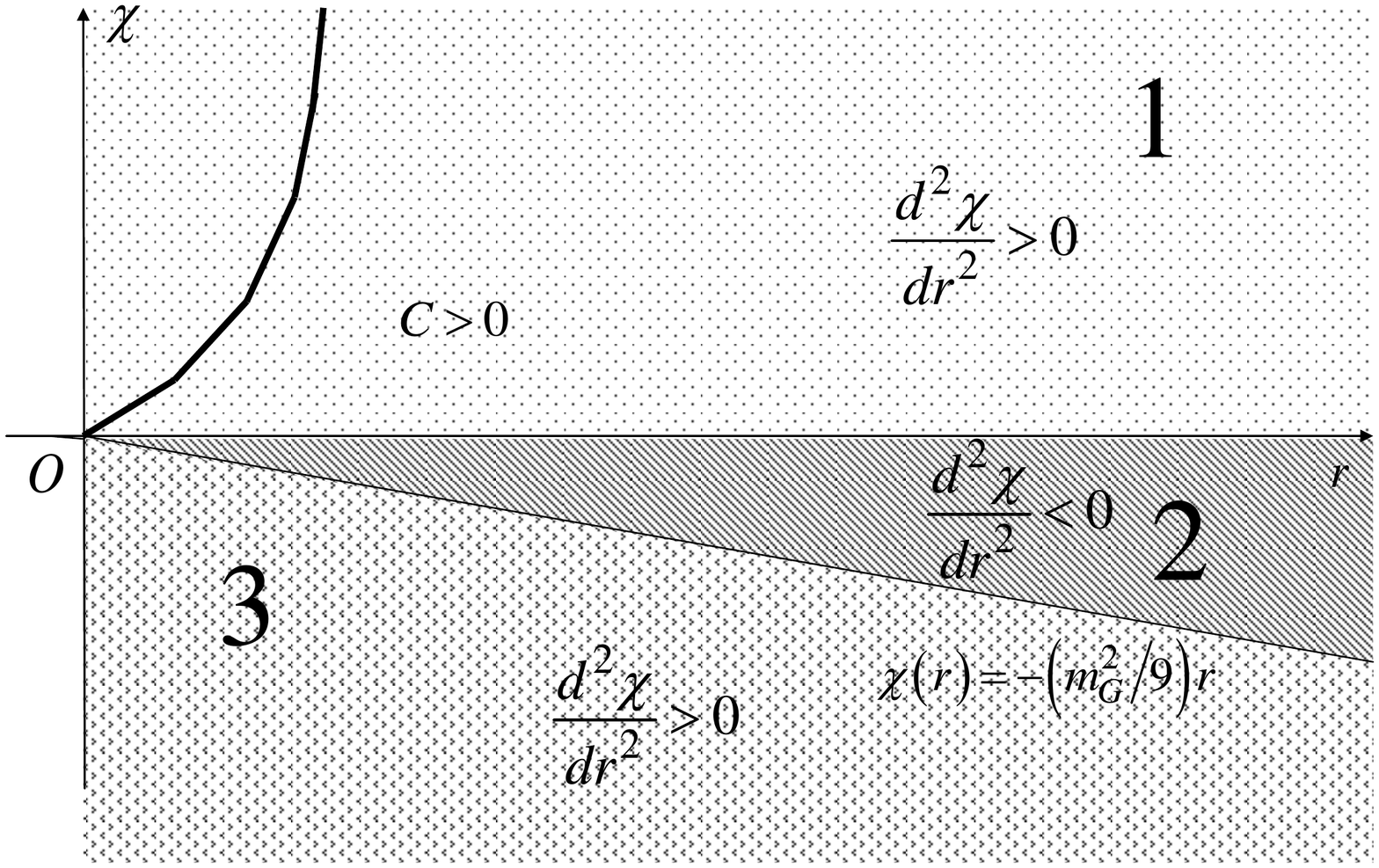}}
 	\vskip-1mm\noindent{\footnotesize Рис. 3. Схематичний вид залежності $\chi \left( r \right)$  для випадку $C>0.$} %
 	\vskip15pt
 \end{figure}  
Ці міркування підтверджуються результатом чисельного розрахунку, який наведено на Рис.4
\begin{figure}
	\center{\includegraphics[scale=0.35]{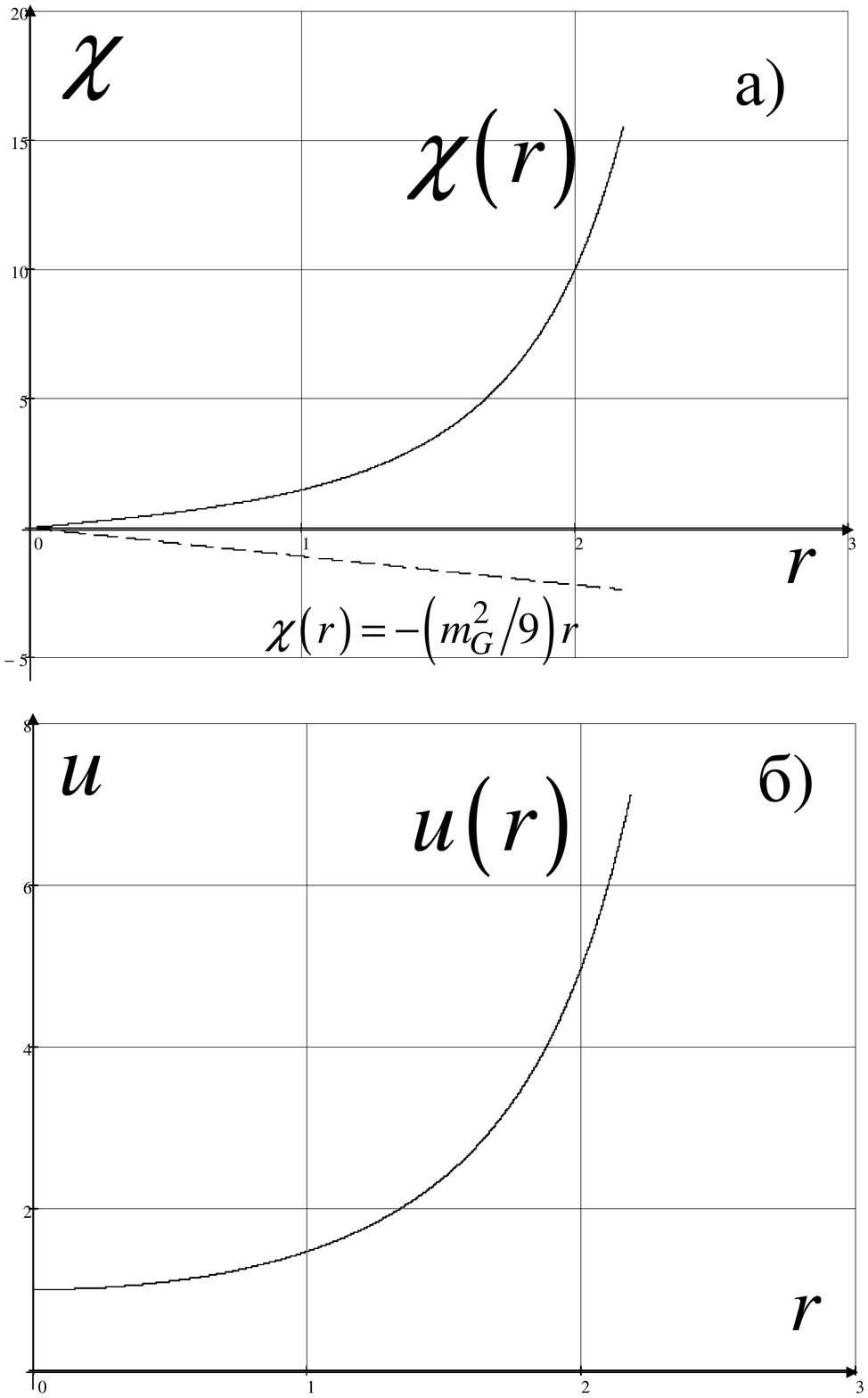}}
	\vskip-1mm\noindent{\footnotesize Рис. 4. Результат чисельного розрахунку залежності $\chi \left( r \right)$  для випадку $C=1.0, {m_{G}^{2}}/{9}\;=1.1$ (а) і відповідної залежності безрозмірного міжкваркового потенціалу $u\left( r \right)$ від безрозмірної відстані $ r.$(б).} %
	\vskip15pt
\end{figure}  
Отже маємо потенціал міжкваркової взаємодії що наближається до плюс нескінченості при наближенні відстані до плюс нескінченності. Отже, \textbf{\textit{розглянута модель описує конфайнмент кварків}}.

У випадку $ -\left( {m_{G}^{2}}/{9}\; \right)r<C<0 $ графік потрапляє в другу область. Оскільки при цьому \mbox{<<прискорення>>} від'ємне, то \mbox{<<швидкість>>} буде залишатися від'ємною і зростати за модулем. Відповідно, \mbox{<<координата>>} також буде залишатися від'ємною і зростати за модулем. Все це призведе до потрапляння графіка в третю область. Після цього можна очікувати коливання навколо розв'язку $ \chi \left( r \right)=-\left( {m_{G}^{2}}/{9}\; \right)r\, $ як це схематично показано на Рис.5.
\begin{figure}
	\center{\includegraphics[scale=0.35]{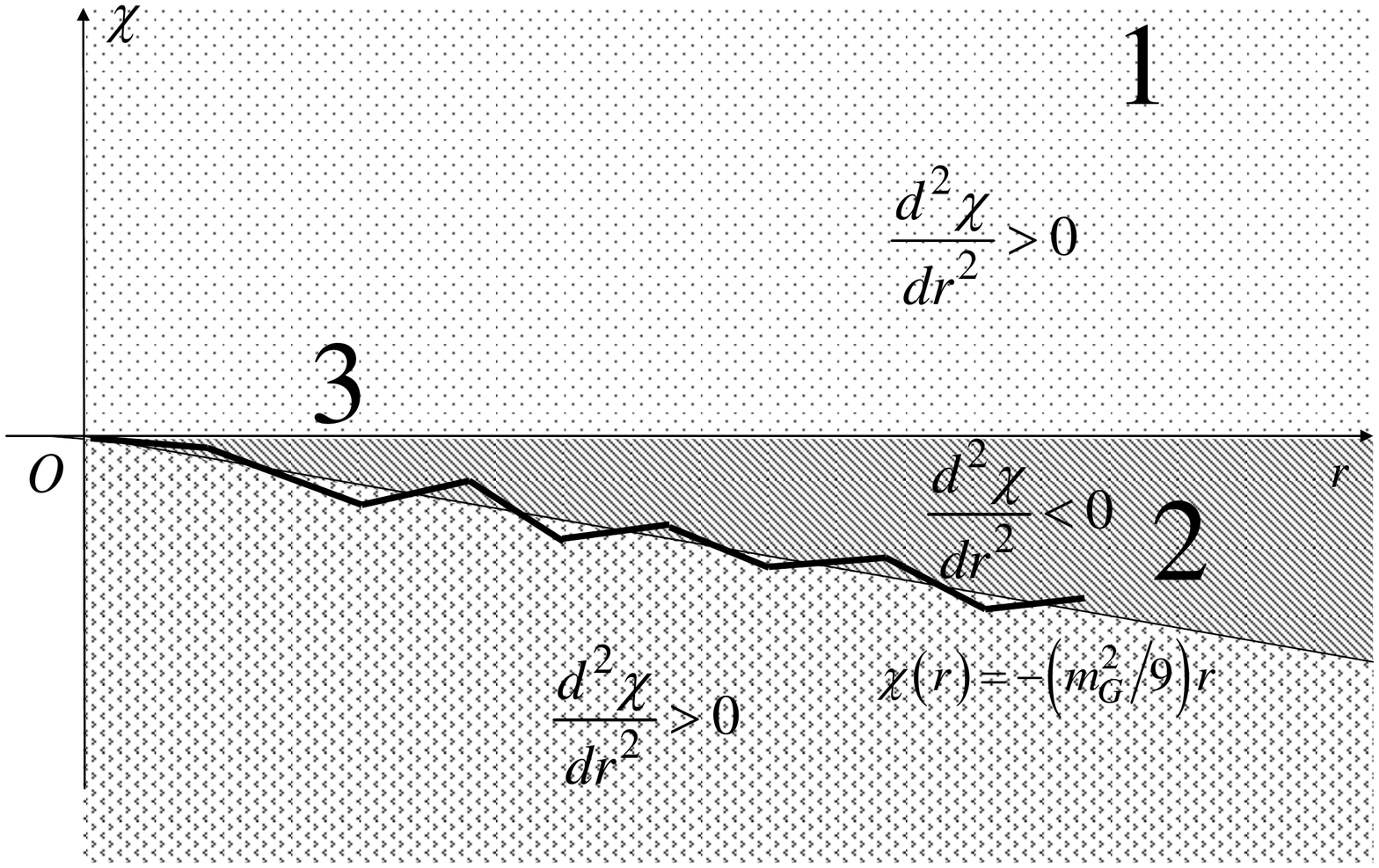}}
	\vskip-1mm\noindent{\footnotesize Рис.5. Схематичний вид залежності $\chi \left( r \right)$  для випадку $-\left( {m_{G}^{2}}/{9}\; \right)r<C<0$ } %
	\vskip15pt
\end{figure}
Результат відповідного чисельного розрахунку наведено на Рис.6.
\begin{figure}
	\center{\includegraphics[scale=0.35]{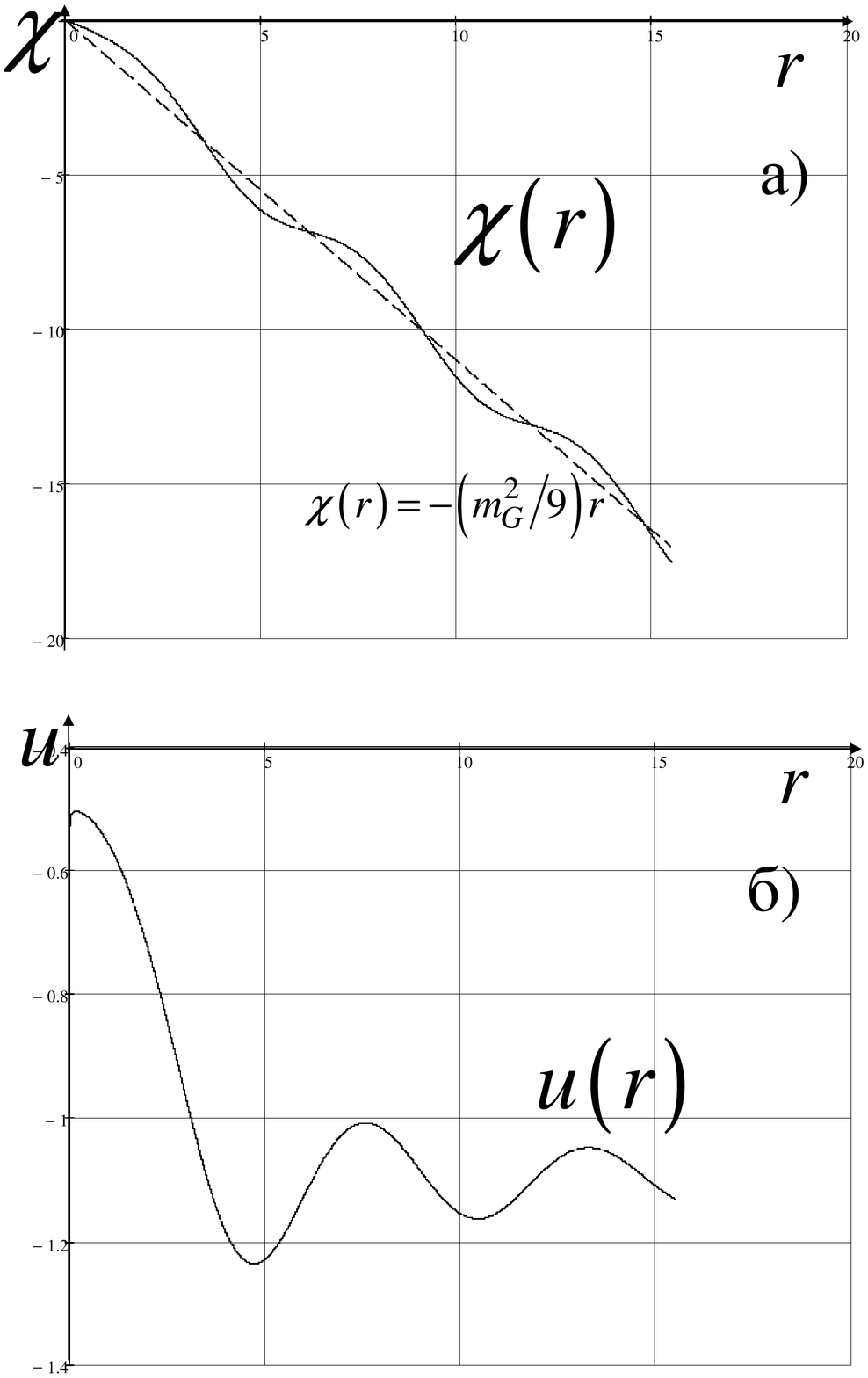}}
	\vskip-1mm\noindent{\footnotesize Рис. 6. Результат чисельного розрахунку залежності $\chi \left( r \right)$  для випадку $C=-0.5, {m_{G}^{2}}/{9}\;=1.1$ (а) і відповідної залежності безрозмірного міжкваркового потенціалу $u\left( r \right)$ від безрозмірної відстані $ r.$(б).} %
	\vskip15pt
\end{figure} 
\begin{figure}
	\center{\includegraphics[scale=0.35]{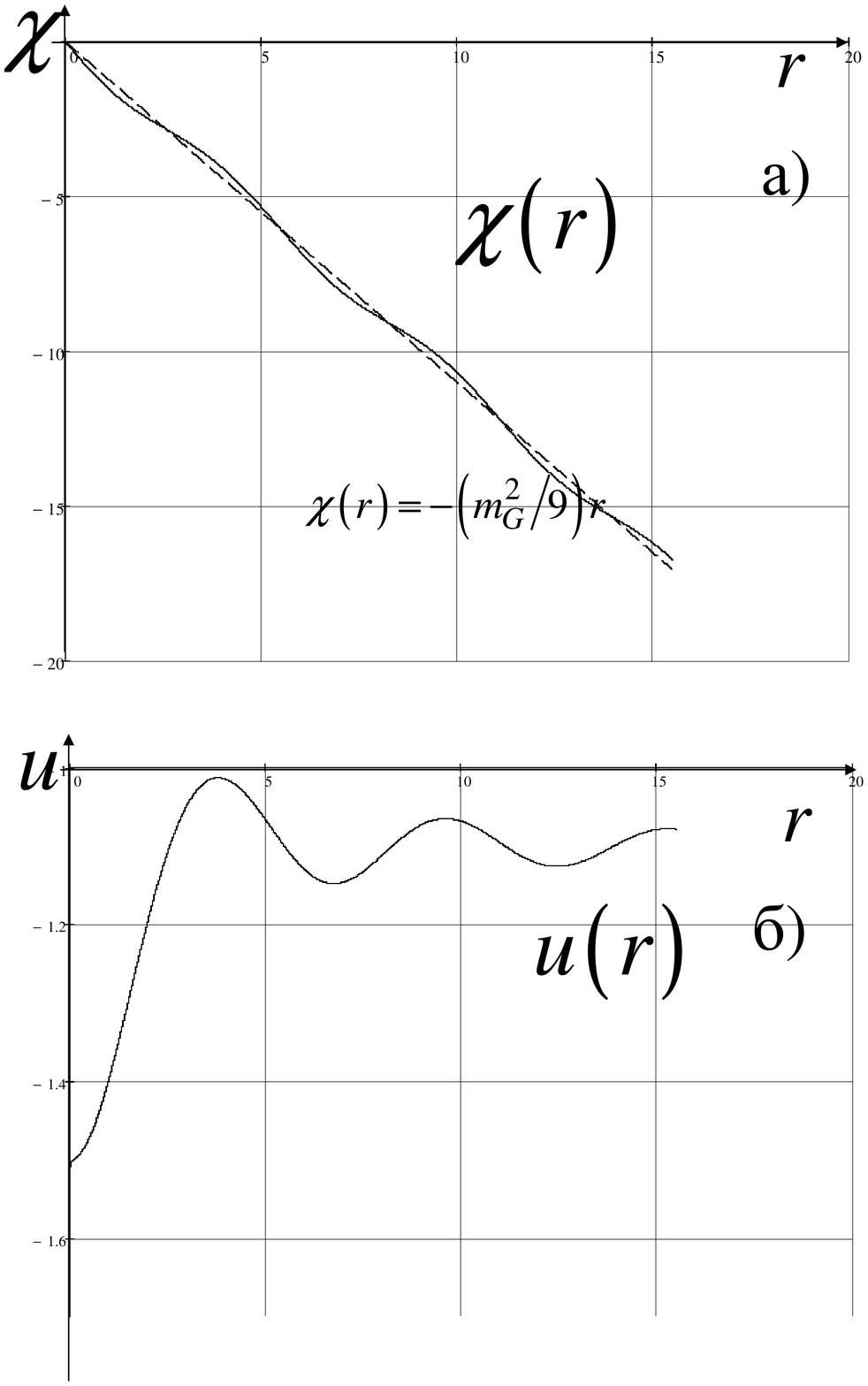}}
	\vskip-1mm\noindent{\footnotesize Рис. 7. Результат чисельного розрахунку залежності $\chi \left( r \right)$  для випадку $C=-1.5, {m_{G}^{2}}/{9}\;=1.1$ (а) і відповідної залежності безрозмірного міжкваркового потенціалу $u\left( r \right)$ від безрозмірної відстані $ r.$(б).} %
	\vskip15pt
	\end{figure}
\begin{figure}
	\center{\includegraphics[scale=0.35]{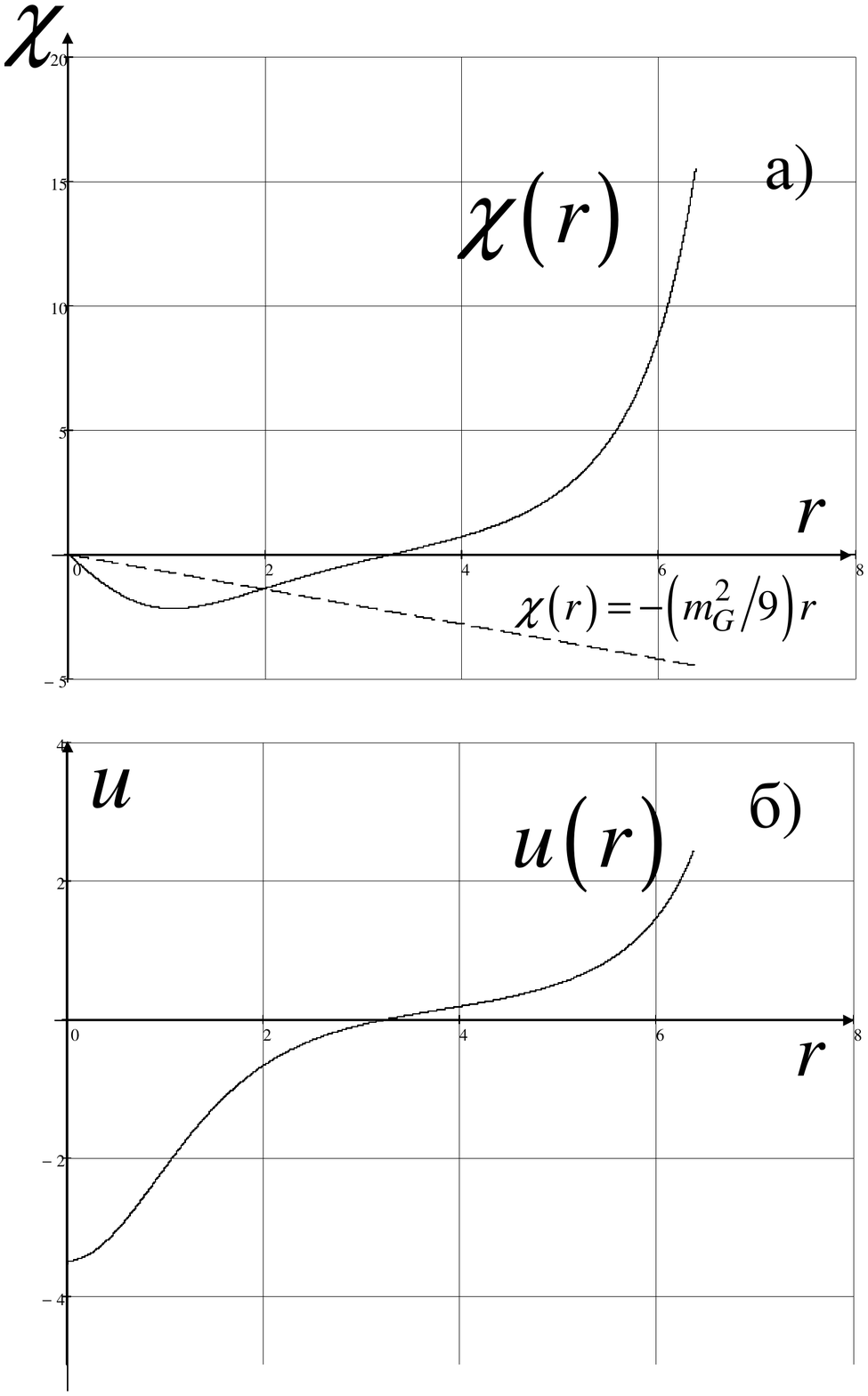}}
	\vskip-1mm\noindent{\footnotesize Рис. 8. Результат чисельного розрахунку залежності $\chi \left( r \right)$  для випадку $C=-3.5, {m_{G}^{2}}/{9}\;=0.7$ (а) і відповідної залежності безрозмірного міжкваркового потенціалу $u\left( r \right)$ від безрозмірної відстані $ r.$(б).} %
	\vskip15pt
\end{figure}
Звернемо увагу на те, що у випадку $-\left( {m_{G}^{2}}/{9}\; \right)r<C<0$ маємо $ \underset{r\to +\infty }{\mathop{\lim }}\,u\left( r \right)=-\left( {m_{G}^{2}}/{9}\; \right)<0 $. Це означає, що внутрішній гамільтоніан з таким потенціалом має від'ємні власні значення, що дозволяє дещо по іншому розглянути відомий механізм Хіггса. За допомогою \mbox{<<старої>>} моделі багаточастинкових полів, це питання розглядалося в роботі \cite{Mercotan_2018JPS}. 

У випадку $ C<-\left( {m_{G}^{2}}/{9}\; \right) $ можливі як коливальний режим, показаний на Рис.7, так і режим конфайнменту, показаний на Рис.8. У випадку, наведеному на Рис. 8, графік \mbox{<<проскакує>>} область 2. Тобто від'ємне \mbox{<<прискореня>>} не встигає змінити знак додатньої \mbox{<<швидкості>>} поки графік перебуває в області 2 і він \mbox{<<проривається>>} в область 1 і далі вже можна зстосовувати  наведений вище аналіз для цієї області. Така ситуація можлива коли величина $ {m_{G}^{2}}/{9}\; $, яка визначє \mbox{<<ширину>>} області 2 є достатньо малою. З цієї точки зору цікаво зістаити ці результати з відомими експериментальними даними. Зокрема з експерименту відомо, що маси скалярних глюболів порядка 1ГеВ \cite{CREDE200974,PhysRevD.98.030001}. Натомість маса бозону Хіггса, який в цій моделі багаточастинкових полів виступає як аналог глюболу, бо розглядається як зв'язаний стан ${{W}^{\pm }}-$бозонів \cite{Mercotan_2018JPS,Hoh:2016}, складає на два порядки більшу величину 125ГеВ \cite{Aad:2015zhl}. Тому якщо припустити, що з якихось причин фізичними є лише граничні умови $ C<-\left( {m_{G}^{2}}/{9}\; \right), $ то для малих $ {m_{G}^{2}}/{9}\; $  матимемо конфайнмент, а для великих - спонтанне порушення симетрії, що відповідає теоретичним уявленням про властивості сильної і слабкої взаємодії.

Розглянемо тепер ту частину лагранжіана (\ref{eq:Z_nulovim_H_internal}) яка містить лише двоглюонну скалярну польову функцію $ V\left( X,\vec{y} \right) :$
\begin{equation}
\begin{split}
  & L=\ldots + \\ 
& +\frac{1}{2}{{g}^{ab,Minc}}\left( {\partial V\left( X,\vec{y} \right)}/{\partial {{X}^{a}}}\; \right)\left( {\partial V\left( X,\vec{y} \right)}/{\partial {{X}^{b}}}\; \right)- \\ 
& -4\sum\limits_{b=1}^{3}{{{\left( {\partial V\left( X,\vec{y} \right)}/{\partial {{y}^{b}}}\; \right)}^{2}}}+ \\ 
& +\frac{3}{2}{{\left( V\left( X,\vec{y} \right) \right)}^{3}}-\frac{1}{2}M_{G}^{2}{{\left( V\left( X,\vec{y} \right) \right)}^{2}}. \\ 
\end{split}
\label{eq:Dvogluonna_chastina_lagrangianu}
\end{equation}
Тут як \mbox{<<трикрапки>>} позначена решта доданків лагранжіана 
 (\ref{eq:Z_nulovim_H_internal}). Ми хочемо враховувати представлення (\ref{eq:Potencial}). Однак зручніше розглянути це представлення не для лагрнжіана (\ref{eq:Dvogluonna_chastina_lagrangianu}), а для динамічного рівняння (\ref{eq:Dinamichne_rivnanna_dla_pola_Votq}), яке породжується суто двоглюонною частиною лагранжіана.
 Виділяючи в ньому залежність від координат центру мас і внутрішніх змінних маємо
 \begin{equation}
 \begin{split}
   & -{{g}^{ca}}\frac{{{\partial }^{2}}V\left( X,\vec{y} \right)}{\partial {{X}^{c}}\partial {{X}^{a}}}+ \\ 
 & +4{{\Delta }_{{\vec{y}}}}V\left( X,\vec{y} \right)-M_{G}^{2}V\left( X,\vec{y} \right)+ \\ 
 & -\frac{9}{2}{{\left( V\left( X,\vec{y} \right) \right)}^{2}}=0. \\ 
  \end{split}
 \label{eq:Dinamichne_rivnanna_dla_VotXy}
 \end{equation}
 Враховуючи підстановку (\ref{eq:Potencial}), а також те що $ {{V}_{0}}\left( {\vec{y}} \right) $ задовольняє рівнянню (\ref{Rivnanna_dla_potencialu}) отримаємо:
 \begin{equation}
 \begin{split}
   & -{{g}^{ca}}\frac{{{\partial }^{2}}{{V}_{1}}\left( X,\vec{y} \right)}{\partial {{X}^{c}}\partial {{X}^{a}}}-M_{G}^{2}{{V}_{1}}\left( X,\vec{y} \right)- \\ 
 & -{{\left( \hat{H}_{G}^{\text{internal}} \right)}^{2}}{{V}_{1}}\left( X,\vec{y} \right)- \\ 
 & -\frac{9}{2}{{\left( {{V}_{1}}\left( X,\vec{y} \right) \right)}^{2}}=0, \\ 
   \end{split}
 \label{eq:Rivnanna_dla_Glubolu_z_confimentom}
 \end{equation}
 де введене позначення 
 \begin{equation}
 \begin{split}
    & {{\left( \hat{H}_{G}^{\text{internal}} \right)}^{2}}{{V}_{1}}\left( X,\vec{y} \right)\equiv  \\ 
  & \equiv -4{{\Delta }_{{\vec{y}}}}{{V}_{1}}\left( X,\vec{y} \right)+ \\ 
  & +9{{V}_{0}}\left( {\vec{y}} \right){{V}_{1}}\left( X,\vec{y} \right). \\ 
  \end{split}
 \label{eq:Kvadrat_vnutrichnogo_gamiltonianu_glubolu}
 \end{equation}
 Якщо тепер перейти до безрозмірних змінних (\ref{eq:Bezrazmer}) то оператор   $ {{\left( \hat{H}_{G}^{\text{internal}} \right)}^{2}} $ формально співпаде з безрозмірним нерелятивістським одночастинковим гамільтоніаном з потенційною енергією $9u\left( {\vec{r}} \right)$, умноженим на константу що забезпечує \mbox{<<правильну>>} розмірність квадрата енергії. Якщо потенціал $9u\left( {\vec{r}} \right)$ наближається до плюс нескінченості при наближенні $\left| {\vec{r}} \right|$ до плюс нескінченості, то всі власні функції такого гамільтоніану наближатимуться до нуля на нескінченості, тобто описуватимуть зв'язані стани глюонів. Як і в роботі \cite{Chudak:2016} бачимо, що та ж сама функція $ {{V}_{0}}\left( {\vec{y}} \right), $ що описує конфайнмент кварків, описує й конфайнмент глюонів. Якщо затравочну масу ${{M}_{G}}$ покласти рівною нулю, то як видно, з попереднього аналізу всі розв'язки рівняння (\ref{Newton}) призводитимуть до конфайнменту. При цьому власні значення оператора (\ref{eq:Kvadrat_vnutrichnogo_gamiltonianu_glubolu}) будуть входити в рівняння (\ref{eq:Rivnanna_dla_Glubolu_z_confimentom}) як квадрат \mbox{<<істинної>>} маси глюбола.   
 
%



\section{Висновки/Conclusions}
В запропонованій моделі сильна взаємодія між кварками, що складють адрони, здійснюється за рахунок обміну звязаними станми глюонів - глюболами. Поле $V\left( X,\vec{y} \right)$, що відповідає глюболам, може бути представлено в виді суми двох доданків $ {{V}_{0}}\left( {\vec{y}} \right) $ і $ {{V}_{1}}\left( X,\vec{y} \right). $ Поле $ {{V}_{0}}\left( {\vec{y}} \right) $ не квантується і описує сильну взаємодію кварків і глюонів всередині мезонів і глюболів. Це поле задовольняє динамічном рівнянню, яке за певних граничних умов має розв'язки які описують асимптотичну свободу і конфайнмент кварків і глюонів. За нульового значення затравочної маси глюболу всі розв'язки цього рівняння, незалежно від граничних умов призводитимуть до конфайнменту. Поле $ {{V}_{1}}\left( X,\vec{y} \right) $ може бути проквантоване. В цій роботі ми не розглядали процедуру квантування багаточастинкових полів. Однак вона нічим не відрізняється від процедури описаної в роботі \cite{Chudak:2016}. Отримані після квантування оператори описуватимуть процеси народження і знищення глюболів, як показано в \cite{Chudak:2016}. Відповідно квантування розглянутого в роботі мезонного поля призводить до операторів народження і знищення мезонів. Відповідно процеси взаємодії мезонів за рахунок взаємодії складових кварків можна описати як процеси обміну скалярними глюболами. Такий підхід вільний від основного недоліку, властивого одночастинковим полям, бо в ньому ми отримаємо закон збереження енергії-імпульсу саме для енергій-імпульсів адронів, а не частинок які їх складають. При цьому на відміну від одночастинкових полів в асимптотичних станах \mbox{<<вмикається>>} і \mbox{<<вимикається>>} лише та частина взаємодії, яка описується полем $ {{V}_{1}}\left( X,\vec{y} \right). $ Внутрішня взаємодія, що описується полем $ {{V}_{0}}\left( {\vec{y}} \right) $ в асимптотичних станах залишається, що з фізичної точки зору є, на наш погляд, більш задовільним ніж розрахунки величин, що описують перехід із одного стану вільних кварків і глюонів (як, наприклад, в партонній моделі, або при ґраткових розрахунках) до іншого стану вільних же частинок. Той факт, що в одночастинковій польовій теорії обов'язково  потрібно \mbox{<<вмикаєти>>} і \mbox{<<вимикати>>} взаємодію, на нашу думку, ставить під сумнів принципову можливість описати стадію адронізації в межах одночастинкового підходу. Бо яка ж може бути адронізація із \mbox{<<вимкненою >>} взаємодією ? Натомість той факт що при бгаточастинковому підході \mbox{<<вимикається>>} лише частина взаємодії, а частина залишається дає надію на можливість опису адронізації методом багаточастинкових полів.

\bibliographystyle{aipnum4-1}
\bibliography{references-windows1251}

\begin{thebibliography}{26}%
\makeatletter
\providecommand \@ifxundefined [1]{%
 \@ifx{#1\undefined}
}%
\providecommand \@ifnum [1]{%
 \ifnum #1\expandafter \@firstoftwo
 \else \expandafter \@secondoftwo
 \fi
}%
\providecommand \@ifx [1]{%
 \ifx #1\expandafter \@firstoftwo
 \else \expandafter \@secondoftwo
 \fi
}%
\providecommand \natexlab [1]{#1}%
\providecommand \enquote  [1]{``#1''}%
\providecommand \bibnamefont  [1]{#1}%
\providecommand \bibfnamefont [1]{#1}%
\providecommand \citenamefont [1]{#1}%
\providecommand \href@noop [0]{\@secondoftwo}%
\providecommand \href [0]{\begingroup \@sanitize@url \@href}%
\providecommand \@href[1]{\@@startlink{#1}\@@href}%
\providecommand \@@href[1]{\endgroup#1\@@endlink}%
\providecommand \@sanitize@url [0]{\catcode `\\12\catcode `\$12\catcode
  `\&12\catcode `\#12\catcode `\^12\catcode `\_12\catcode `\%12\relax}%
\providecommand \@@startlink[1]{}%
\providecommand \@@endlink[0]{}%
\providecommand \url  [0]{\begingroup\@sanitize@url \@url }%
\providecommand \@url [1]{\endgroup\@href {#1}{\urlprefix }}%
\providecommand \urlprefix  [0]{URL }%
\providecommand \Eprint [0]{\href }%
\providecommand \doibase [0]{http://dx.doi.org/}%
\providecommand \selectlanguage [0]{\@gobble}%
\providecommand \bibinfo  [0]{\@secondoftwo}%
\providecommand \bibfield  [0]{\@secondoftwo}%
\providecommand \translation [1]{[#1]}%
\providecommand \BibitemOpen [0]{}%
\providecommand \bibitemStop [0]{}%
\providecommand \bibitemNoStop [0]{.\EOS\space}%
\providecommand \EOS [0]{\spacefactor3000\relax}%
\providecommand \BibitemShut  [1]{\csname bibitem#1\endcsname}%
\let\auto@bib@innerbib\@empty
\bibitem [{\citenamefont {Yukawa}(1949)}]{Yukawa_1949_PhysRev.76.300.2}%
  \BibitemOpen
  \bibfield  {author} {\bibinfo {author} {\bibfnamefont {H.}~\bibnamefont
  {Yukawa}},\ }\href {\doibase 10.1103/PhysRev.76.300.2} {\bibfield  {journal}
  {\bibinfo  {journal} {Phys. Rev.}\ }\textbf {\bibinfo {volume} {76}},\
  \bibinfo {pages} {300} (\bibinfo {year} {1949})}\BibitemShut {NoStop}%
\bibitem [{\citenamefont
  {Yukawa}(1950{\natexlab{a}})}]{Yukawa_p1_PhysRev.77.219}%
  \BibitemOpen
  \bibfield  {author} {\bibinfo {author} {\bibfnamefont {H.}~\bibnamefont
  {Yukawa}},\ }\href {\doibase 10.1103/PhysRev.77.219} {\bibfield  {journal}
  {\bibinfo  {journal} {Phys. Rev.}\ }\textbf {\bibinfo {volume} {77}},\
  \bibinfo {pages} {219} (\bibinfo {year} {1950}{\natexlab{a}})}\BibitemShut
  {NoStop}%
\bibitem [{\citenamefont
  {Yukawa}(1950{\natexlab{b}})}]{Yukawa_p2_PhysRev.80.1047}%
  \BibitemOpen
  \bibfield  {author} {\bibinfo {author} {\bibfnamefont {H.}~\bibnamefont
  {Yukawa}},\ }\href {\doibase 10.1103/PhysRev.80.1047} {\bibfield  {journal}
  {\bibinfo  {journal} {Phys. Rev.}\ }\textbf {\bibinfo {volume} {80}},\
  \bibinfo {pages} {1047} (\bibinfo {year} {1950}{\natexlab{b}})}\BibitemShut
  {NoStop}%
\bibitem [{\citenamefont {Dirac}(1949)}]{Dirac:1949cp}%
  \BibitemOpen
  \bibfield  {author} {\bibinfo {author} {\bibfnamefont {P.~A.~M.}\
  \bibnamefont {Dirac}},\ }\href {\doibase 10.1103/RevModPhys.21.392}
  {\bibfield  {journal} {\bibinfo  {journal} {Rev. Mod. Phys.}\ }\textbf
  {\bibinfo {volume} {21}},\ \bibinfo {pages} {392} (\bibinfo {year}
  {1949})}\BibitemShut {NoStop}%
\bibitem [{\citenamefont {Heinzl}(2001)}]{Heinzl:2000ht}%
  \BibitemOpen
  \bibfield  {author} {\bibinfo {author} {\bibfnamefont {T.}~\bibnamefont
  {Heinzl}},\ }\bibfield  {booktitle} {\emph {\bibinfo {booktitle} {{Methods of
  quantization. Proceedings, 39. Internationale Universit?tswochen f?r Kern-
  und Teilchenphysik, IUKT 39: Schladming, Austria, February 26-March 4,
  2000}}},\ }\href {\doibase 10.1007/3-540-45114-5_2} {\bibfield  {journal}
  {\bibinfo  {journal} {Lect. Notes Phys.}\ }\textbf {\bibinfo {volume}
  {572}},\ \bibinfo {pages} {55} (\bibinfo {year} {2001})},\ \Eprint
  {http://arxiv.org/abs/hep-th/0008096} {arXiv:hep-th/0008096 [hep-th]}
  \BibitemShut {NoStop}%
\bibitem [{\citenamefont {Logunov}\ and\ \citenamefont
  {Tavkhelidze}(1963)}]{Logunov1963_NovCim}%
  \BibitemOpen
  \bibfield  {author} {\bibinfo {author} {\bibfnamefont {A.}~\bibnamefont
  {Logunov}}\ and\ \bibinfo {author} {\bibfnamefont {A.}~\bibnamefont
  {Tavkhelidze}},\ }\href {\doibase 10.1007/BF02750359} {\bibfield  {journal}
  {\bibinfo  {journal} {Il Nuovo Cimento Series 10}\ }\textbf {\bibinfo
  {volume} {29}},\ \bibinfo {pages} {380} (\bibinfo {year} {1963})}\BibitemShut
  {NoStop}%
\bibitem [{\citenamefont {Logunov}\ \emph {et~al.}(1963)\citenamefont
  {Logunov}, \citenamefont {Tavkhelidze}, \citenamefont {Todorov},\ and\
  \citenamefont {Khrustalev}}]{Logunov1955_NovCim}%
  \BibitemOpen
  \bibfield  {author} {\bibinfo {author} {\bibfnamefont {A.}~\bibnamefont
  {Logunov}}, \bibinfo {author} {\bibfnamefont {A.}~\bibnamefont
  {Tavkhelidze}}, \bibinfo {author} {\bibfnamefont {I.}~\bibnamefont
  {Todorov}}, \ and\ \bibinfo {author} {\bibfnamefont {O.}~\bibnamefont
  {Khrustalev}},\ }\href {\doibase 10.1007/BF02750754} {\bibfield  {journal}
  {\bibinfo  {journal} {Il Nuovo Cimento Series 10}\ }\textbf {\bibinfo
  {volume} {30}},\ \bibinfo {pages} {134} (\bibinfo {year} {1963})}\BibitemShut
  {NoStop}%
\bibitem [{\citenamefont {Faustov}(1973)}]{Faustov1973176}%
  \BibitemOpen
  \bibfield  {author} {\bibinfo {author} {\bibfnamefont {R.}~\bibnamefont
  {Faustov}},\ }\href {\doibase http://dx.doi.org/10.1016/0003-4916(73)90007-9}
  {\bibfield  {journal} {\bibinfo  {journal} {Annals of Physics}\ }\textbf
  {\bibinfo {volume} {78}},\ \bibinfo {pages} {176 } (\bibinfo {year}
  {1973})}\BibitemShut {NoStop}%
\bibitem [{\citenamefont {Tomonaga}(1946)}]{Tomonaga}%
  \BibitemOpen
  \bibfield  {author} {\bibinfo {author} {\bibfnamefont {S.}~\bibnamefont
  {Tomonaga}},\ }\href {\doibase 10.1143/PTP.1.27} {\bibfield  {journal}
  {\bibinfo  {journal} {Progress of Theoretical Physics}\ }\textbf {\bibinfo
  {volume} {1}},\ \bibinfo {pages} {27} (\bibinfo {year} {1946})}\BibitemShut
  {NoStop}%
\bibitem [{\citenamefont {Dirac}, \citenamefont {Fock},\ and\ \citenamefont
  {Podolsky}(1932)}]{DiracFockPodolsky}%
  \BibitemOpen
  \bibfield  {author} {\bibinfo {author} {\bibfnamefont {P.}~\bibnamefont
  {Dirac}}, \bibinfo {author} {\bibfnamefont {W.}~\bibnamefont {Fock}}, \ and\
  \bibinfo {author} {\bibfnamefont {B.}~\bibnamefont {Podolsky}},\ }\href@noop
  {} {\bibfield  {journal} {\bibinfo  {journal} {Phys. Zs. Sowjet.}\ }\textbf
  {\bibinfo {volume} {2}},\ \bibinfo {pages} {468} (\bibinfo {year}
  {1932})}\BibitemShut {NoStop}%
\bibitem [{\citenamefont {Petrat}\ and\ \citenamefont
  {Tumulka}(2014{\natexlab{a}})}]{PETRAT201417}%
  \BibitemOpen
  \bibfield  {author} {\bibinfo {author} {\bibfnamefont {S.}~\bibnamefont
  {Petrat}}\ and\ \bibinfo {author} {\bibfnamefont {R.}~\bibnamefont
  {Tumulka}},\ }\href {\doibase https://doi.org/10.1016/j.aop.2014.03.004}
  {\bibfield  {journal} {\bibinfo  {journal} {Annals of Physics}\ }\textbf
  {\bibinfo {volume} {345}},\ \bibinfo {pages} {17 } (\bibinfo {year}
  {2014}{\natexlab{a}})}\BibitemShut {NoStop}%
\bibitem [{\citenamefont {Чудак}\ \emph {et~al.}(2016)\citenamefont {Чудак},
  \citenamefont {Меркотан}, \citenamefont {Пташинський}, \citenamefont
  {Потієнко}, \citenamefont {Делієргієв}, \citenamefont {Тихонов},
  \citenamefont {Сохранний}, \citenamefont {Жарова}, \citenamefont
  {Березовський}, \citenamefont {Войтенко}, \citenamefont {Шарф},\ and\
  \citenamefont {Русов}}]{Chudak_2016UJP}%
  \BibitemOpen
  \bibfield  {author} {\bibinfo {author} {\bibfnamefont {Н.~О.}\ \bibnamefont
  {Чудак}}, \bibinfo {author} {\bibfnamefont {К.~К.}\ \bibnamefont {Меркотан}},
  \bibinfo {author} {\bibfnamefont {Д.~А.}\ \bibnamefont {Пташинський}},
  \bibinfo {author} {\bibfnamefont {О.~С.}\ \bibnamefont {Потієнко}}, \bibinfo
  {author} {\bibfnamefont {М.~А.}\ \bibnamefont {Делієргієв}}, \bibinfo
  {author} {\bibfnamefont {А.~В.}\ \bibnamefont {Тихонов}}, \bibinfo {author}
  {\bibfnamefont {Г.~О.}\ \bibnamefont {Сохранний}}, \bibinfo {author}
  {\bibfnamefont {О.~В.}\ \bibnamefont {Жарова}}, \bibinfo {author}
  {\bibfnamefont {О.~Д.}\ \bibnamefont {Березовський}}, \bibinfo {author}
  {\bibfnamefont {В.~В.}\ \bibnamefont {Войтенко}}, \bibinfo {author}
  {\bibfnamefont {І.~В.}\ \bibnamefont {Шарф}}, \ and\ \bibinfo {author}
  {\bibfnamefont {В.~Д.}\ \bibnamefont {Русов}},\ }\href@noop {} {\bibfield
  {journal} {\bibinfo  {journal} {УФЖ}\ }\textbf {\bibinfo {volume} {61}},\
  \bibinfo {pages} {1039} (\bibinfo {year} {2016})}\BibitemShut {NoStop}%
\bibitem [{\citenamefont {Marx}(1972)}]{Marx1972}%
  \BibitemOpen
  \bibfield  {author} {\bibinfo {author} {\bibfnamefont {E.}~\bibnamefont
  {Marx}},\ }\href {\doibase 10.1007/BF01258729} {\bibfield  {journal}
  {\bibinfo  {journal} {International Journal of Theoretical Physics}\ }\textbf
  {\bibinfo {volume} {6}},\ \bibinfo {pages} {359} (\bibinfo {year}
  {1972})}\BibitemShut {NoStop}%
\bibitem [{\citenamefont {Sazdjian}(1986)}]{SazdjianPhysRevD.33.3401}%
  \BibitemOpen
  \bibfield  {author} {\bibinfo {author} {\bibfnamefont {H.}~\bibnamefont
  {Sazdjian}},\ }\href {\doibase 10.1103/PhysRevD.33.3401} {\bibfield
  {journal} {\bibinfo  {journal} {Phys. Rev. D}\ }\textbf {\bibinfo {volume}
  {33}},\ \bibinfo {pages} {3401} (\bibinfo {year} {1986})}\BibitemShut
  {NoStop}%
\bibitem [{\citenamefont {Petrat}\ and\ \citenamefont
  {Tumulka}(2014{\natexlab{b}})}]{Petrat20130632}%
  \BibitemOpen
  \bibfield  {author} {\bibinfo {author} {\bibfnamefont {S.}~\bibnamefont
  {Petrat}}\ and\ \bibinfo {author} {\bibfnamefont {R.}~\bibnamefont
  {Tumulka}},\ }\href {\doibase 10.1098/rspa.2013.0632} {\bibfield  {journal}
  {\bibinfo  {journal} {Proceedings of the Royal Society of London A:
  Mathematical, Physical and Engineering Sciences}\ }\textbf {\bibinfo {volume}
  {470}},\ \bibinfo {pages} {1364} (\bibinfo {year}
  {2014}{\natexlab{b}})}\BibitemShut {NoStop}%
\bibitem [{\citenamefont {Дирак}(1979)}]{DiracPrincipi}%
  \BibitemOpen
  \bibfield  {author} {\bibinfo {author} {\bibfnamefont {П.}~\bibnamefont
  {Дирак}},\ }\href
  {http://gen.lib.rus.ec/book/index.php?md5=01E551E20717D41E8B653F068339F48D}
  {\emph {\bibinfo {title} {Принципы квантовой механики}}}\ (\bibinfo {year}
  {1979})\BibitemShut {NoStop}%
\bibitem [{\citenamefont {Chudak}\ \emph {et~al.}(2016)\citenamefont {Chudak},
  \citenamefont {Deliyergiyev}, \citenamefont {Merkotan}, \citenamefont
  {Potiienko}, \citenamefont {Ptashynskyi}, \citenamefont {Shabatura},
  \citenamefont {Sokhrannyi}, \citenamefont {Tykhonov}, \citenamefont
  {Volkotrub}, \citenamefont {Sharph},\ and\ \citenamefont
  {Rusov}}]{Chudak:2016}%
  \BibitemOpen
  \bibfield  {author} {\bibinfo {author} {\bibfnamefont {N.}~\bibnamefont
  {Chudak}}, \bibinfo {author} {\bibfnamefont {M.}~\bibnamefont
  {Deliyergiyev}}, \bibinfo {author} {\bibfnamefont {K.}~\bibnamefont
  {Merkotan}}, \bibinfo {author} {\bibfnamefont {O.}~\bibnamefont {Potiienko}},
  \bibinfo {author} {\bibfnamefont {D.}~\bibnamefont {Ptashynskyi}}, \bibinfo
  {author} {\bibfnamefont {Y.}~\bibnamefont {Shabatura}}, \bibinfo {author}
  {\bibfnamefont {G.}~\bibnamefont {Sokhrannyi}}, \bibinfo {author}
  {\bibfnamefont {A.}~\bibnamefont {Tykhonov}}, \bibinfo {author}
  {\bibfnamefont {Y.}~\bibnamefont {Volkotrub}}, \bibinfo {author}
  {\bibfnamefont {I.}~\bibnamefont {Sharph}}, \ and\ \bibinfo {author}
  {\bibfnamefont {V.}~\bibnamefont {Rusov}},\ }\href@noop {} {\bibfield
  {journal} {\bibinfo  {journal} {Physics Journal}\ }\textbf {\bibinfo {volume}
  {2}},\ \bibinfo {pages} {181} (\bibinfo {year} {2016})}\BibitemShut {NoStop}%
\bibitem [{\citenamefont {Боголюбов}\ and\ \citenamefont
  {Ширков}(1984)}]{Bogolyubov_rus}%
  \BibitemOpen
  \bibfield  {author} {\bibinfo {author} {\bibfnamefont {Н.}~\bibnamefont
  {Боголюбов}}\ and\ \bibinfo {author} {\bibfnamefont {Д.}~\bibnamefont
  {Ширков}},\ }\href@noop {} {{\selectlanguage {russian}\emph {\bibinfo {title}
  {Введение в теорию квантованных полей}}}},\ \bibinfo {edition} {4th}\ ed.\
  (\bibinfo  {publisher} {Наука},\ \bibinfo {year} {1984})\BibitemShut
  {NoStop}%
\bibitem [{\citenamefont {Березин}(1986)}]{Berezin:1986VtorKv}%
  \BibitemOpen
  \bibfield  {author} {\bibinfo {author} {\bibfnamefont {Ф.~А.}\ \bibnamefont
  {Березин}},\ }\href@noop {} {{\selectlanguage {russian}\emph {\bibinfo
  {title} {Метод вторичного квантования.}}}}\ (\bibinfo  {publisher} {Наука},\
  \bibinfo {year} {1986})\BibitemShut {NoStop}%
\bibitem [{\citenamefont {Merkotan}\ \emph {et~al.}(2017)\citenamefont
  {Merkotan}, \citenamefont {Zelentsova}, \citenamefont {Chudak} \emph
  {et~al.}}]{Mercotan2017}%
  \BibitemOpen
  \bibfield  {author} {\bibinfo {author} {\bibfnamefont {K.~K.}\ \bibnamefont
  {Merkotan}}, \bibinfo {author} {\bibfnamefont {T.~M.}\ \bibnamefont
  {Zelentsova}}, \bibinfo {author} {\bibfnamefont {N.~O.}\ \bibnamefont
  {Chudak}},  \emph {et~al.},\ }\href@noop {} {\  (\bibinfo {year} {2017})},\
  \Eprint {http://arxiv.org/abs/1711.01914} {arXiv:1711.01914 [physics.gen-ph]}
  \BibitemShut {NoStop}%
\bibitem [{\citenamefont {Patrignani}\ \emph {et~al.}(2016)\citenamefont
  {Patrignani} \emph {et~al.}}]{Olive:2016xmw}%
  \BibitemOpen
  \bibfield  {author} {\bibinfo {author} {\bibfnamefont {C.}~\bibnamefont
  {Patrignani}} \emph {et~al.} (\bibinfo {collaboration} {Particle Data
  Group}),\ }\href {\doibase 10.1088/1674-1137/40/10/100001} {\bibfield
  {journal} {\bibinfo  {journal} {Chin. Phys.}\ }\textbf {\bibinfo {volume}
  {C40}},\ \bibinfo {pages} {100001} (\bibinfo {year} {2016})}\BibitemShut
  {NoStop}%
\bibitem [{\citenamefont {Меркотан}\ \emph {et~al.}(2018)\citenamefont
  {Меркотан}, \citenamefont {Зеленцова}, \citenamefont {Чудак}, \citenamefont
  {Пташинський}, \citenamefont {Урбаневич}, \citenamefont {Потієнко},
  \citenamefont {Березовський}, \citenamefont {Войтенко}, \citenamefont
  {Шарф},\ and\ \citenamefont {Русов}}]{Mercotan_2018JPS}%
  \BibitemOpen
  \bibfield  {author} {\bibinfo {author} {\bibfnamefont {К.~К.}\ \bibnamefont
  {Меркотан}}, \bibinfo {author} {\bibfnamefont {Т.}~\bibnamefont {Зеленцова}},
  \bibinfo {author} {\bibfnamefont {Н.~О.}\ \bibnamefont {Чудак}}, \bibinfo
  {author} {\bibfnamefont {Д.~А.}\ \bibnamefont {Пташинський}}, \bibinfo
  {author} {\bibfnamefont {В.}~\bibnamefont {Урбаневич}}, \bibinfo {author}
  {\bibfnamefont {О.~С.}\ \bibnamefont {Потієнко}}, \bibinfo {author}
  {\bibfnamefont {О.~Д.}\ \bibnamefont {Березовський}}, \bibinfo {author}
  {\bibfnamefont {В.~В.}\ \bibnamefont {Войтенко}}, \bibinfo {author}
  {\bibfnamefont {І.~В.}\ \bibnamefont {Шарф}}, \ and\ \bibinfo {author}
  {\bibfnamefont {В.~Д.}\ \bibnamefont {Русов}},\ }\href@noop {} {\bibfield
  {journal} {\bibinfo  {journal} {Журнал фізичних досліджень}\ }\textbf
  {\bibinfo {volume} {22}},\ \bibinfo {pages} {3001} (\bibinfo {year}
  {2018})}\BibitemShut {NoStop}%
\bibitem [{\citenamefont {Crede}\ and\ \citenamefont
  {Meyer}(2009)}]{CREDE200974}%
  \BibitemOpen
  \bibfield  {author} {\bibinfo {author} {\bibfnamefont {V.}~\bibnamefont
  {Crede}}\ and\ \bibinfo {author} {\bibfnamefont {C.}~\bibnamefont {Meyer}},\
  }\href {\doibase https://doi.org/10.1016/j.ppnp.2009.03.001} {\bibfield
  {journal} {\bibinfo  {journal} {Progress in Particle and Nuclear Physics}\
  }\textbf {\bibinfo {volume} {63}},\ \bibinfo {pages} {74 } (\bibinfo {year}
  {2009})}\BibitemShut {NoStop}%
\bibitem [{\citenamefont {Tanabashi}\ \emph {et~al.}(2018)\citenamefont
  {Tanabashi}, \citenamefont {Hagiwara}, \citenamefont {Hikasa}, \citenamefont
  {Nakamura}, \citenamefont {Sumino}, \citenamefont {Takahashi}, \citenamefont
  {Tanaka}, \citenamefont {Agashe}, \citenamefont {Aielli}, \citenamefont
  {Amsler}, \citenamefont {Antonelli}, \citenamefont {Asner}, \citenamefont
  {Baer}, \citenamefont {Banerjee}, \citenamefont {Barnett}, \citenamefont
  {Basaglia}, \citenamefont {Bauer}, \citenamefont {Beatty}, \citenamefont
  {Belousov}, \citenamefont {Beringer}, \citenamefont {Bethke}, \citenamefont
  {Bettini}, \citenamefont {Bichsel}, \citenamefont {Biebel}, \citenamefont
  {Black}, \citenamefont {Blucher}, \citenamefont {Buchmuller}, \citenamefont
  {Burkert}, \citenamefont {Bychkov}, \citenamefont {Cahn}, \citenamefont
  {Carena}, \citenamefont {Ceccucci}, \citenamefont {Cerri}, \citenamefont
  {Chakraborty}, \citenamefont {Chen}, \citenamefont {Chivukula}, \citenamefont
  {Cowan}, \citenamefont {Dahl}, \citenamefont {D'Ambrosio}, \citenamefont
  {Damour}, \citenamefont {de~Florian}, \citenamefont {de~Gouv\^ea},
  \citenamefont {DeGrand}, \citenamefont {de~Jong}, \citenamefont {Dissertori},
  \citenamefont {Dobrescu}, \citenamefont {D'Onofrio}, \citenamefont {Doser},
  \citenamefont {Drees}, \citenamefont {Dreiner}, \citenamefont {Dwyer},
  \citenamefont {Eerola}, \citenamefont {Eidelman}, \citenamefont {Ellis},
  \citenamefont {Erler}, \citenamefont {Ezhela}, \citenamefont {Fetscher},
  \citenamefont {Fields}, \citenamefont {Firestone}, \citenamefont {Foster},
  \citenamefont {Freitas}, \citenamefont {Gallagher}, \citenamefont {Garren},
  \citenamefont {Gerber}, \citenamefont {Gerbier}, \citenamefont {Gershon},
  \citenamefont {Gershtein}, \citenamefont {Gherghetta}, \citenamefont
  {Godizov}, \citenamefont {Goodman}, \citenamefont {Grab}, \citenamefont
  {Gritsan}, \citenamefont {Grojean}, \citenamefont {Groom}, \citenamefont
  {Gr\"unewald}, \citenamefont {Gurtu}, \citenamefont {Gutsche}, \citenamefont
  {Haber}, \citenamefont {Hanhart}, \citenamefont {Hashimoto}, \citenamefont
  {Hayato}, \citenamefont {Hayes}, \citenamefont {Hebecker}, \citenamefont
  {Heinemeyer}, \citenamefont {Heltsley}, \citenamefont {Hern\'andez-Rey},
  \citenamefont {Hisano}, \citenamefont {H\"ocker}, \citenamefont {Holder},
  \citenamefont {Holtkamp}, \citenamefont {Hyodo}, \citenamefont {Irwin},
  \citenamefont {Johnson}, \citenamefont {Kado}, \citenamefont {Karliner},
  \citenamefont {Katz}, \citenamefont {Klein}, \citenamefont {Klempt},
  \citenamefont {Kowalewski}, \citenamefont {Krauss}, \citenamefont {Kreps},
  \citenamefont {Krusche}, \citenamefont {Kuyanov}, \citenamefont {Kwon},
  \citenamefont {Lahav}, \citenamefont {Laiho}, \citenamefont {Lesgourgues},
  \citenamefont {Liddle}, \citenamefont {Ligeti}, \citenamefont {Lin},
  \citenamefont {Lippmann}, \citenamefont {Liss}, \citenamefont {Littenberg},
  \citenamefont {Lugovsky}, \citenamefont {Lugovsky}, \citenamefont {Lusiani},
  \citenamefont {Makida}, \citenamefont {Maltoni}, \citenamefont {Mannel},
  \citenamefont {Manohar}, \citenamefont {Marciano}, \citenamefont {Martin},
  \citenamefont {Masoni}, \citenamefont {Matthews}, \citenamefont
  {Mei\ss{}ner}, \citenamefont {Milstead}, \citenamefont {Mitchell},
  \citenamefont {M\"onig}, \citenamefont {Molaro}, \citenamefont {Moortgat},
  \citenamefont {Moskovic}, \citenamefont {Murayama}, \citenamefont {Narain},
  \citenamefont {Nason}, \citenamefont {Navas}, \citenamefont {Neubert},
  \citenamefont {Nevski}, \citenamefont {Nir}, \citenamefont {Olive},
  \citenamefont {Pagan~Griso}, \citenamefont {Parsons}, \citenamefont
  {Patrignani}, \citenamefont {Peacock}, \citenamefont {Pennington},
  \citenamefont {Petcov}, \citenamefont {Petrov}, \citenamefont {Pianori},
  \citenamefont {Piepke}, \citenamefont {Pomarol}, \citenamefont {Quadt},
  \citenamefont {Rademacker}, \citenamefont {Raffelt}, \citenamefont
  {Ratcliff}, \citenamefont {Richardson}, \citenamefont {Ringwald},
  \citenamefont {Roesler}, \citenamefont {Rolli}, \citenamefont {Romaniouk},
  \citenamefont {Rosenberg}, \citenamefont {Rosner}, \citenamefont {Rybka},
  \citenamefont {Ryutin}, \citenamefont {Sachrajda}, \citenamefont {Sakai},
  \citenamefont {Salam}, \citenamefont {Sarkar}, \citenamefont {Sauli},
  \citenamefont {Schneider}, \citenamefont {Scholberg}, \citenamefont
  {Schwartz}, \citenamefont {Scott}, \citenamefont {Sharma}, \citenamefont
  {Sharpe}, \citenamefont {Shutt}, \citenamefont {Silari}, \citenamefont
  {Sj\"ostrand}, \citenamefont {Skands}, \citenamefont {Skwarnicki},
  \citenamefont {Smith}, \citenamefont {Smoot}, \citenamefont {Spanier},
  \citenamefont {Spieler}, \citenamefont {Spiering}, \citenamefont {Stahl},
  \citenamefont {Stone}, \citenamefont {Sumiyoshi}, \citenamefont {Syphers},
  \citenamefont {Terashi}, \citenamefont {Terning}, \citenamefont {Thoma},
  \citenamefont {Thorne}, \citenamefont {Tiator}, \citenamefont {Titov},
  \citenamefont {Tkachenko}, \citenamefont {T\"ornqvist}, \citenamefont
  {Tovey}, \citenamefont {Valencia}, \citenamefont {Van~de Water},
  \citenamefont {Varelas}, \citenamefont {Venanzoni}, \citenamefont {Verde},
  \citenamefont {Vincter}, \citenamefont {Vogel}, \citenamefont {Vogt},
  \citenamefont {Wakely}, \citenamefont {Walkowiak}, \citenamefont {Walter},
  \citenamefont {Wands}, \citenamefont {Ward}, \citenamefont {Wascko},
  \citenamefont {Weiglein}, \citenamefont {Weinberg}, \citenamefont {Weinberg},
  \citenamefont {White}, \citenamefont {Wiencke}, \citenamefont {Willocq},
  \citenamefont {Wohl}, \citenamefont {Womersley}, \citenamefont {Woody},
  \citenamefont {Workman}, \citenamefont {Yao}, \citenamefont {Zeller},
  \citenamefont {Zenin}, \citenamefont {Zhu}, \citenamefont {Zhu},
  \citenamefont {Zimmermann}, \citenamefont {Zyla}, \citenamefont {Anderson},
  \citenamefont {Fuller}, \citenamefont {Lugovsky},\ and\ \citenamefont
  {Schaffner}}]{PhysRevD.98.030001}%
  \BibitemOpen
  \bibfield  {author} {\bibinfo {author} {\bibfnamefont {M.}~\bibnamefont
  {Tanabashi}}, \bibinfo {author} {\bibfnamefont {K.}~\bibnamefont {Hagiwara}},
  \bibinfo {author} {\bibfnamefont {K.}~\bibnamefont {Hikasa}}, \bibinfo
  {author} {\bibfnamefont {K.}~\bibnamefont {Nakamura}}, \bibinfo {author}
  {\bibfnamefont {Y.}~\bibnamefont {Sumino}}, \bibinfo {author} {\bibfnamefont
  {F.}~\bibnamefont {Takahashi}}, \bibinfo {author} {\bibfnamefont
  {J.}~\bibnamefont {Tanaka}}, \bibinfo {author} {\bibfnamefont
  {K.}~\bibnamefont {Agashe}}, \bibinfo {author} {\bibfnamefont
  {G.}~\bibnamefont {Aielli}}, \bibinfo {author} {\bibfnamefont
  {C.}~\bibnamefont {Amsler}}, \bibinfo {author} {\bibfnamefont
  {M.}~\bibnamefont {Antonelli}}, \bibinfo {author} {\bibfnamefont {D.~M.}\
  \bibnamefont {Asner}}, \bibinfo {author} {\bibfnamefont {H.}~\bibnamefont
  {Baer}}, \bibinfo {author} {\bibfnamefont {S.}~\bibnamefont {Banerjee}},
  \bibinfo {author} {\bibfnamefont {R.~M.}\ \bibnamefont {Barnett}}, \bibinfo
  {author} {\bibfnamefont {T.}~\bibnamefont {Basaglia}}, \bibinfo {author}
  {\bibfnamefont {C.~W.}\ \bibnamefont {Bauer}}, \bibinfo {author}
  {\bibfnamefont {J.~J.}\ \bibnamefont {Beatty}}, \bibinfo {author}
  {\bibfnamefont {V.~I.}\ \bibnamefont {Belousov}}, \bibinfo {author}
  {\bibfnamefont {J.}~\bibnamefont {Beringer}}, \bibinfo {author}
  {\bibfnamefont {S.}~\bibnamefont {Bethke}}, \bibinfo {author} {\bibfnamefont
  {A.}~\bibnamefont {Bettini}}, \bibinfo {author} {\bibfnamefont
  {H.}~\bibnamefont {Bichsel}}, \bibinfo {author} {\bibfnamefont
  {O.}~\bibnamefont {Biebel}}, \bibinfo {author} {\bibfnamefont {K.~M.}\
  \bibnamefont {Black}}, \bibinfo {author} {\bibfnamefont {E.}~\bibnamefont
  {Blucher}}, \bibinfo {author} {\bibfnamefont {O.}~\bibnamefont {Buchmuller}},
  \bibinfo {author} {\bibfnamefont {V.}~\bibnamefont {Burkert}}, \bibinfo
  {author} {\bibfnamefont {M.~A.}\ \bibnamefont {Bychkov}}, \bibinfo {author}
  {\bibfnamefont {R.~N.}\ \bibnamefont {Cahn}}, \bibinfo {author}
  {\bibfnamefont {M.}~\bibnamefont {Carena}}, \bibinfo {author} {\bibfnamefont
  {A.}~\bibnamefont {Ceccucci}}, \bibinfo {author} {\bibfnamefont
  {A.}~\bibnamefont {Cerri}}, \bibinfo {author} {\bibfnamefont
  {D.}~\bibnamefont {Chakraborty}}, \bibinfo {author} {\bibfnamefont {M.-C.}\
  \bibnamefont {Chen}}, \bibinfo {author} {\bibfnamefont {R.~S.}\ \bibnamefont
  {Chivukula}}, \bibinfo {author} {\bibfnamefont {G.}~\bibnamefont {Cowan}},
  \bibinfo {author} {\bibfnamefont {O.}~\bibnamefont {Dahl}}, \bibinfo {author}
  {\bibfnamefont {G.}~\bibnamefont {D'Ambrosio}}, \bibinfo {author}
  {\bibfnamefont {T.}~\bibnamefont {Damour}}, \bibinfo {author} {\bibfnamefont
  {D.}~\bibnamefont {de~Florian}}, \bibinfo {author} {\bibfnamefont
  {A.}~\bibnamefont {de~Gouv\^ea}}, \bibinfo {author} {\bibfnamefont
  {T.}~\bibnamefont {DeGrand}}, \bibinfo {author} {\bibfnamefont
  {P.}~\bibnamefont {de~Jong}}, \bibinfo {author} {\bibfnamefont
  {G.}~\bibnamefont {Dissertori}}, \bibinfo {author} {\bibfnamefont {B.~A.}\
  \bibnamefont {Dobrescu}}, \bibinfo {author} {\bibfnamefont {M.}~\bibnamefont
  {D'Onofrio}}, \bibinfo {author} {\bibfnamefont {M.}~\bibnamefont {Doser}},
  \bibinfo {author} {\bibfnamefont {M.}~\bibnamefont {Drees}}, \bibinfo
  {author} {\bibfnamefont {H.~K.}\ \bibnamefont {Dreiner}}, \bibinfo {author}
  {\bibfnamefont {D.~A.}\ \bibnamefont {Dwyer}}, \bibinfo {author}
  {\bibfnamefont {P.}~\bibnamefont {Eerola}}, \bibinfo {author} {\bibfnamefont
  {S.}~\bibnamefont {Eidelman}}, \bibinfo {author} {\bibfnamefont
  {J.}~\bibnamefont {Ellis}}, \bibinfo {author} {\bibfnamefont
  {J.}~\bibnamefont {Erler}}, \bibinfo {author} {\bibfnamefont {V.~V.}\
  \bibnamefont {Ezhela}}, \bibinfo {author} {\bibfnamefont {W.}~\bibnamefont
  {Fetscher}}, \bibinfo {author} {\bibfnamefont {B.~D.}\ \bibnamefont
  {Fields}}, \bibinfo {author} {\bibfnamefont {R.}~\bibnamefont {Firestone}},
  \bibinfo {author} {\bibfnamefont {B.}~\bibnamefont {Foster}}, \bibinfo
  {author} {\bibfnamefont {A.}~\bibnamefont {Freitas}}, \bibinfo {author}
  {\bibfnamefont {H.}~\bibnamefont {Gallagher}}, \bibinfo {author}
  {\bibfnamefont {L.}~\bibnamefont {Garren}}, \bibinfo {author} {\bibfnamefont
  {H.-J.}\ \bibnamefont {Gerber}}, \bibinfo {author} {\bibfnamefont
  {G.}~\bibnamefont {Gerbier}}, \bibinfo {author} {\bibfnamefont
  {T.}~\bibnamefont {Gershon}}, \bibinfo {author} {\bibfnamefont
  {Y.}~\bibnamefont {Gershtein}}, \bibinfo {author} {\bibfnamefont
  {T.}~\bibnamefont {Gherghetta}}, \bibinfo {author} {\bibfnamefont {A.~A.}\
  \bibnamefont {Godizov}}, \bibinfo {author} {\bibfnamefont {M.}~\bibnamefont
  {Goodman}}, \bibinfo {author} {\bibfnamefont {C.}~\bibnamefont {Grab}},
  \bibinfo {author} {\bibfnamefont {A.~V.}\ \bibnamefont {Gritsan}}, \bibinfo
  {author} {\bibfnamefont {C.}~\bibnamefont {Grojean}}, \bibinfo {author}
  {\bibfnamefont {D.~E.}\ \bibnamefont {Groom}}, \bibinfo {author}
  {\bibfnamefont {M.}~\bibnamefont {Gr\"unewald}}, \bibinfo {author}
  {\bibfnamefont {A.}~\bibnamefont {Gurtu}}, \bibinfo {author} {\bibfnamefont
  {T.}~\bibnamefont {Gutsche}}, \bibinfo {author} {\bibfnamefont {H.~E.}\
  \bibnamefont {Haber}}, \bibinfo {author} {\bibfnamefont {C.}~\bibnamefont
  {Hanhart}}, \bibinfo {author} {\bibfnamefont {S.}~\bibnamefont {Hashimoto}},
  \bibinfo {author} {\bibfnamefont {Y.}~\bibnamefont {Hayato}}, \bibinfo
  {author} {\bibfnamefont {K.~G.}\ \bibnamefont {Hayes}}, \bibinfo {author}
  {\bibfnamefont {A.}~\bibnamefont {Hebecker}}, \bibinfo {author}
  {\bibfnamefont {S.}~\bibnamefont {Heinemeyer}}, \bibinfo {author}
  {\bibfnamefont {B.}~\bibnamefont {Heltsley}}, \bibinfo {author}
  {\bibfnamefont {J.~J.}\ \bibnamefont {Hern\'andez-Rey}}, \bibinfo {author}
  {\bibfnamefont {J.}~\bibnamefont {Hisano}}, \bibinfo {author} {\bibfnamefont
  {A.}~\bibnamefont {H\"ocker}}, \bibinfo {author} {\bibfnamefont
  {J.}~\bibnamefont {Holder}}, \bibinfo {author} {\bibfnamefont
  {A.}~\bibnamefont {Holtkamp}}, \bibinfo {author} {\bibfnamefont
  {T.}~\bibnamefont {Hyodo}}, \bibinfo {author} {\bibfnamefont {K.~D.}\
  \bibnamefont {Irwin}}, \bibinfo {author} {\bibfnamefont {K.~F.}\ \bibnamefont
  {Johnson}}, \bibinfo {author} {\bibfnamefont {M.}~\bibnamefont {Kado}},
  \bibinfo {author} {\bibfnamefont {M.}~\bibnamefont {Karliner}}, \bibinfo
  {author} {\bibfnamefont {U.~F.}\ \bibnamefont {Katz}}, \bibinfo {author}
  {\bibfnamefont {S.~R.}\ \bibnamefont {Klein}}, \bibinfo {author}
  {\bibfnamefont {E.}~\bibnamefont {Klempt}}, \bibinfo {author} {\bibfnamefont
  {R.~V.}\ \bibnamefont {Kowalewski}}, \bibinfo {author} {\bibfnamefont
  {F.}~\bibnamefont {Krauss}}, \bibinfo {author} {\bibfnamefont
  {M.}~\bibnamefont {Kreps}}, \bibinfo {author} {\bibfnamefont
  {B.}~\bibnamefont {Krusche}}, \bibinfo {author} {\bibfnamefont {Y.~V.}\
  \bibnamefont {Kuyanov}}, \bibinfo {author} {\bibfnamefont {Y.}~\bibnamefont
  {Kwon}}, \bibinfo {author} {\bibfnamefont {O.}~\bibnamefont {Lahav}},
  \bibinfo {author} {\bibfnamefont {J.}~\bibnamefont {Laiho}}, \bibinfo
  {author} {\bibfnamefont {J.}~\bibnamefont {Lesgourgues}}, \bibinfo {author}
  {\bibfnamefont {A.}~\bibnamefont {Liddle}}, \bibinfo {author} {\bibfnamefont
  {Z.}~\bibnamefont {Ligeti}}, \bibinfo {author} {\bibfnamefont {C.-J.}\
  \bibnamefont {Lin}}, \bibinfo {author} {\bibfnamefont {C.}~\bibnamefont
  {Lippmann}}, \bibinfo {author} {\bibfnamefont {T.~M.}\ \bibnamefont {Liss}},
  \bibinfo {author} {\bibfnamefont {L.}~\bibnamefont {Littenberg}}, \bibinfo
  {author} {\bibfnamefont {K.~S.}\ \bibnamefont {Lugovsky}}, \bibinfo {author}
  {\bibfnamefont {S.~B.}\ \bibnamefont {Lugovsky}}, \bibinfo {author}
  {\bibfnamefont {A.}~\bibnamefont {Lusiani}}, \bibinfo {author} {\bibfnamefont
  {Y.}~\bibnamefont {Makida}}, \bibinfo {author} {\bibfnamefont
  {F.}~\bibnamefont {Maltoni}}, \bibinfo {author} {\bibfnamefont
  {T.}~\bibnamefont {Mannel}}, \bibinfo {author} {\bibfnamefont {A.~V.}\
  \bibnamefont {Manohar}}, \bibinfo {author} {\bibfnamefont {W.~J.}\
  \bibnamefont {Marciano}}, \bibinfo {author} {\bibfnamefont {A.~D.}\
  \bibnamefont {Martin}}, \bibinfo {author} {\bibfnamefont {A.}~\bibnamefont
  {Masoni}}, \bibinfo {author} {\bibfnamefont {J.}~\bibnamefont {Matthews}},
  \bibinfo {author} {\bibfnamefont {U.-G.}\ \bibnamefont {Mei\ss{}ner}},
  \bibinfo {author} {\bibfnamefont {D.}~\bibnamefont {Milstead}}, \bibinfo
  {author} {\bibfnamefont {R.~E.}\ \bibnamefont {Mitchell}}, \bibinfo {author}
  {\bibfnamefont {K.}~\bibnamefont {M\"onig}}, \bibinfo {author} {\bibfnamefont
  {P.}~\bibnamefont {Molaro}}, \bibinfo {author} {\bibfnamefont
  {F.}~\bibnamefont {Moortgat}}, \bibinfo {author} {\bibfnamefont
  {M.}~\bibnamefont {Moskovic}}, \bibinfo {author} {\bibfnamefont
  {H.}~\bibnamefont {Murayama}}, \bibinfo {author} {\bibfnamefont
  {M.}~\bibnamefont {Narain}}, \bibinfo {author} {\bibfnamefont
  {P.}~\bibnamefont {Nason}}, \bibinfo {author} {\bibfnamefont
  {S.}~\bibnamefont {Navas}}, \bibinfo {author} {\bibfnamefont
  {M.}~\bibnamefont {Neubert}}, \bibinfo {author} {\bibfnamefont
  {P.}~\bibnamefont {Nevski}}, \bibinfo {author} {\bibfnamefont
  {Y.}~\bibnamefont {Nir}}, \bibinfo {author} {\bibfnamefont {K.~A.}\
  \bibnamefont {Olive}}, \bibinfo {author} {\bibfnamefont {S.}~\bibnamefont
  {Pagan~Griso}}, \bibinfo {author} {\bibfnamefont {J.}~\bibnamefont
  {Parsons}}, \bibinfo {author} {\bibfnamefont {C.}~\bibnamefont {Patrignani}},
  \bibinfo {author} {\bibfnamefont {J.~A.}\ \bibnamefont {Peacock}}, \bibinfo
  {author} {\bibfnamefont {M.}~\bibnamefont {Pennington}}, \bibinfo {author}
  {\bibfnamefont {S.~T.}\ \bibnamefont {Petcov}}, \bibinfo {author}
  {\bibfnamefont {V.~A.}\ \bibnamefont {Petrov}}, \bibinfo {author}
  {\bibfnamefont {E.}~\bibnamefont {Pianori}}, \bibinfo {author} {\bibfnamefont
  {A.}~\bibnamefont {Piepke}}, \bibinfo {author} {\bibfnamefont
  {A.}~\bibnamefont {Pomarol}}, \bibinfo {author} {\bibfnamefont
  {A.}~\bibnamefont {Quadt}}, \bibinfo {author} {\bibfnamefont
  {J.}~\bibnamefont {Rademacker}}, \bibinfo {author} {\bibfnamefont
  {G.}~\bibnamefont {Raffelt}}, \bibinfo {author} {\bibfnamefont {B.~N.}\
  \bibnamefont {Ratcliff}}, \bibinfo {author} {\bibfnamefont {P.}~\bibnamefont
  {Richardson}}, \bibinfo {author} {\bibfnamefont {A.}~\bibnamefont
  {Ringwald}}, \bibinfo {author} {\bibfnamefont {S.}~\bibnamefont {Roesler}},
  \bibinfo {author} {\bibfnamefont {S.}~\bibnamefont {Rolli}}, \bibinfo
  {author} {\bibfnamefont {A.}~\bibnamefont {Romaniouk}}, \bibinfo {author}
  {\bibfnamefont {L.~J.}\ \bibnamefont {Rosenberg}}, \bibinfo {author}
  {\bibfnamefont {J.~L.}\ \bibnamefont {Rosner}}, \bibinfo {author}
  {\bibfnamefont {G.}~\bibnamefont {Rybka}}, \bibinfo {author} {\bibfnamefont
  {R.~A.}\ \bibnamefont {Ryutin}}, \bibinfo {author} {\bibfnamefont {C.~T.}\
  \bibnamefont {Sachrajda}}, \bibinfo {author} {\bibfnamefont {Y.}~\bibnamefont
  {Sakai}}, \bibinfo {author} {\bibfnamefont {G.~P.}\ \bibnamefont {Salam}},
  \bibinfo {author} {\bibfnamefont {S.}~\bibnamefont {Sarkar}}, \bibinfo
  {author} {\bibfnamefont {F.}~\bibnamefont {Sauli}}, \bibinfo {author}
  {\bibfnamefont {O.}~\bibnamefont {Schneider}}, \bibinfo {author}
  {\bibfnamefont {K.}~\bibnamefont {Scholberg}}, \bibinfo {author}
  {\bibfnamefont {A.~J.}\ \bibnamefont {Schwartz}}, \bibinfo {author}
  {\bibfnamefont {D.}~\bibnamefont {Scott}}, \bibinfo {author} {\bibfnamefont
  {V.}~\bibnamefont {Sharma}}, \bibinfo {author} {\bibfnamefont {S.~R.}\
  \bibnamefont {Sharpe}}, \bibinfo {author} {\bibfnamefont {T.}~\bibnamefont
  {Shutt}}, \bibinfo {author} {\bibfnamefont {M.}~\bibnamefont {Silari}},
  \bibinfo {author} {\bibfnamefont {T.}~\bibnamefont {Sj\"ostrand}}, \bibinfo
  {author} {\bibfnamefont {P.}~\bibnamefont {Skands}}, \bibinfo {author}
  {\bibfnamefont {T.}~\bibnamefont {Skwarnicki}}, \bibinfo {author}
  {\bibfnamefont {J.~G.}\ \bibnamefont {Smith}}, \bibinfo {author}
  {\bibfnamefont {G.~F.}\ \bibnamefont {Smoot}}, \bibinfo {author}
  {\bibfnamefont {S.}~\bibnamefont {Spanier}}, \bibinfo {author} {\bibfnamefont
  {H.}~\bibnamefont {Spieler}}, \bibinfo {author} {\bibfnamefont
  {C.}~\bibnamefont {Spiering}}, \bibinfo {author} {\bibfnamefont
  {A.}~\bibnamefont {Stahl}}, \bibinfo {author} {\bibfnamefont {S.~L.}\
  \bibnamefont {Stone}}, \bibinfo {author} {\bibfnamefont {T.}~\bibnamefont
  {Sumiyoshi}}, \bibinfo {author} {\bibfnamefont {M.~J.}\ \bibnamefont
  {Syphers}}, \bibinfo {author} {\bibfnamefont {K.}~\bibnamefont {Terashi}},
  \bibinfo {author} {\bibfnamefont {J.}~\bibnamefont {Terning}}, \bibinfo
  {author} {\bibfnamefont {U.}~\bibnamefont {Thoma}}, \bibinfo {author}
  {\bibfnamefont {R.~S.}\ \bibnamefont {Thorne}}, \bibinfo {author}
  {\bibfnamefont {L.}~\bibnamefont {Tiator}}, \bibinfo {author} {\bibfnamefont
  {M.}~\bibnamefont {Titov}}, \bibinfo {author} {\bibfnamefont {N.~P.}\
  \bibnamefont {Tkachenko}}, \bibinfo {author} {\bibfnamefont {N.~A.}\
  \bibnamefont {T\"ornqvist}}, \bibinfo {author} {\bibfnamefont {D.~R.}\
  \bibnamefont {Tovey}}, \bibinfo {author} {\bibfnamefont {G.}~\bibnamefont
  {Valencia}}, \bibinfo {author} {\bibfnamefont {R.}~\bibnamefont {Van~de
  Water}}, \bibinfo {author} {\bibfnamefont {N.}~\bibnamefont {Varelas}},
  \bibinfo {author} {\bibfnamefont {G.}~\bibnamefont {Venanzoni}}, \bibinfo
  {author} {\bibfnamefont {L.}~\bibnamefont {Verde}}, \bibinfo {author}
  {\bibfnamefont {M.~G.}\ \bibnamefont {Vincter}}, \bibinfo {author}
  {\bibfnamefont {P.}~\bibnamefont {Vogel}}, \bibinfo {author} {\bibfnamefont
  {A.}~\bibnamefont {Vogt}}, \bibinfo {author} {\bibfnamefont {S.~P.}\
  \bibnamefont {Wakely}}, \bibinfo {author} {\bibfnamefont {W.}~\bibnamefont
  {Walkowiak}}, \bibinfo {author} {\bibfnamefont {C.~W.}\ \bibnamefont
  {Walter}}, \bibinfo {author} {\bibfnamefont {D.}~\bibnamefont {Wands}},
  \bibinfo {author} {\bibfnamefont {D.~R.}\ \bibnamefont {Ward}}, \bibinfo
  {author} {\bibfnamefont {M.~O.}\ \bibnamefont {Wascko}}, \bibinfo {author}
  {\bibfnamefont {G.}~\bibnamefont {Weiglein}}, \bibinfo {author}
  {\bibfnamefont {D.~H.}\ \bibnamefont {Weinberg}}, \bibinfo {author}
  {\bibfnamefont {E.~J.}\ \bibnamefont {Weinberg}}, \bibinfo {author}
  {\bibfnamefont {M.}~\bibnamefont {White}}, \bibinfo {author} {\bibfnamefont
  {L.~R.}\ \bibnamefont {Wiencke}}, \bibinfo {author} {\bibfnamefont
  {S.}~\bibnamefont {Willocq}}, \bibinfo {author} {\bibfnamefont {C.~G.}\
  \bibnamefont {Wohl}}, \bibinfo {author} {\bibfnamefont {J.}~\bibnamefont
  {Womersley}}, \bibinfo {author} {\bibfnamefont {C.~L.}\ \bibnamefont
  {Woody}}, \bibinfo {author} {\bibfnamefont {R.~L.}\ \bibnamefont {Workman}},
  \bibinfo {author} {\bibfnamefont {W.-M.}\ \bibnamefont {Yao}}, \bibinfo
  {author} {\bibfnamefont {G.~P.}\ \bibnamefont {Zeller}}, \bibinfo {author}
  {\bibfnamefont {O.~V.}\ \bibnamefont {Zenin}}, \bibinfo {author}
  {\bibfnamefont {R.-Y.}\ \bibnamefont {Zhu}}, \bibinfo {author} {\bibfnamefont
  {S.-L.}\ \bibnamefont {Zhu}}, \bibinfo {author} {\bibfnamefont
  {F.}~\bibnamefont {Zimmermann}}, \bibinfo {author} {\bibfnamefont {P.~A.}\
  \bibnamefont {Zyla}}, \bibinfo {author} {\bibfnamefont {J.}~\bibnamefont
  {Anderson}}, \bibinfo {author} {\bibfnamefont {L.}~\bibnamefont {Fuller}},
  \bibinfo {author} {\bibfnamefont {V.~S.}\ \bibnamefont {Lugovsky}}, \ and\
  \bibinfo {author} {\bibfnamefont {P.}~\bibnamefont {Schaffner}} (\bibinfo
  {collaboration} {Particle Data Group}),\ }\href {\doibase
  10.1103/PhysRevD.98.030001} {\bibfield  {journal} {\bibinfo  {journal} {Phys.
  Rev. D}\ }\textbf {\bibinfo {volume} {98}},\ \bibinfo {pages} {030001}
  (\bibinfo {year} {2018})}\BibitemShut {NoStop}%
\bibitem [{\citenamefont {Hoh}(2016)}]{Hoh:2016}%
  \BibitemOpen
  \bibfield  {author} {\bibinfo {author} {\bibfnamefont {F.}~\bibnamefont
  {Hoh}},\ }\href {\doibase 10.4236/jmp.2016.711115} {\bibfield  {journal}
  {\bibinfo  {journal} {Journal of Modern Physics}\ }\textbf {\bibinfo {volume}
  {7}},\ \bibinfo {pages} {36} (\bibinfo {year} {2016})}\BibitemShut {NoStop}%
\bibitem [{\citenamefont {Aad}\ \emph {et~al.}(2015)\citenamefont {Aad} \emph
  {et~al.}}]{Aad:2015zhl}%
  \BibitemOpen
  \bibfield  {author} {\bibinfo {author} {\bibfnamefont {G.}~\bibnamefont
  {Aad}} \emph {et~al.} (\bibinfo {collaboration} {ATLAS, CMS}),\ }\bibfield
  {booktitle} {\emph {\bibinfo {booktitle} {{Proceedings, Meeting of the APS
  Division of Particles and Fields (DPF 2015): Ann Arbor, Michigan, USA, 4-8
  Aug 2015}}},\ }\href {\doibase 10.1103/PhysRevLett.114.191803} {\bibfield
  {journal} {\bibinfo  {journal} {Phys. Rev. Lett.}\ }\textbf {\bibinfo
  {volume} {114}},\ \bibinfo {pages} {191803} (\bibinfo {year} {2015})},\
  \Eprint {http://arxiv.org/abs/1503.07589} {arXiv:1503.07589 [hep-ex]}
  \BibitemShut {NoStop}%
\end{thebibliography}%
%
%
%
%

%
%
%
%

\end{document}